\documentclass[12pt]{article}
\pdfoutput=1

\usepackage{geometry,amsmath,amsfonts}
\usepackage{slashed}
\usepackage{epsfig}
\usepackage{latexsym}
\usepackage{graphicx}
\usepackage{amssymb}
\usepackage[caption=false]{subfig}
\usepackage{color}
\usepackage{multirow}
\usepackage{color}
\usepackage{rotating}
\usepackage{ifthen}
\usepackage{epsfig}
\usepackage{lineno}
\usepackage[low-sup]{subdepth}
\numberwithin{equation}{section}
\usepackage[linktocpage=true]{hyperref}
\hypersetup{
colorlinks=true,%
linkcolor=[rgb]{0,0,1}, 
citecolor=[rgb]{0,0,1}
}

\newcommand{\beqs}{\begin{subequations}}
\newcommand{\eeqs}{\end{subequations}}
\newcommand{\beq}{\begin{equation}}
\newcommand{\eeq}{\end{equation}}
\newcommand{\bea}{\begin{eqnarray}}
\newcommand{\eea}{\end{eqnarray}}

\newcommand{\gsim}{\lower.7ex\hbox{$\;\stackrel{\textstyle>}{\sim}\;$}}
\newcommand{\lsim}{\lower.7ex\hbox{$\;\stackrel{\textstyle<}{\sim}\;$}}

\newcommand{\tb}{\tan\!\beta}

\newcommand{\ti}{\tilde}

\def\hs{\hspace*{0.3mm}}
\def\hsx{\hspace*{0.5mm}}
\def\hsm{\hspace*{-0.3mm}}

\def\to{\rightarrow}
\def\ito{\!\rightarrow\!}

\begin{document}


\thispagestyle{empty}

\begin{center}

{\large\bf  The hMSSM with a Light Gaugino/Higgsino Sector: }
\\[2mm]
{\large\bf  Implications for Collider and Astroparticle Physics }

\vspace*{9mm}

{\sc Giorgio Arcadi$\hs^{1}$},\,  
{\sc Abdelhak~Djouadi$\hs^{2,3}$},\, 
{\sc Hong-Jian He$\hs^{4,5}$},\, 
\\[2mm]
{\sc Jean-Loic Kneur$\hs^6$} \,and\,  
{\sc Rui-Qing Xiao$\hs^{4,7}$}

\vspace*{9mm}

{\small

$^1$ Dipartimento di Scienze Matematiche e Informatiche, Scienze Fisiche e Scienze della Terra, Universita degli Studi di Messina, Via Ferdinando Stagno d’Alcontres 31, I-98166 Messina, Italy 
\vspace{0.2cm}

$^2$ Centro Andaluz de F\'isica de Particulas Elementales \& Departamento de F\'isica Te\'orica y del Cosmos, Universidad de Granada, E--18071 Granada, Spain 
\vspace{0.2cm}

$^3$ NICPB, R{\"a}vala pst.\ 10, 10143 Tallinn, Estonia \vspace{0.2cm}

$^4$  T.~D.~Lee Institute \& School of Physics and Astronomy,
Key Laboratory for Particle Astrophysics and Cosmology\,(MOE),
Shanghai Jiao Tong University, Shanghai, China
\vspace{0.2cm}

$^5$ Institute of Modern Physics and Physics Department,
Tsinghua University, Beijing, China;\\
Center for High Energy Physics, Peking University,
Beijing, China
\vspace{0.2cm}

$^6$ Laboratoire Charles Coulomb (L2C), UMR 5221 CNRS-Universit\'e de Montpellier II, \\ F-34095 Montpellier, France
\vspace{0.2cm}

$^7$\,Department of Physics, King's College London, Strand, London WC2R 2LS, UK

}

\end{center}

\vspace*{4mm}

\begin{abstract}
\noindent 
The hMSSM is a special parameterization of the minimal supersymmetric extension of the Standard Model (MSSM) in which the mass of the lightest Higgs boson is automatically set to the LHC measured value, $M_h\!\!=\!\! 125$\,GeV, by adjusting the supersymmetric particle spectrum such that it provides the required amount of radiative corrections to the Higgs boson masses.\ The latter spectrum was in general assumed to be very heavy, as indicated by the present exclusion limits of the LHC, not to affect the phenomenology of the Higgs sector.\ In this work, we investigate the impact on the hMSSM by a light gaugino and higgsino sector, that is allowed by the present LHC data.\ In particular, we  discuss the radiative corrections due to charginos and neutralinos to the Higgs boson masses and couplings and show that an hMSSM can still be realized in this context.\ We first describe how this scenario is implemented in the package SuSpect that generates the MSSM Higgs and supersymmetric spectra.\ We then analyze the possible impact of Higgs boson decays into these new states, as well as the reverse cascade channels with Higgs bosons in the final states, for the constraints on the MSSM Higgs sector at the LHC.\ We further explore the cosmological constraints on the hMSSM with a light gaugino--higgsino spectrum.\ We analyze the relic abundance of the lightest neutralino as a candidate of the dark matter in the Universe and the constraints on its mass and couplings 
by the present and future astroparticle physics experiments.
\end{abstract}

\newpage
\tableofcontents

\section{Introduction} 
\label{sec:1}

The search for Higgs bosons, in addition to the one discovered by the ATLAS and CMS collaborations in 2012\,\cite{H-discovery} which completed the particle spectrum of the Standard Model (SM), is one of the main missions of the LHC experiments. Such particles are predicted in a plethora of extensions of the SM. This is particularly the case of supersymmetric theories \cite{SUSY,HaberKane} which address one of the main theoretical issues of the SM Higgs sector, namely the hierarchy problem and the instability of the Higgs boson mass against very high energy scales.  In its most economical version, the minimal supersymmetric SM (MSSM) \cite{HaberKane,pMSSM}, the theory requires the existence of two Higgs doublet fields that lead to five Higgs states in the particle spectrum: two CP-even $h$ and $H$, a CP-odd or pseudoscalar $A$ and two charged $H^\pm$ states \cite{HHG,Anatomy2}. While the $h$ boson is identified with the one which has been observed at the LHC and measured to have a mass of $M_h\!=\!125.09$\,GeV 
and SM-like couplings to the known fermions and weak gauge bosons\,\cite{H-couplings}, the other Higgs particles are expected to be heavy enough as to escape detection at the current stage of the experiments\,\cite{LHC-searches}. 


One of the attractive features of the MSSM is that, despite of its complexity, the Higgs sector can be described by two free parameters at the tree level, compared, for instance, to the 7 parameters needed in the case of a general two-Higgs doublet model\,\cite{2HDM}.\ This allows the model to have some predictability despite of its rich Higgs sector and to allow for rather straightforward phenomenological analyses.\ Unfortunately, such a simple picture is spoiled when radiative corrections, which turn out to be quite important in the Higgs sector \cite{Reviews-cor}, are taken into account.\  In this case, the numerous SUSY parameters enter the characterization of the Higgs sector and make any analysis a daunting task.\ One solution to ease the problem is to resort to benchmark scenarios that are representative of the phenomenology of the model, in which one keeps free the two basic input parameters and fixes all the additional ones entering the loop corrections.\ These benchmarks have been proposed quite early\,\cite{Benchmarks}\cite{Bagnaschi-WG} and have been used for a long time by theorists and experimentalists in MSSM Higgs studies, in particular before the discovery of the $h$ boson.      

One drawback of these benchmark scenarios is that, when varying the relevant input  parameters, the mass of the $h$ Higgs state which is extremely sensitive to radiative corrections has to also vary and might exceed significantly the experimental value of 
$M_h\! \simeq \!125$\,GeV. One way out of this problem, would be to enforce the $h$ mass to be equal to its true value and, hence, to use it as an input from the very beginning; this is the essence of the hMSSM scenario proposed a decade 
ago\,\cite{Habemus1}\cite{Habemus2}.\ 
This is made feasible by the fact that by far the dominant correction that enters in the MSSM Higgs sector and, hence, in the CP-even Higgs masses, appears in one single 
block\,\cite{Leading-cor}\cite{Leading-cor-high} 
that can be traded against the mass parameter $M_h^2$. 
Hence, to an approximation which has been shown to be 
very good\,\cite{Bagnaschi-WG}\cite{Habemus1}\cite{Habemus2} 
(see also Ref.\,\cite{Wagner+Lee}), one can still describe the radiatively corrected MSSM Higgs sector with only two parameters as at the tree level, while keeping the $h$ mass at its measured value.\  This is particularly true for very large masses of the superpartners of the SM fermions, in particular those of the top squark which have the largest Higgs couplings and hence contribute mostly to the loop corrections\footnote{For a recent different $M_h$ input setup in the MSSM, that does reduce the number of input parameters but rather determines semi-analytically the trilinear coupling $A_t(M_h)$, see Ref.\,\cite{h-inverse}.}.   

\vspace*{1mm}

The hMSSM is currently used by the ATLAS and CMS collaborations in their study of the MSSM Higgs sector and in setting the constraints on its parameter space from their searches of the heavier Higgs
states $H,A$ and $H^\pm$, in the various discovery channels.\ One implicit assumption of the hMSSM, though, is that all SUSY particles should be heavy enough (or should  couple very weakly) to these Higgs states.\ While this can be certainly true in the case of the strongly interacting superparticles, the squarks and the gluinos, which should be (have been) constrained to have masses above a few TeV (at least 2\,TeV) from the negative LHC 
searches\,\cite{Adam:2021rrw}, 
this is not the case of the weakly interacting superparticles.\  Nevertheless, for the ${\cal O}(1\,$TeV) masses that are being currently probed at the LHC, sleptons generate small contributions to the radiative corrections and have a limited impact on the MSSM Higgs sector in general.\ In turn, the charginos and neutralinos, the mixtures of the gauginos and higgsinos, the superpartners of the gauge and Higgs bosons, are expected to be lighter and might couple substantially to the Higgs particles. 

\vspace*{1mm}

One should therefore allow, in the context of the hMSSM,  for the description of the phenomenology of these particles and their interplay with the Higgs sector.\ This is particularly important as in the next Run\,II and the higher luminosity 
LHC runs\,\cite{HL-LHC}, a large data sample could be collected which would allow to have sensitivity to the channels in which the MSSM Higgs bosons could decay into charginos and neutralinos.\  In fact, this is also important for the reverse processes in which it is these weakly interacting superparticles that could decay into the various Higgs bosons, including the heavier Higgs states, which  will be closely studied by the up coming experiments.  


From a rather different perspective, another very interesting and attractive feature of supersymmetry in general, and the MSSM in particular, is that the lightest SUSY particle (LSP) identified with the lightest of the four neutralinos, can be made absolutely stable by virtue of a discrete symmetry called R-parity\,\cite{R-parity}.\  This particle might thus form the dark matter (DM) in the Universe and in fact,  for a long time it was thought as the best DM candidate\,\cite{DM-Drees}.\ If the sfermions are very heavy as indicated by current LHC data, the neutral MSSM Higgs states can serve as the privileged portals to the DM neutralino in a large area of the parameter space\,\cite{DM-PhysRep}.\ The hMSSM should therefore describe these astroparticle aspects too. 


Hence, it is necessary to extend the original hMSSM scenario in  order to cope with these interesting possibilities for collider searches in the context of an intertwined MSSM Higgs and gaugino-higgsino sectors, and to address the astrophysical issues in which the dark matter formed by the lightest neutralino interacts through the MSSM Higgs portal. This is the main scope of this paper.\ We will extend the hMSSM in which the Higgs sector is described as usual by two parameters, taken to be the CP-odd Higgs mass and the ratio of vacuum expectation values (vevs) of the two Higgs doublets fields $\tan\hsm\beta\hs$, and we link it to the chargino and neutralinos sectors which are  described by three additional parameters, namely the two soft-SUSY breaking gaugino mass parameters $M_1$ and $M_2$ and the (supersymmetric) higgsino mass parameter $\mu$  \cite{GunionHaber12}.\ The interplay of the two sectors of the theory will be explored in detail. 


After discussing how such a scenario is implemented in one of the major numerical packages that generate the supersymmetric spectrum, the program SuSpect \cite{Suspect}, we will analyze the impact of the radiative corrections of the supersymmetric sector to the Higgs boson masses and couplings, including the so-called direct corrections to the Higgs-fermion couplings.\ We will then analyze the contributions of these superparticles to loop induced Higgs decays, the impact of the chargino and neutralinos in MSSM Higgs decays and the decays of these SUSY states into Higgs bosons.\ The impact of these new channels on the LHC constraints on the Higgs sector will be highlighted.\ Finally, we perform an updated analysis of the phenomenology of the lightest neutralino DM in the hMSSM, study its relic abundance and the constraints and prospects for direct and indirect detection experiments and the complementarity of these searches in connection to those performed at colliders.

\vspace*{1mm}

The paper is organized as follows. In the section\,\ref{sec:2}, we present the hMSSM including the chargino and neutralino sector and its impact on the radiative corrections to the Higgs masses and couplings, and discuss how the model is implemented in the program SuSpect.\ In section\,\ref{sec:3}, we analyze the LHC constraints on the gaugino-higgsino sector and, in section\,\ref{sec:4}, the collider constraints on the Higgs sector when these new particles are included in the analyses.\ Section\,\ref{sec:5} studies systematically the dark matter implications in the hMSSM and how it is tested by the present astroparticle physics experiments.\ Finally, we present our conclusions in section\,\ref{sec:6}.


\section{Theoretical Setup}
\label{sec:2}
\vspace*{1.5mm}

\subsection{Formulation of the hMSSM}
\label{sec:2.1}
\vspace*{1.5mm}

In the MSSM, two doublets of complex scalar fields are necessary to spontaneously break the electroweak symmetry, leading to a spectrum of five Higgs states: two CP--even $h$ and $H$ bosons, a pseudoscalar $A$ boson and two charged $H^\pm$ bosons \cite{HHG,Anatomy2}. At tree-level, the Higgs sector can be simply described by two basic parameters, taken usually to be the ratio of vevs  $\tan\beta$ and the mass $M_A$ of the pseudoscalar state. For a large value of the latter,  $M_A \gg M_Z$, one is in the so called decoupling regime where the three heavy Higgs states are almost degenerate in mass while the lighter one reaches it maximal mass value, 
\beq
M_H \approx M_{H^\pm} \approx M_A~~~~~{\rm and}~~~~ M_h \simeq  M_Z|\cos2 \beta| <M_Z \, . 
\eeq
The Higgs couplings to fermions and gauge bosons are given in terms of $\tb$ and the angle $\alpha$ that diagonalises the $2\times 2$ mass matrix of the two CP--even $h$ and $H$ states
\begin{eqnarray}
{\cal M}^2 = \left[ \begin{array}{cc} 
M_Z^2 \cos^2\beta+ M_A^2 \sin^2\beta & -(M_Z^2+ M_A^2) \sin\beta \cos\beta \\ -(M_Z^2+ M_A^2) \sin \beta \cos \beta  & 
M_Z^2 \sin^2\beta+ M_A^2 \cos^2\beta \end{array} \right] \, . 
\label{mhiggs2-tree}
\end{eqnarray}

Compared to their SM values, the Yukawa couplings $g^\phi_f$ for isospin up- and down-type fermions and the reduced Higgs couplings to the $V=W,Z$ bosons are given by \cite{HHG,Anatomy2}
\begin{eqnarray}
g^h_u  =  \frac{\cos\alpha}{\sin\beta}\;,\quad 
g^h_d =  -\frac{\sin\alpha}{\cos\beta}\;, \quad 
g^h_V =  \sin(\beta -\alpha) \;, \nonumber \\ 
g^H_u  = ~ \frac{\sin\alpha}{\sin\beta}\;,\quad
g^H_d =   \frac{\cos\alpha}{\cos\beta}\;, \quad 
g^H_V =  \cos(\beta -\alpha) \;, \nonumber \\ 
g^A_u  = \cot\beta\;,  \quad ~~ g^A_d =  \tan\beta\;, \quad  ~g^A_V =0 \, .  \hspace*{1.9cm}
\label{eq:gfhlo}
\end{eqnarray}
In the decoupling limit $M_A \gg M_Z$ in which one has $\alpha \to \beta-\frac\pi2$, the $h$ couplings reduce to those of the SM Higgs boson, $g^h_X \to 1$ with $X=u,d,V$, while those of the $H$ state (as well as the ones of the two $H^\pm$ states) follow asymptotically the $A$ couplings, $g^H_X \to g_X^A$.  

This simple description of the MSSM Higgs sector with only two basic parameters is as well known spoiled by radiative corrections in which all MSSM parameters in principle enter \cite{Reviews-cor}. Nevertheless, the by far largest correction is contained into one single term: the correction to one of the elements of the CP-even Higgs mass matrix eq.~(\ref{mhiggs2-tree}), $\Delta {\cal M}_{22}^2$. The dominant contribution to this element, coming from the top-stop sector, is quadratic in the top quark mass or Yukawa coupling $\lambda_t\! = \! \sqrt 2 m_t/v \sin \beta$ and involves the logarithm of the SUSY scale, defined in terms of the two stop squark masses $M_S\! =\! \sqrt{  m_{\tilde t_1} m_{\tilde t_2} }$. In particular, in the decoupling limit, one has for this large correction at one-loop \cite{Leading-cor} 
\begin{eqnarray} 
\label{higgscorr}  
 \Delta {\cal M}_{22}^2 \sim \! 
\frac{3 \,\bar{m}_t^4}{\,2\pi^2 v^2\sin^2\!\beta\,} \left[ \log \frac{M_S^2}{\bar{m}_t^2} \!+ \! \frac{X_t^2}{M_S^2} 
\left(\! 1 \! - \! \frac{X_t^2}{\,12M_S^2\,} \!\right) \right]\!,
\end{eqnarray}  
where $X_{t}$ is the stop mixing parameter given, in terms of the trilinear coupling $A_t$ in the stop sector\footnote{To ease the discussion, we use in the expressions above and below, the simplification $X_t \approx A_t$ which occurs for sufficiently low $\mu$ and/or high $\tan\beta$ values.} and the higgsino mass parameter $\mu$, by  $X_t\!= \! A_t\! - \! \mu/\tb$  which maximizes the radiative corrections for the special value $X_t = \sqrt {6} M_S$; $\bar m_t$ is the running ${\rm \overline{MS}}$ top quark mass introduced to account for the leading two-loop radiative corrections in a renormalisation-group (RG) improved approach.  The two entries $\Delta {\cal M}_{11}^2$ and $\Delta {\cal M}_{12}^2$ of the radiatively corrected Higgs mass matrix extension of eq.~(\ref{mhiggs2-tree}) do not involve large ${\cal O}(m_t^4)$ radiative corrections and one has in general
\beq
\label{hMSSM-def}
\Delta {\cal M}_{11}^2,   \Delta {\cal M}_{ 12}^2 \ll \Delta {\cal M}_{22}^2 \, .  
\eeq


The non--leading corrections \cite{Reviews-cor} enter in all $\Delta {\cal M}_{ij}^2$ terms of the correction matrix such as those controlled by the $b$-quark Yukawa coupling $\lambda_b\! = \! m_b/v \cos\beta$ which at large values of $\tb$ becomes relevant, 
the corrections 
proportional to $\lambda_t^2$ or $\lambda_b^2$ or those originating from the gaugino sector which, in addition,  introduce  a dependence on the gaugino mass parameters $M_1,M_2,M_3$ and the higgsino mass parameter $\mu$. However, these contributions are much smaller than the leading one stemming from the top-stop contribution and they can be ignored in general \cite{Reviews-cor}. This will be explicitly discussed in section 2.3 for the corrections stemming from the gaugino-higgsino sector.  

In the approximation above, the maximal value of the $h$ boson mass is given by 
\beq
\label{Mh-max}
M_h^2 \to  M_Z^2 \cos^2 2 \beta + \sin^2\beta \Delta {\cal M}_{22}^2 \, ,
\eeq
close to the decoupling regime with $M_A\! \sim \mathcal{O}$(TeV) and the experimentally measured value $M_h \simeq 125$ GeV  is  obtained if the following conditions are met (see Ref.~\cite{Arbey} for instance):  have relatively high $\tb$ values, $\tb \gsim 5$, such that $\cos^2 2 \beta \approx 1$, have very heavy stop squarks, $M_S\!  \gsim \!1$--3 TeV to generate the large logarithmic corrections and, eventually, have a stop mixing parameter $X_t \approx \sqrt{6}M_S$ to be in the so--called ma\-xi\-mal  mixing scenario as to  maximize the stop loop contributions \cite{Benchmarks}. 

One should note that the system defined for instance in the approximation of eq.~(\ref{higgscorr}) is more constrained in the decoupling regime in which the $M_h$ limit of eq.~(\ref{Mh-max}) is obtained. Indeed,  for a given $\tb$ value and in a given scenario for stop mixing, such as the no stop mixing $X_t\!=\!0$ or maximal mixing $X_t/M_S\!=\! \sqrt 6$ scenarios, the lightest Higgs mass is related to $\log (M_S/\bar m_t)$.  In particular, to achieve the $h$ mass measured value $M_h\! \simeq \!125$ GeV, a very high SUSY scale, $M_S  \approx  500$ TeV, is required in the optimistic case of maximal mixing to comply with low $\tb$ values, $\tb  \approx  1$ \cite{Habemus2}. This SUSY scale reduces to $M_S \approx 1$ TeV for $\tb \gsim 10 $. In the no-mixing case, the resulting scale $M_S$ in this approximation should be raised by several orders of magnitude to allow for values $\tb \approx 1$ and raised by a factor of three, $M_S \approx 3$ TeV,  to allow for $\tb \gsim 10$.  

In any case, using
the eigenvalue equation for the $2\! \times \!2$ symmetric CP-even Higgs mass matrix ${\cal M}_{ij}$, one may always formally trade e.g. the full matrix element ${\cal M}^2_{22}$ by $M_h^2$ when the measured Higgs mass value $M_h \simeq 125$ GeV is properly taken into account. Consequently, provided that $\Delta {\cal M}_{11}^2, \Delta {\cal M}_{12}^2$ are sufficiently small, and more precisely \cite{Pietro} 
\beq
\Delta {\cal M}_{11}^2 \ll  {\cal M}_{11}^2|_{\rm tree} \ ,  \ \  \Delta {\cal M}_{12}^2 \ll  {\cal M}_{12}^2|_{\rm tree} \, , 
\label{condition11+12}
\eeq
$\Delta {\cal M}_{22}^2$ can be traded against $M_h$ and the MSSM Higgs  sector can be, as at tree--level, again described with only two free parameters such as $\tb$ and $M_A$ \cite{Habemus1,Habemus2,Habemus-early}. Indeed, one can then write 
\beq
\label{DelM22}
\Delta {\cal M}_{22}^2 = \frac{M_h^2(M_A^2+M_Z^2-M_h^2)-M_A^2 M_Z^2 \cos^2(2\beta)}
{M_Z^2 \cos^2\!\beta +M_A^2\sin^2\!\beta-M_h^2},
\eeq
with $ \Delta {\cal M}_{22}^2 $ being the full correction and not only the approximation given in eq.~(\ref{higgscorr}), and obtain the $H$ boson mass and the mixing angle $\alpha$, which will be simply given by 
\begin{eqnarray} 
M_{H}^2 &= & \frac{\,(M_{A}^2+M_{Z}^2-M_{h}^2)(M_{Z}^2\cos^2\!{\beta}
+M_{A}^2\sin^2\!{\beta}) - M_{A}^2 M_{Z}^2 \cos^2\!{2\beta}\,} 
{M_{Z}^2 \cos^2\!{\beta}+M_{A}^2 \sin^2\!{\beta} - M_{h}^2}\, , 
\nonumber \\
\ \ \  
\alpha &=& -\arctan\!\left(\frac{ (M_{Z}^2+M_{A}^2) 
\cos\!{\beta} \sin\!{\beta}} {M_{Z}^2\cos^2\!{\beta}+M_{A}^2 \sin^{2}\!{\beta} - M_{h}^2}\right)
\, , 
\label{eq:hMSSM} 
\end{eqnarray}
while the mass of the charged Higgs state, which is not significantly affected by radiative corrections, can be still 
expressed by the tree-level relation, namely \cite{H+mass}   \beq
M_{H^\pm} \simeq \sqrt { M_A^2 + M_W^2} \, .
\eeq

This approach, which was dubbed hMSSM in Ref.~\cite{Habemus1,Habemus2}, has been shown to provide a very good approximation of the MSSM Higgs sector when sfermions and in particular stop squarks, are rather heavy; see also the related analyses of Refs.~\cite{Bagnaschi-WG,Wagner+Lee,Pietro}.

In our study here,  we consider the usual hMSSM for the Higgs sector with only two inputs $\tan\beta$ and $M_A$ and assume the sfermions to be too heavy to have a phenomenological impact on it, i.e. they will be integrated out and decoupled from the low energy spectrum. To cope with the severe LHC bounds \cite{Adam:2021rrw}, the gluino mass parameter will be also assumed to be large, $m_{\tilde g} \approx M_3 \gsim 3\,$TeV. In turn, the gaugino parameters $M_1, M_2$ and the higgsino one $\mu$ that enter the  electroweak sector will be assumed to be in the few 100 GeV range, leading to charginos and neutralinos that could be accessible at the LHC. 

To describe their full impact on the Higgs sector, we would thus need five basic input parameters in our set-up. We will nevertheless assume some relation between the wino and bino mass parameters, defined at the electroweak scale, 
to reduce the number of inputs. Besides the GUT relation $M_1 \! \simeq \! \frac12 M_2$, we will mainly study the possibilities of wino and bino masses such that  $M_2\!=\! M_1$,  $M_2\! =\! 10 M_1$ and $M_2\! =\! \frac{1}{10} M_1$  that lead to  interesting phenomenological implications.   

\vspace*{1mm}
\subsection{Light Neutralinos and Charginos in the hMSSM} 
\label{sec:2.2}
\vspace*{1.5mm}

Let us now briefly introduce and discuss the  gaugino and higgsino sectors of the theory and summarize their interplay, when they involve a relatively light spectrum, with the MSSM Higgs sector.  The bino, the three winos and the four higgsinos  mix to generate the physical states, namely the four chargino $\chi^\pm_{1,2}$ and the four neutralino  $\chi^0_{1\!-\!4}$ Majorana particles. The lightest neutralino $\chi^0_{1}$ is in general (and definitely in our case) the lightest  SUSY particle or LSP which  is assumed to be stable and constitutes the dark matter candidate. The inputs in this sector are taken to be  $\mu, M_1, M_2$ together with $\tb$. While the real  parameter $\mu$ takes both signs, $M_2$ and $M_1$ are assumed to be real and positive. 

The chargino mass matrix, in terms of the input parameters above, simply reads \cite{HaberKane}
\begin{eqnarray}
{\cal M}_C = \left[ \begin{array}{cc} M_2 & \sqrt{2}M_W \sin \beta
\\[1.5mm]
\sqrt{2}M_W \cos\beta & \mu \end{array} \right] \, . 
\end{eqnarray}
The two chargino (and their CP-conjugate) eigenstates  $\chi_1^\pm$ and  $\chi_2^\pm$ as well as their masses are determined via a unitary transformation $U^* {\cal M}_C V^{-1} \!=\! {\rm diag} (m_{\chi_{1}^\pm}, m_{\chi_{2}^\pm})$ with $U,V$ being unitary matrices. In the interesting limit $|\mu| \gg M_2$, the lightest (heaviest) charginos correspond to pure winos (higgsinos) with masses $m_{\chi_{1}^\pm}\!  \simeq \! M_2$ ($m_{\chi_{2}^\pm} = |\mu|$). In the opposite  $M_2 \gg |\mu|$ limit,  the roles of the states $\chi_{1}^\pm$ and $\chi_{2}^\pm$ are simply reversed.

In the case of the neutralinos, the four-dimensional mass matrix  depends on the same three parameters above, namely $\mu$, $M_2$ and $\tb$,  and in addition, on the bino mass $M_1$.  In the basis $(-i\tilde{B}, -i\tilde{W}_3, \tilde{H}^0_1,$ $\tilde{H}^0_2)$, with  the mixing angles $\beta$ and $\theta_W$,  it has the form  \cite{HaberKane} 
\begin{eqnarray} 
\label{eq:chi-mass-matrix}
{\cal M}_N\! = \!\!\left[\!\!\! \begin{array}{cccc}
 M_1 & 0  & -M_Z \sin \theta_W \cos\beta & M_Z  \sin \theta_W \sin\beta \\
 0   & M_2  & M_Z \cos \theta_W \cos\beta & -M_Z  \cos \theta_W \sin\beta \\
 -M_Z \sin \theta_W \cos\beta  & M_Z  \cos \theta_W \cos\beta  & 0 & -\mu \\
 M_Z \sin \theta_W \sin\beta    & -M_Z  \sin \theta_W \sin\beta  & -\mu & 0
\end{array} \!\!\!\right]\!\!.~~~~
\end{eqnarray}
The neutralino eigenstates $\chi_{1,2,3,4}^0$ and their masses are determined with a transformation $Z^T {\cal M}_N Z^{ -1 } = {\rm diag} (m_{\chi_1^0}, m_{\chi_2^0}, m_{\chi_3^0}, m_{\chi_4^0})$ where, again $Z$ is  a unitary matrix. In this case also, for $|\mu| \gg M_2 \! > \! M_1$, two neutralinos will be pure gauginos with masses $m_{\chi_{1}^0}  \simeq M_1$ and $m_{\chi_{2}^0} \simeq M_2$, while the two others will be pure  higgsinos with masses $m_{\chi_{3}^0} \simeq m_{\chi_{4}^0} \simeq |\mu|$. In the opposite limit, the roles are again reversed and $m_{\chi_{1}^0} \simeq m_{\chi_{2}^0} \simeq |\mu|, m_{\chi_{3}^0} \simeq M_1$ and $m_{\chi_{4}^0} \simeq M_2$. When $M_2 \! < \! M_1$ the role of the two gauginos is reversed. 

The couplings of the neutralinos and charginos to the MSSM Higgs bosons, as well as the couplings to the massive gauge bosons, are given in terms of the matrices $U,V$ and $Z$ and we briefly summarize them below.  Denoting the Higgs bosons by $H_k$ with $k=1,2,3$, corresponding respectively to $H,h, A$, and $H_4=H^\pm$ and normalizing to the electric charge $e$, the
Higgs couplings to chargino and neutralino pairs can conveniently be written as  \cite{GunionHaber12}:
\begin{eqnarray}
g^{L,R}_{\chi^-_i \chi^+_j H_k} = g^{L,R}_{ijk} & {\rm with} & 
\Big\{\!\!
\begin{array}{l} 
g^L_{ijk}= \frac{1}{\sqrt{2}\sin\theta_W} \left[ e_k V_{j1}U_{i2}-d_k V_{j2}U_{i1}
\right] \\[2mm]
g^R_{ijk}= \frac{1}{\sqrt{2}\sin\theta_W} \left[ e_k V_{i1}U_{j2}-d_k V_{i2}U_{j1}
\right] \epsilon_k 
\end{array} , \\
g^{L,R}_{\chi^0_i \chi^0_j H_k} = g^{L,R}_{ijk} & {\rm with} & 
\Big\{\!\!
\begin{array}{l} 
g^L_{ijk} = \frac{1}{2 \sin\theta_W} \left( Z_{j2}- \tan\theta_W Z_{j1} \right) 
\left(e_k Z_{i3} + d_kZ_{i4} \right) \ + \ i \leftrightarrow j
\\[2mm]
g^R_{ijk} = \frac{1}{2 \sin\theta_W}  \left( Z_{j2}- \tan\theta_W Z_{j1} 
\right) \left(e_k Z_{i3} + d_kZ_{i4} \right) \epsilon_k \ + \ i 
\leftrightarrow j 
\end{array} , \ \ \nonumber \\
g^{L,R}_{\chi^0_i \chi^+_j H_4}= g^{L,R}_{ij4} & {\rm with} &
\Big\{\!\!
\begin{array}{l} 
g^L_{ij4} = \frac{\cos\beta}{\sin\theta_W} \big[ Z_{j4} V_{i1} + \frac{1}{\sqrt{2}} 
\left( Z_{j2} + \tan \theta_W Z_{j1} \right) V_{i2} \big] 
\\[2mm]
g^R_{ij4} = \frac{\sin\beta}{\sin\theta_W} \big[ Z_{j3} U_{i1} - \frac{1}{\sqrt{2}}
\left( Z_{j2} + \tan \theta_W Z_{j1} \right) U_{i2} \big]  
\label{cp:inos1}
\end{array} , \nonumber
\label{cp:inos}
\end{eqnarray}
with the convention $\epsilon_{1,2}=-\epsilon_3 =1$.  The coefficients $e_k$ and $d_k$ for a given $H_k$ state, together with their limiting values in the decoupling regime $M_A \gg M_Z$, are
\beqs 
\begin{align}
e_1^{} &=+ \cos \alpha \to \sin\beta ,~~ \ e_2=- \sin \alpha \to \cos\beta ,~~  \ e_3=- \sin\beta ,  
\\[2mm] 
d_1^{} &=  -\sin\alpha \to \cos\beta ,~~ 
\ d_2= -\cos\alpha \to \sin\beta ,~~  \ d_3= +  \cos\beta .
\label{ed-coefficients}
\end{align}
\eeqs 
Note that the Higgs couplings to pairs of the dark matter states $\chi_1^0$, recalling that $Z_{11}, Z_{12}$ ($Z_{13},Z_{14}$) are the gaugino (higgsino) components of the $Z$ matrix, vanish if the LSP neutralino is a pure gaugino or higgsino. This feature can be in fact  generalized to all the couplings of MSSM Higgs bosons to the various neutralinos and charginos: the Higgs bosons couple only to higgsino-gaugino mixtures or states and do not couple to pure gauginos or pure higgsinos. This makes that Higgs couplings to mixed heavy and light chargino/neutralino states are maximal in the pure gaugino or higgsino  regions, while the couplings involving only heavy or light gaugino or higgsino states are suppressed by powers of $M_i/|\mu|$ ($|\mu|/M_i$)  for $|\mu| \gg M_i$ ($|\mu| \ll M_i$). Note that some of the Higgs  couplings to neutralinos can also accidentally vanish outside the decoupling limit for certain values of $\tb$ and $M_A$ which enter in the coefficients $d_k$ and $e_k$ given previously.  

Finally, we will need the couplings of the charginos and neutralinos to the massive gauge bosons.   Using the same ingredients as above, they are given by \cite{GunionHaber12}
{ 
\begin{align}
g^L_{\chi^0_i \chi^+_j W} &=  \frac{c_W}{\sqrt{2}s_W} 
[-Z_{i4} V_{j2}+\sqrt{2}Z_{i2} V_{j1}] ,~~
& g^R_{ \chi^0_i \chi^+_j W} &=  \frac{c_W}{\sqrt{2}s_W} 
[Z_{i3} U_{j2}+ \sqrt{2} Z_{i2} U_{j1}] , 
\nonumber \\
g^L_{\chi^0_i \chi^0_j Z} &= - \frac{1}{2s_W} 
[Z_{i3} Z_{j3} - Z_{i4} Z_{j4}],~~
& g^R_{\chi^0_i \chi^0_j Z} &= + \frac{1}{2s_W} [Z_{i3} 
Z_{j3} - Z_{i4} Z_{j4} ] , 
\label{cp:V-gauginos}
\\
g^L_{\chi^-_i \chi^+_j Z} & = \frac{1}{c_W} \left[\delta_{ij}s_W^2 - \frac{1}{2} 
V_{i2} V_{j2} - V_{i1} V_{j1} \right]\! ,~~
& g^R_{\chi^-_i \chi^+_j Z} &= \frac{1}{c_W} \left[\delta_{ij}s_W^2 - \frac{1}{2} 
U_{i2} U_{j2} - U_{i1} U_{j1} \right]\! .~~~~ 
\nonumber
\end{align}
}
In contrast to  the couplings of the Higgs bosons,  the gauge boson couplings to charginos and neutralinos are important only for higgsino-- or gaugino--like states. 

With all these elements posed, we can now  study the impact of the charginos and neutralinos to the MSSM Higgs sector, starting with their contributions to the radiative corrections to the Higgs mass matrix eq.~(\ref{mhiggs2-tree}) that we discuss in the next subsection\footnote{Strictly speaking,  one should also include the contributions of the pure electroweak and Higgs corrections to this mass matrix. Nevertheless, these corrections are rather small and can be safely neglected, and we will not consider them here.}. 

\subsection{Radiative Corrections to the Higgs Boson Masses}
\label{sec:2.3}
The gaugino and higgsino states enter the radiative corrections to the MSSM Higgs sector and, in particular, contribute to the CP-even Higgs mass matrix elements  ${\cal M}^2_{ij}$. However, contrary to squarks which mainly contribute to the entry $\Delta {\cal M}^2_{22}$ as discussed in section 2.1, these will more equally contribute to all entries. {With the defined hMSSM prescription given in the previous subsection, the chargino/neutralino contributions $\Delta {\cal M}^2_{22}|_\chi$ are implicitly entering the total contribution which is traded against $M_h$ as done in eq.~(\ref{DelM22}), and one needs in principle to insure that eq.~(\ref{hMSSM-def}) is verified. 
 and strictly speaking, for the trade of eq.~(\ref{DelM22}) to be valid, the entries $\Delta {\cal M}^2_{11}|_\chi$ and $\Delta {\cal M}^2_{12}|_\chi$ should be small as compared to the corresponding tree-level values, eq.~(\ref{condition11+12}). 
However there are two caveats with this option. A first one is that at high $\tan\beta$, the entry ${\cal M}_{12}^2|_{\rm tree}$ vanishes\footnote{In fact, this problem also occurs in the case of the dominant top-stop loop contributions. However, the contributions to ${\cal M}_{12}^2$ scale like $\mu A_t/M_S^2$ and are small for very large $M_S$ values or, eventually, for small $\mu$ values. We thank Pietro Slavich for pointing this peculiarity to us.}, while it is not necessarily the case for $\Delta {\cal M}^2_{12}|_\chi$. Another problem is that, in principle, we do not have a concrete measure of how $\Delta {\cal M}^2_{11}|_\chi$ and $\Delta {\cal M}^2_{12}|_\chi)$ should be small as compared to the tree-level values, to have a result that we can consider to be accurate, especially if some of them are accidentally small. {This is particularly delicate at small $\tan\beta$ values since there is no tree  level contribution to $M_h^2$, as e.g. shown in eq.~(\ref{Mh-max}), and the $h$ mass is entirely generated by radiative corrections}. 


{Accordingly,
another possible approach, that we will also investigate in this paper, is to evaluate the impact of the full radiative corrections stemming from charginos and neutralinos to the MSSM Higgs sector and check that they are small enough to be ignored. 
The chargino/neutralino radiative contributions will act as perturbations which will modify the original hMSSM relations. In particular, these will give rise to extra contributions to the mass $M_h$ and, hence, modify its input value in the hMSSM, $M_h\!=\!125$ GeV. Nevertheless, if these additional corrections are at the level of a few percent at most, they could be acceptable since, as is well known, there is an intrinsic uncertainty in the determination of $M_h$ in the MSSM which is a result of experimental and theoretical errors\footnote{The experimental errors are mostly due to the uncertainties on the values of the top quark mass $m_t$ and the strong coupling constant $\alpha_s$ while the theoretical ones come from missing higher order effects.}. This uncertainty  has been estimated to be of the order of $\Delta M_h \simeq \pm 3$ GeV \cite{Reviews-cor}. } 

The additional radiative corrections will also modify the values of the parameters $\alpha$ and $M_H$ obtained through eq.\,(\ref{eq:hMSSM}) and we also need to check that the  modifications are rather modest and do not exceed the few percent level. We should note that apart from the corrections that are due to charginos and neutralinos, other radiative contributions from gauge plus heavy Higgs bosons,  which are very moderate,  will be neglected here.

\vspace*{1mm}



 In the second option with {a modified prescription with respect to the strict hMSSM definition given previously}, we incorporate within the CP-even Higgs mass matrix, on top of eq.~(\ref{DelM22}), the contributions of neutralinos and charginos in the 
 relevant matrix elements $\Delta M^2_{ij}$ with exact one-loop expressions, as given e.g. in Ref.~\cite{BPMZ}, and we recalculate consistently from this corrected mass matrix,  the CP-even Higgs boson pole masses.
More precisely, we consider the mass eigenvalues as obtained from the corrected CP-even Higgs squared mass matrix:
\begin{eqnarray}
{\cal M}^2 (p^2) \!=\! \left[ \begin{array}{cc}   M_Z^2 c^2_\beta\!+\!  M_A^2 s^2_\beta \!+\! \Delta M^2_{11}|_\chi (p^2)  & -(M_Z^2\!+\! M_A^2) s_\beta c_\beta \!+\! \Delta M^2_{12}|_\chi (p^2) \\ \!-\!(M_Z^2\!+\! M_A^2) s_\beta c_\beta \!+\! \Delta M^2_{21}|_\chi (p^2)  & M_Z^2 s^2_\beta\!+\! M_A^2 c^2_\beta +\! \Delta M^2_{22} \!+\! \Delta M^2_{22}|_\chi (p^2)  \end{array} \right]\! ,~ 
\label{mhiggs2}
\end{eqnarray}
where we have used the abbreviations; $s_\beta\!=\!\sin\beta$, $c_\beta\!=\!\cos\beta$. 
$\Delta M^2_{22}$ is given by eq.~(\ref{DelM22}), 
while the neutralino and chargino contributions entering  $\Delta M^2_{ij}|_\chi$ are given by
\begin{eqnarray}
\Delta M^2_{11}|_\chi (p^2) &=& -\Pi^{\chi}_{s_1 s_1}(p^2)+t_1/v_1
+c_\beta^2 \Delta M_Z^2 + s_\beta^2 \Delta M_A^2 \, , \nonumber \\
\Delta M^2_{22}|_\chi (p^2) &=& -\Pi^{\chi}_{s_2 s_2}(p^2)+t_2/v_2
+s_\beta^2 \Delta M_Z^2 + c_\beta^2 \Delta M_A^2 \, , \nonumber \\
\Delta M^2_{12}|_\chi (p^2) &=& -\Pi^{\chi}_{s_1 s_2}(p^2) -
(\Delta M_Z^2 + \Delta M_A^2) s_\beta c_\beta \, , 
\end{eqnarray}
with $\Delta M_Z^2$ and $\Delta M_A^2$ the corrections to the $Z,A$ masses, given in terms of the running masses $\overline M_Z$ and  $\overline M_A$, defined at momentum squared $p^2\!=\! M_Z^2$ and $M_A^2$ respectively, by
\begin{eqnarray}
\Delta M_Z^2 &=& \overline M_Z^2 -M_Z^2= \Pi_{ZZ}(p^2) \, , \nonumber \\
\Delta M_A^2 &=& \overline M_A^2 -M_A^2= \Pi_{AA}(M_A^2) -s_\beta^2 t_1/v_1- c_\beta^2 t_2/v_2 \, .  
\end{eqnarray}
At a given squared momentum transfer of $p^2$, one explicitly has 
\bea
\hspace*{-7mm}
(4\pi)^2 \Pi^{\chi}_{s_k s_l}(p^2) \!\!\!&=&\!\!\!
\frac{1}{2} \sum_{i,j=1}^4 
\left[ f^0_{ij s_{kl}} G(p^2,m_{\chi_i^0}, m_{\chi_j^0})
\!-\!2 g^0_{ij s_{kl} } m_{\chi_i^0} m_{\chi_j^0} B_0(p^2,m_{ \chi_i^0},m_{ \chi_j^0}) \right] \ \ \ \ \nonumber\\
\!\!\!&&\!\!\! +\sum_{i,j\!=\!1}^2 
\left[ f^+_{ij s_{kl}} G(p^2,m_{\chi_i^+}, m_{\chi_j^+})\!-\!2 g^+_{ij s_{kl} } m_{\chi_i^+} m_{\chi_j^+} B_0(p^2,m_{ \chi_i^+},m_{\chi_j^+}) \right]\!\!,~~~~
\label{NCself}
\eea
for the Higgs two-point functions or self-energies, and for the tadpole {$t_1^{}/v_1^{}\,$,}
\bea
\hspace*{-4mm}
(4\pi)^2 \frac{t_1}{v_1} \!=\! \frac{-g_2^2}{M_W c_\beta}\!\!
 \left[\sum_{i\!=\!1}^4\!\! Z_{i3}(Z_{i2}\!-\!\tan\theta_W Z_{i1}) 
m_{\chi_i^0} A_0(m_{\chi_i^0}) \right. 
\left. \!+\!\sqrt{2} \sum_{i\!=\!1}^2\!  V_{i1} U_{i2}
m_{\chi_i^+} A_0(m_{\chi_i^+})\!\right]\!\!.~
\eea
and a similar expression for $t_2/v_2$ with the following replacements $Z_{i3}\to Z_{i4}$,  $V_{i1} \to V_{i2}$, {$c_\beta\to s_\beta$}.\ 
In the above, 
$g_2^{}$ is the SU(2) coupling constant and we have used the function 
\beq
G(p^2,m_1,m_2) = (p^2-m_1^2-m_2^2) B_0(p^2,m_1,m_2)-A_0(m_1)-A_0(m_2)\,, 
\eeq
where $B_0$ and $A_0$  are the standard Passarino-Veltman \cite{PV} two- and one-point loop functions, respectively. In eq.(\ref{NCself}), the relevant couplings in the $(s_1^{}, s_2^{})$ basis are 
\begin{eqnarray}
f_{ij s_{kl}}=& g^L_{\chi_i \chi_j s_k} g^L_{\chi_i \chi_j s_l}
+ g^R_{\chi_i \chi_j s_k} g^R_{\chi_i \chi_j s_l}\;, \nonumber \\
g_{ij s_{kl}}=& g^L_{\chi_i \chi_j s_k} g^R_{\chi_i \chi_j s_l}
+ g^R_{\chi_i \chi_j s_k} g^L_{\chi_i \chi_j s_l} , 
\label{slk2LR}
\end{eqnarray}
where in terms of the physical Higgs boson couplings of eq.(\ref{cp:inos}), one has
\bea
&g^{L,R}_{\chi_i \chi_j s_1}=
  \cos\alpha\, g^{L,R}_{\chi_i \chi_j H_1}
  -\sin\alpha\, g^{L,R}_{\chi_i \chi_j H_2} , \nonumber \\ 
&g^{L,R}_{\chi_i \chi_j s_2} =
\sin\alpha\, g^{L,R}_{\chi_i \chi_j H_1}
  +\cos\alpha\, g^{L,R}_{\chi_i \chi_j H_2} .
\eea

Similarly, the chargino and neutralino contributions to the $\Pi_{AA}$ self-energy {is also given by eq.~(\ref{NCself})} 
with couplings $ f^0_{ij A}, g^0_{ij A}$ and $f^+_{ij A}, g^+_{ij A}$ defined similarly as in eq.~(\ref{slk2LR}). 

\vspace*{1mm} 

We can now move to the quantitative evaluation of these corrections and the discussion of their impact on the hMSSM. For this, we fix $\tan\beta$ to two representative values: a high $\tan\beta=30$ and a relatively low $\tan\beta=3$ value, and  perform scans on the additional higgsino and wino mass parameters $\mu$ and $M_2$ in the range between 0 and 3 TeV, allowing for both signs of $\mu$. The bino mass parameter $M_1$ will  be related to $M_2$, $M_1\!=\! M_2$ or $\frac{1}{10} M_2$,   while the gluino mass parameter $M_3$ is fixed such that the gluino mass is 
$m_{\tilde g} \! = \! 3$\,TeV.

\vspace*{1mm}

We will use the program Suspect in which the hMSSM prescriptions have  been implemented. Since it is necessary to specify all parameters in the program and since we kept the SUSY scale $M_S$ as an input in the code,  we need to fix its value. Indeed, although eq.(\ref{DelM22})  does not depend on $M_S$ explicitly, in such an hMSSM prescription and as was already mentioned, one should take $M_S$ values such that eq.(\ref{DelM22}) and eq.(\ref{higgscorr})  are at least approximately consistent. This implicitly sets the sfermion scale $M_S$ to depend on $\tan\beta$ choices.\ 
 In the illustrations to be given below, 
we will choose the sample scenarios 
of $M_S\! \sim \! 10$\,TeV for $\tan\!\beta\!=\!30$ 
(corresponding to a small mixing)
and of $M_S \! \sim \! 100$\,TeV 
for $\tan\!\beta\!=\!3$ (corresponding to a large mixing),
respectively.

In addition, since we are considering only the corrections from gauginos and higgsinos,  the electroweak symmetry breaking scale will be taken to be $\,Q^2_{\rm EWSB}= \frac12 (\mu^2+M_2^2)$. 
Finally, the pseudoscalar Higgs mass  $M_A$ will be fixed to $M_A=1$\,TeV.\  
It is interesting to note that the actual value of $M_A$ used in eq.~(\ref{DelM22}) is almost irrelevant for moderate and large $\tan\beta$ and sufficiently large $M_A$. Indeed,  one would have in this case
\beq
\Delta M^2_{22} \simeq M^2_h-M^2_Z + [M^2_h+3 M^2_Z + {\cal O}(M^4_Z/M^2_A)]/ {\tan^2\!\beta} +{\cal O}(1/\tan^3\!\beta) \, ,
\eeq
which remains a very good approximation, only departing by a few percent at most from eq.~(\ref{DelM22}), even for moderate $\tan\beta \gtrsim 3$ and $M_A \gtrsim 2 M_h$ values.

\vspace*{1mm}

At a first step, we should verify that the radiative corrections  are small enough in order that the defining assumptions of the hMSSM  remain valid.  This is in principle insured by requiring that the two corrections $\Delta M^2_{11}|_\chi$ and $ \Delta M^2_{12}|_\chi$ are much smaller than the corresponding tree--level values, eq.~(\ref{condition11+12}).

\vspace*{1mm}

This is illustrated in Fig.\ref{DMij} where the two relative contributions $\Delta M^2_{11}|_\chi/ M^2_{11}|_{\rm tree}$ (left panel) and  $\Delta M^2_{12}|_\chi/ M^2_{12}|_{\rm tree}$ (right panel) are shown as a function of $\mu$ for the two values of $\tan\!\beta$ that we have adopted, $\tan\!\beta=30$ and $ \tan\!\beta=3$. Here, a scan is performed on the gaugino and Higgsino mass parameters 
$(M_2, \mu)$, assuming a bino mass parameter that is related by  
$M_2=M_1$ for $\tan\beta=30$ and $M_2=10 M_1$ for $\tan\beta=3$; the corrections only slightly differ for other values of the ratio $M_1/M_2$ as will be observed shortly. As can be seen, for these two $\tan\beta$ values, the relative corrections are small as the ratios $\Delta M^2_{1i}|_\chi/ M^2_{1i}|_{\rm tree}$ for 
$i\!=\!\!1,2$ remain below 1\% in absolute value in the entire scanned parameter space.

\begin{figure}[!h]
\vspace*{-2mm}
\centerline{
~\epsfig{figure=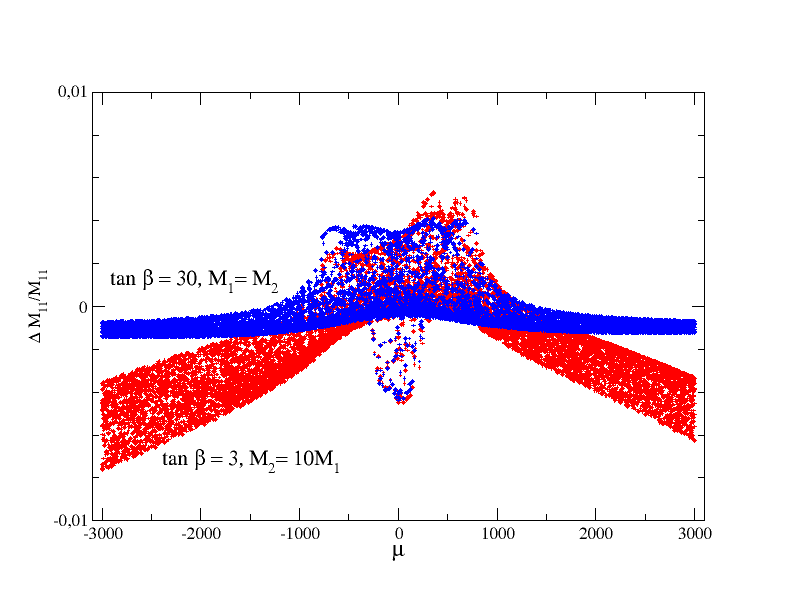,width=8.6cm} \hspace*{-10mm}
\epsfig{figure=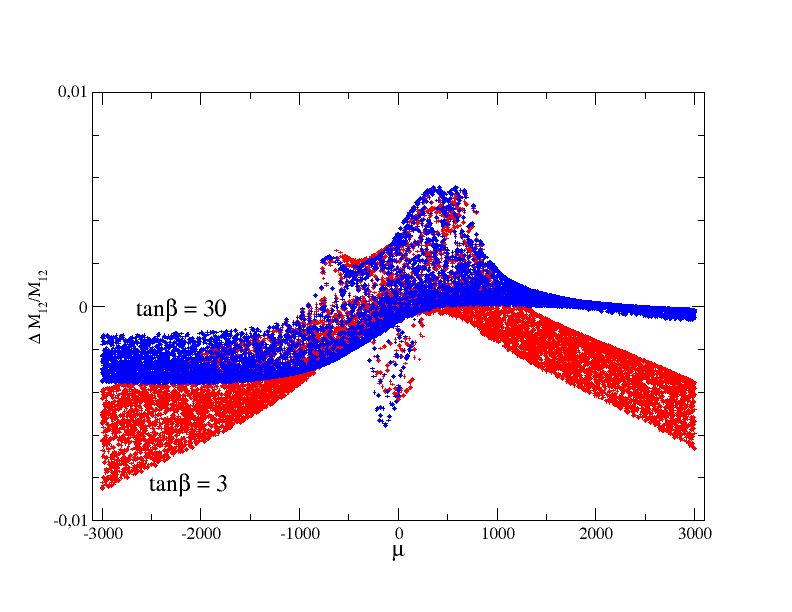,width=8.6cm}
}
\vspace*{-5mm}
\caption{
The relative contributions $\Delta M^2_{11}|_\chi/M^2_{11}|_{\rm tree}$ (left), $\Delta M^2_{12}|_\chi/M^2_{12}|_{\rm tree}$ (right) as a function of the higgsino mass $\mu$ 
for $\tan\beta =30$, $M_2=M_1$ (blue dots) and $\tan\beta =3$, $M_2=10M_1$ (red dots) when a scan on the parameters $M_2, \mu$ is performed.} 
\label{DMij}
\end{figure}

Hence, according to the discussion above, one would thus expect very moderate deviations from the hMSSM  when the chargino/neutralino radiative corrections are taken into account: the correction $\Delta M^2_{22} |_\chi$ is by definition  implicitly included in the complete correction $\Delta M^2_{22}$ in eq.~(\ref{mhiggs2}), which is traded against the value of $M_h$, eq.~(\ref{DelM22}), while the corrections $\Delta M^2_{11}|_\chi$ and $\Delta M^2_{12}|_\chi$ will act as small perturbations which will marginally modify the input value of the $h$ boson mass $M_h=125$ GeV. 

To check this explicitly, we perform  scans over the relevant  parameters $\mu$ and $M_2$, for the two representative choices of $\tan\beta$ previously used, $\tan\beta= 3$ and $30$, to illustrate the impact of the chargino-neutralino corrections on the value of $M_h$. We do this first in the context of the strict hMSSM in which the contribution of the correction  $\Delta M^2_{22} |_\chi$ is by definition simply included in the full radiative correction $\Delta M^2_{22}$ in  eq.~(\ref{mhiggs2}). 

The outcome is shown in Fig.~\ref{Fig:scan-hMSSM+} where the recalculated $h$ boson mass for the two values  $\tan\beta=3$ (red points) and  $\tan\beta=30$ (blue points) as  functions of $\mu$ (left panel) and $M_2$ (right panel) is displayed; the bino mass parameter is set to $M_1= \frac{1}{10}M_2$ in the first case and to $M_1= M_2$ in the second one. 

One can see that the deviation from the input value $M_h=125$ GeV is rather small in the case of $\tan\beta=30$, less than a hundred MeV, but it can be much larger in the case of $\tan\beta=3$. Indeed, while the corrections modify the value of $M_h$ by less than 2 GeV at $|\mu|$ and $M_2$ values smaller than 1 TeV, they are positive at large values of the latter two parameters and, for $|\mu| \approx 3$ TeV or $M_2 \approx 2$ TeV, the corrections  increase the value of $M_h$ by slightly more than 3 GeV.   

\begin{figure}[!h]
\vspace*{-5mm}
\centerline{
~\epsfig{figure=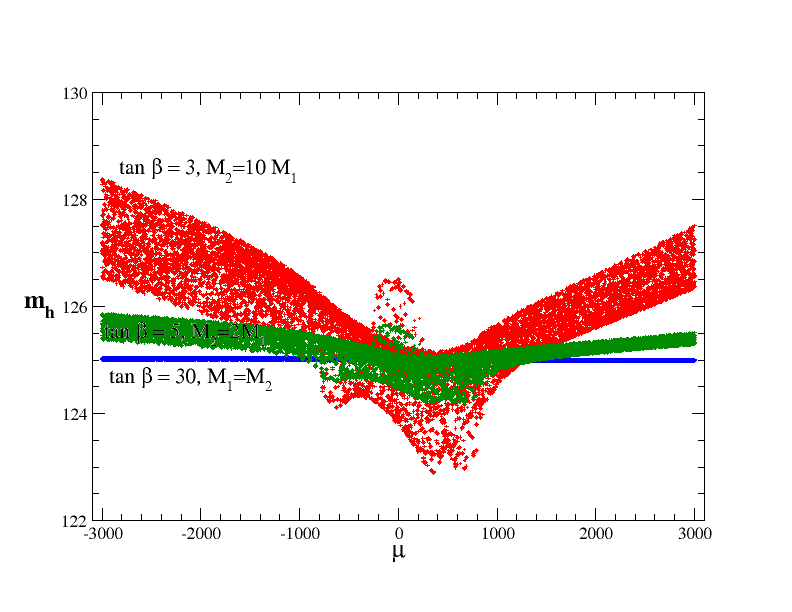,width=8.6cm} \hspace*{-10mm}
\epsfig{figure=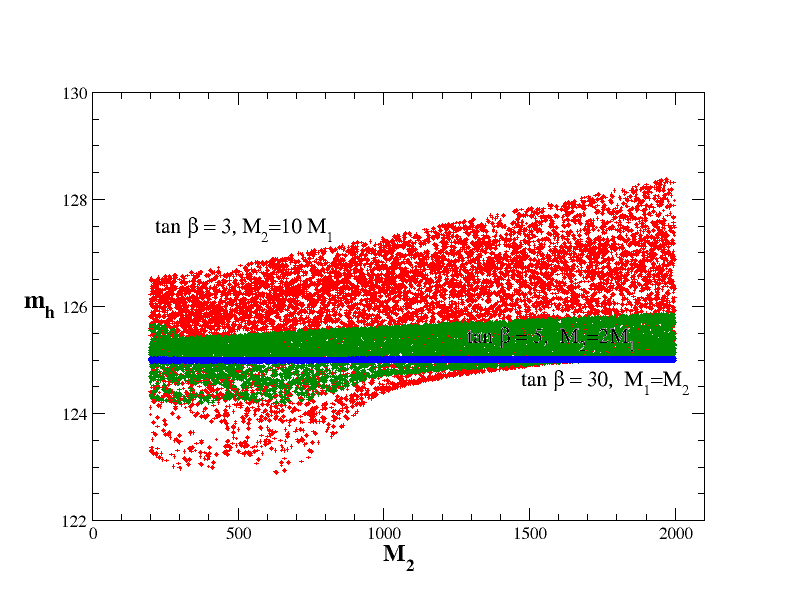,width=8.6cm}
}
\vspace*{-5mm}
\caption{
The recalculated mass $M_h$ in the conventional hMSSM, with chargino and neutralino radiative corrections added for $\tan\beta\!=\!3$, $M_1\! =\! M_2/10$ (red points) and $\tan\beta \! =\! 30$,
$M_1\! =\! M_2$ (blue points) as a function of $\mu$ (left) and as a function of $M_2$ (right) when scanning on $M_2, \mu$. We have added the case $\tan\beta \! =\! 5$ with $M_2\! =\! 2M_1$ in green.}
\label{Fig:scan-hMSSM+}
\end{figure}



In fact,  one should note that for $M_A^2 \gg M^2_Z$, and at first order in $\Delta M^2_{ij}|_\chi$, one has
\beq 
\Delta M^2_h \simeq  c_\beta^2 \Delta M^2_{11}|_\chi+ 2s_\beta c_\beta \Delta M^2_{12}|_\chi + {\cal O} (M_Z^2/M_A^2) \, ,  
\label{DMijlargeMA}
\eeq
so that the separate chargino neutralino contributions $\Delta M^2_{11}|_\chi$ and $\Delta M^2_{12}|_\chi$ happen to be further suppressed or enhanced for respectively large and small $\tan\beta$ values. For $\tan\beta=5$ for instance, the deviation in $M_h$  does not exceed 1 GeV for all values of $\mu, M_2$ in the selected range as is explicitly shown by the additional green points of Fig.~\ref{Fig:scan-hMSSM+}. 

It turns out that the enhancement at small $\tan\beta$ values observed above is partly compensated by corresponding  $\Delta M^2_{22}|_\chi $ contributions when they are not included in the entire correction  $\Delta M^2_{22}$ which, in the hMSSM, is then expressed in terms of the physical masses $M_h$ and $M_A$. These contributions will add a factor $s_\beta^2 \Delta M^2_{22}|_\chi$ to eq.~(\ref{DMijlargeMA}), giving 
a further correction to the input mass $M_h$. 

Hence, by slightly modifying the hMSSM prescription and by assuming that the additional chargino/neutralino correction $\Delta M^2_{22}|_\chi $ is treated separately as in eq.~(\ref{mhiggs2}), therefore added to  $\Delta M^2_{11,12}|_\chi$
that modify $M_h$, the entire chargino/neutralino extra contributions become moderate also for small $\tan\beta$ values.
This is exemplified  in Fig.~\ref{Fig:scan-hMSSM-} where the same exercise that led to 
Fig.~\ref{Fig:scan-hMSSM+} is repeated but this time, when the contribution with $\Delta M^2_{22}|_\chi$ is added on top of $\Delta M^2_{22}$ and, hence, also enters the corrections that modify the value of $M_h$. This is, in  principle, an implicit double counting of the contribution $\Delta M^2_{22}|_\chi$, but it 
allows cancellations against the $\Delta M^2_{11,12}|_\chi$ contributions
and it is acceptable in practice as we have $\Delta M^2_{22}|_\chi \ll \Delta M^2_{22}$.  

\begin{figure}[!h]
\vspace*{-5mm}
\centerline{
~\epsfig{figure=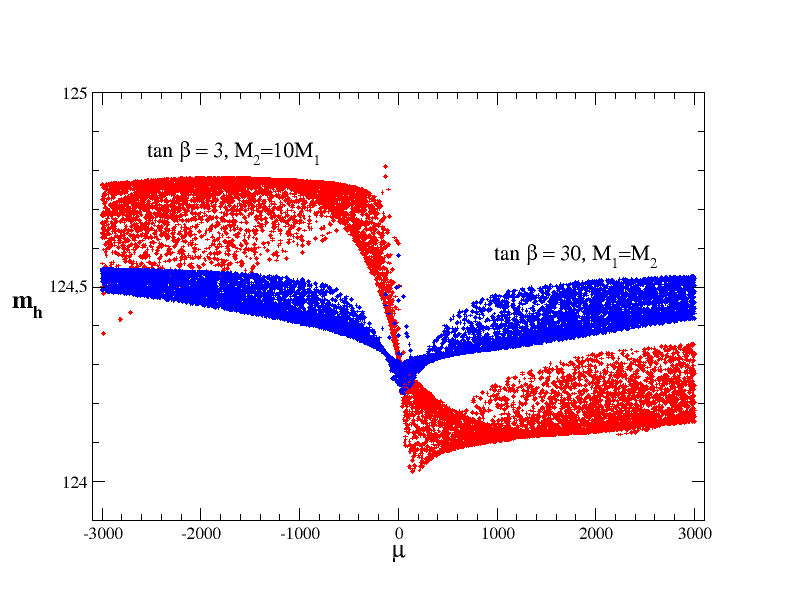,width=8.6cm}\hspace*{-8mm}
\epsfig{figure=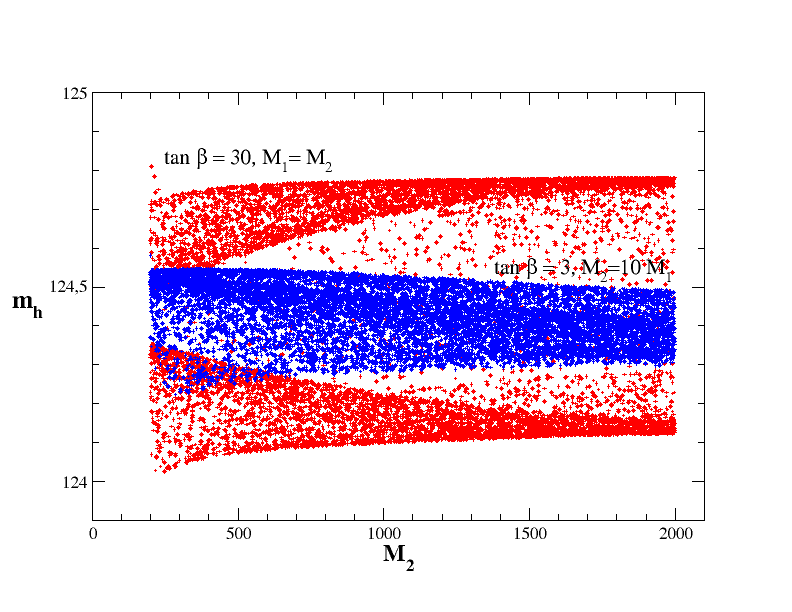,width=8.6cm}
}
\vspace*{-6mm}
\caption{
The recalculated lightest CP-even $h$ mass $M_h$ in the slightly modified hMSSM prescription with $\Delta M^2_{22}|_\chi$ not included in $\Delta M^2_{22}$, with chargino and neutralino radiative corrections added for $\tan\beta\!=\!3, M_1\!=\!\frac{1}{10} M_2$ (red points) and $\tan\beta\!=\!30, M_1=M_2$ (blue points) as a function of $\mu$ (left) and as a function of $M_2$ (right) when scanning on the mass parameters $\mu, M_2$.}
\label{Fig:scan-hMSSM-}
\end{figure}

As it can be seen, the chargino plus neutralino contributions to the lightest Higgs mass remain rather moderate for most of the parameter space and stay below the present theoretical uncertainty on $M_h$, namely  $\Delta M_h \simeq 3$ GeV.   The deviations reach a maximum} of $\approx -1\;$GeV for, not too surprisingly, very low $|\mu|$ and $M_2$ values (close to the 100 GeV limit which is excluded by LEP2 data/searches as will be seen later), and another maximum of $\approx 0.5$ GeV for large and negative $|\mu|$ values (as they depend on the sign of this parameter) as well as large $M_2$ values.  The latter is explained by non-decoupling logarithmic dependencies $\propto {\rm log}( m_{\chi^+_i}/M_h), {\rm log}( m_{\chi^0_i}/M_h)$ that remain moderate as long as the gaugino spectrum is not too heavy. These deviations are not  very sensitive to the $M_2/M_1$ ratio nor to the different values of $\tan\beta$ as long as $\tan\beta \gsim 5$.

\begin{figure}[!h]
\vspace*{-1mm}
\centerline{
\epsfig{figure=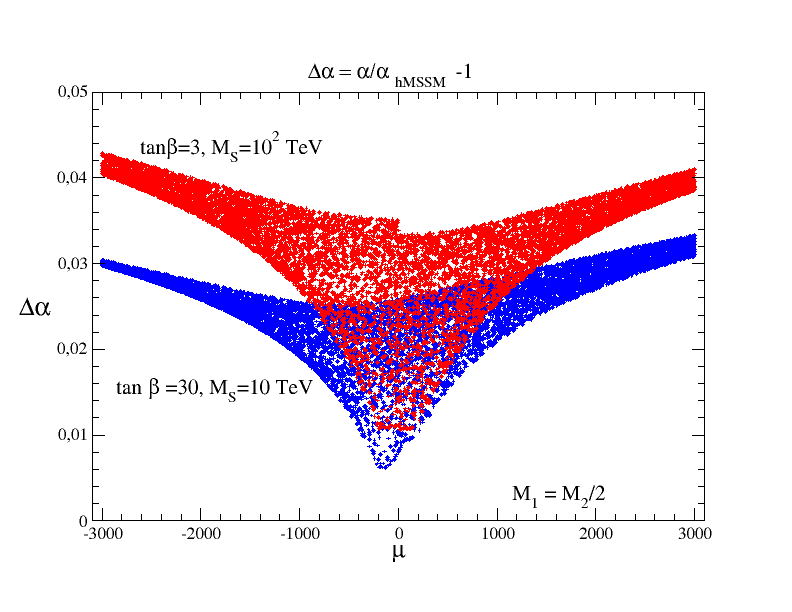,width=8.6cm} \hspace*{-10mm}
\epsfig{figure=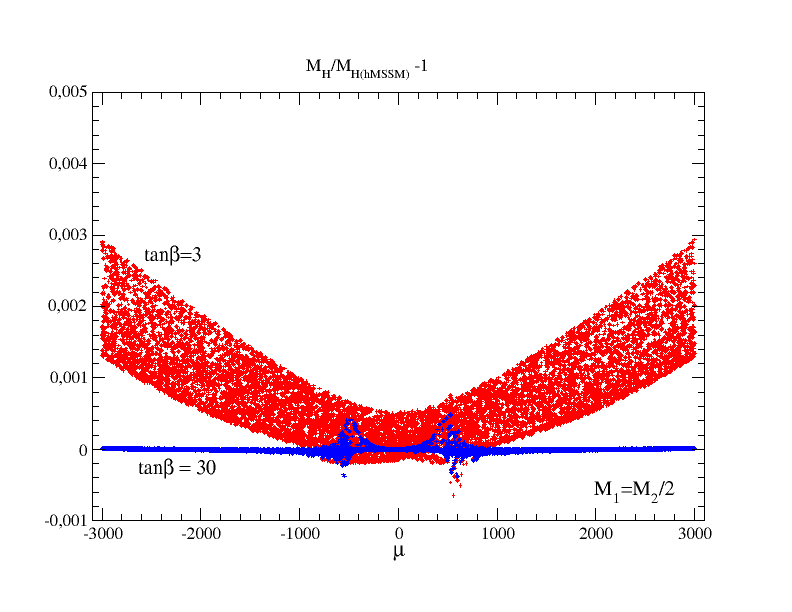,width=8.6cm}
} 
\vspace*{-4mm}
\caption{The relative difference of the recalculated  mixing angle $\alpha$ (left) and CP-even Higgs mass $M_H$  with chargino and neutralino radiative corrections included with respect to the simple hMSSM value with the prescription eq.~(\ref{mhiggs2}),
as a function of the parameter $\mu$ for fixed choices $\tan\beta=3, 30$ and  $M_2= \frac12 M_1$. }
\label{alpha_mh}
\vspace*{-1mm}
\end{figure}

Still using the hMSSM prescription, namely with 
$\Delta M^2_{22}|_\chi$ removed from the mass matrix eq.~(\ref{mhiggs2}) and hence 
implicitly added to the $\Delta M^2_{22}$ corrections which is defined by eq.~(\ref{DelM22}), we finally also illustrate in Fig.~\ref{alpha_mh} the relative difference, with respect to eq.~(\ref{eq:hMSSM}), of the recalculated CP-even Higgs mixing angle $\alpha$ (the left-hand side) and heavy CP-even $H$ boson mass $M_H$ (the right-hand side), when the chargino and neutralino radiative corrections are added. For the mixing angle $\alpha$, the effects remain moderate, reaching a maximum slightly above $\Delta \alpha = 0.04$ for $\tan\beta=3$ and $0.03 $ for $\tan\beta=30$ at extremely high $\mu$ values, $\mu= \pm 3$ TeV. For small $\mu$, the correction is about $0.01$ in both cases.  

The difference between the recalculated  heavy CP-even Higgs mass $M_H$ and the one given in eq.~(\ref{eq:hMSSM}) is even smaller, being at the few permille level for our considered range of $\tan\beta$ as well as $\mu, M_2$ values  as it is  illustrated in the right panel of Fig.~\ref{alpha_mh}. The relatively more pronounced sensitivities that one can observe  around $|\mu| \approx 500$  GeV correspond to artificial threshold effects, from chargino or neutralino contributions to the $H$ boson self-energies, i.e. when $M_H \sim 1\, \mbox{TeV} \simeq 2 m_{\chi^+_i}$ and/or $M_H\simeq 2 m_{\chi^0_i}$. Even in this case, the corrections are well below the permille level.



\subsection{Direct Corrections to the Higgs-Fermion Couplings} 
\label{sec:2.4}
\vspace*{1.5mm}

Another type of radiative corrections which is  important in the context of the MSSM Higgs sector, is the so-called direct corrections which appear in Higgs decays into third generation fermion pairs. Analogous corrections for Higgs decays into muon and strange quark decays also occur, but they do not play an important role in general and will be ignored here. In the case of Higgs decays into quarks, one encounters large QCD corrections,    while the electroweak corrections are in general rather small \cite{hffselw}. The dominant part of these QCD corrections can be absorbed in the running fermion Yukawa couplings when defined in the $\overline{\rm MS}$ scheme and evaluated at the scale of the corresponding Higgs boson mass \cite{hffqcd}. But there also pure SUSY--QCD corrections mediated by gluino-squark exchange \cite{BPMZ,Deltab1+Deltab2} and SUSY-electroweak corrections that involve squark and chargino/neutralino exchange that cannot be absorbed in the running quark masses and should be therefore considered separately  \cite{hffselw,Deltab1+Deltab2}. 

These SUSY direct corrections for third generation quarks and leptons,  called $\Delta_f\;  (f\!=\!t,b,\tau$ in the notation of third generation) corrections, modify the effective top, bottom and $\tau$ Yukawa couplings $\tilde g^\phi_f$ (with $\phi\!=\!h,H,A)$ of the neutral CP-even $h,H$ and the CP-odd $A$ states,  in the  following way
\cite{Deltab1+Deltab2}
\begin{eqnarray}
\tilde g^h_f =  \frac{g^h_f}{1\!+\!\Delta_f}\left[ 1 \!-\!
\frac{\Delta_f}{\tan\alpha\tan\beta}  \right]\!, 
\tilde g^H_f  =  \frac{g^H_f}{1\!+\!\Delta_f}\left[ 1 \!+\! \Delta_f
\frac{\tan\alpha}{\tan\beta} \right]\!, 
\tilde g^A_f =  \frac{g^A_f}{1\!+\!\Delta_f}\left[ 1 \!-\!
\frac{\Delta_f}{\tan^2\beta} \right]\!, \ \ 
\label{eq:gqresum}
\end{eqnarray}
with the reduced Yukawa couplings $g^\phi_f$ (i.e. normalized to the SM values) for isospin up- and down-type fermions given in eq.~(\ref{eq:gfhlo}).  

Considering first  the Higgs couplings to bottom quarks, the $\Delta_b$ correction at one-loop order is given by $\Delta_b  =  \Delta_b^{\rm QCD} + \Delta_b^{\rm EW}$ with the individual QCD and electroweak contributions reading \cite{Deltab1+Deltab2}
\begin{eqnarray}
\label{Delb}
&& \Delta_b^{\rm QCD} \! =\! 
\frac{2}{3}~\frac{\alpha_s}{\pi}~m_{\tilde g}~\mu~\tan\beta~
I(m^2_{\tilde b_1},m^2_{\tilde b_2},m^2_{\tilde g})\;, 
 \hspace*{-5cm}
\\
&& \Delta_b^{\rm EW} \! =\! 
\frac{\lambda_t^2}{(4\pi)^2}~A_t~\mu~\tan\beta~
I(m_{\tilde t_1}^2,m_{\tilde t_2}^2,\mu^2)\; \!-\!\frac{1}{12\pi}\mu\tan\beta \bigg\{ \nonumber
\\
&& \alpha_1 M_1 \bigg[ \frac{1}{3}I(m_{\tilde b_1}^2,
m_{\tilde b_2}^2,M_1^2) 
\!+\!\bigg( \frac{c_b^2}{2}\!+\! s_b^2\bigg)I(m_{\tilde b_1}^2,M_1^2,\mu^2) 
\!+\!\bigg( \frac{s_b^2}{2}\!+\!c_b^2\bigg) I(m_{\tilde b_2}^2,M_1^2,\mu^2)
\bigg]\;, \nonumber \\
&\! +\! & 3  \,\alpha_2 M_2 \bigg[
c_t^2 I(m_{\tilde t_1}^2,M_2^2,\mu^2) \!+\! s_t^2 I(m_{\tilde t_2}^2,M_2^2,\mu^2)
\!+\!\frac{c_b^2}{2} I(m_{\tilde b_1}^2,M_2^2,\mu^2)
\!+\!\frac{s_b^2}{2} I(m_{\tilde b_2}^2,M_2^2,\mu^2) \bigg]
\bigg\} \;,  \nonumber
\end{eqnarray}
where $\alpha_s$ denotes the strong coupling constant, $\lambda_t \!=\!\sqrt{2} m_t/(v\sin\beta)$ the top quark Yukawa coupling, $\alpha_{1,2} \!=\! g_{1,2}^2/4\pi$ the U(1),  SU(2) gauge couplings.  $s_{b}/c_{b} = \sin/\cos \theta_{b}$ are the sine and cosine of the sbottom mixing angle $\theta_{b}$, while the function $I$ is given by
\begin{equation}
I(a,b,c) = \frac{\displaystyle ab\log(a/b) + bc\log(b/c)
+ ca\log(c/a)}{(a-b)(b-c)(a-c)} \;.
\end{equation}
At high values of the parameters $\tan\beta$ and $\mu$ and for not too large gluino and squark masses (the corrections are damped by terms max$( \tilde m_q^2, \tilde m_{\ti g}^2)$ from the denominator), the full SUSY--QCD corrections can reach the  level of a factor of two in extreme cases, while the SUSY--electroweak corrections are much smaller and can reach at most 10\% only. 

 In the case of the Higgs coupling to tau leptons,
 there is no contribution from strong interaction and 
  the corresponding $\Delta_\tau$ term receives only contributions from the electroweak gauge couplings $\alpha_{1,2}$.  
The SUSY direct corrections are small in this case, again at the 10\% level at most and in general much  less. 

Similarly to what occurs in the case of the Higgs couplings to $b$-quarks, direct corrections also affect the Higgs couplings to top quarks but, contrary to the case above, they are suppressed by $\tb$ factors and could only be important at low $\tb$ values, $\tb\! \approx\! 1$. The corresponding $\Delta_t\! = \! \Delta_t^{\rm QCD} \!+\! \Delta_t^{\rm EW}$ SUSY-corrections is dominantly given by the much simpler expression
\begin{eqnarray}
\Delta_t =   \mu \cot \beta \left[ \frac{2\alpha_s}{3\pi}m_{\tilde{g}}
 I(m_{\tilde{t}_1}^2,m_{\tilde{t}_2}^2,m_{\tilde{g}}^2) + \frac{\lambda_b^2}{(4\pi)^2} A_b
I(m_{\tilde{b}_1}^2,m_{\tilde{b}_2}^2,\mu^2) \right]\;,
 \label{eq:deltat}
\end{eqnarray}
Hence, only the first term, i.e. the SUSY-QCD correction proportional to $\alpha_S m_{\tilde g}$, gives rise to potentially large contributions because in the electroweak part, $\lambda_b=\sqrt{2}  m_b/(v\cos\beta)$ is expected to be tiny for small $\tb$ values.

\begin{figure}[!h]
\vspace*{-3cm}
\centerline{ 
\epsfig{figure=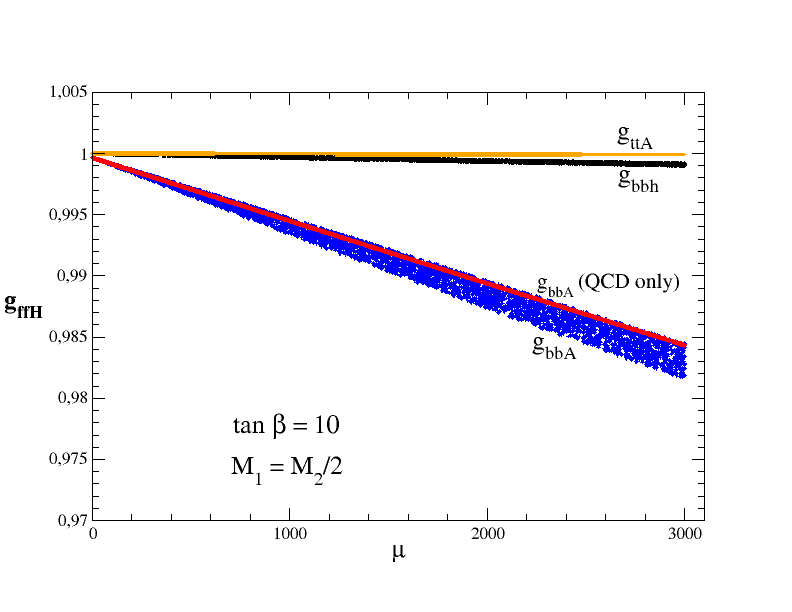,width=8.8cm} \hspace*{-10mm}
\epsfig{figure=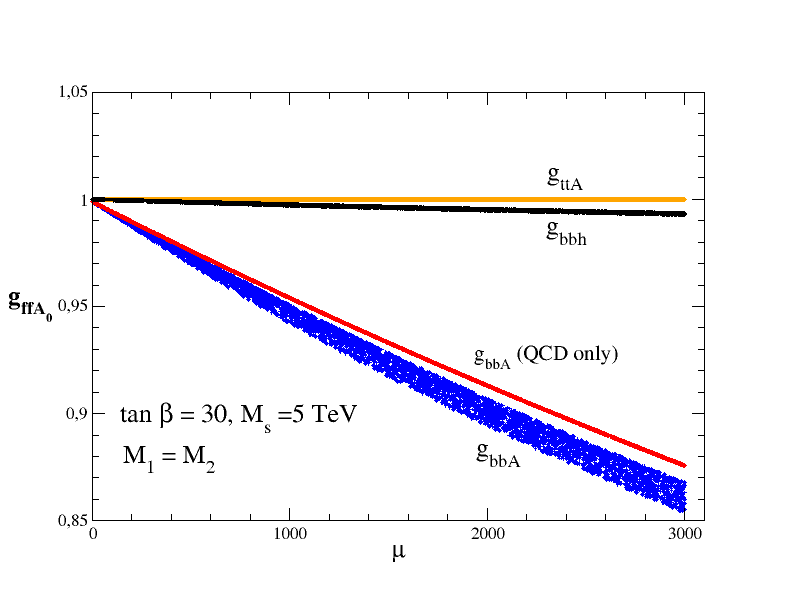,width=8.8cm}
}
\vspace*{-4mm}
\caption{The direct corrections to the coupling $g_{Ab\bar{b}}$  as a function of $\mu$ for $\tan\beta=10$ and  $M_2=2M_1$ (left) and $\tan\beta=30$ and $M_2=M_1$ (right). In red, only the SUSY-QCD correction is displayed for $m_{\tilde g}=3$ TeV,  while in blue, the additional contributions of the electroweak corrections are shown. We also display the very small corrections to the  $g_{ttA}$ (in orange) and $g_{bbh}$ (in black) couplings that do not have the enhancement factors at large $\tan\beta$ values.}
\label{Fig:deltab1}
\end{figure}

The supersymmetric-QCD corrections involving the gluino gives large radiative corrections only to the heavy $A, H$ and $H^\pm$ coupling to heavy quarks. The corresponding corrections in the context of the SM-like $h$ boson decouple and are thus extremely small in general as will be seen later (see also the recent analysis performed in Ref.~\cite{last-SUSY-paper}).  The $\Delta_b$ corrections are also potentially enhanced for large $\tan\beta$ values due to top-charginos contributions.  In Fig.~\ref{Fig:deltab1}, we illustrate the deviations in the coupling $g_{Ab\bar{b}}$ and $g_{Hb\bar{b}}$ which are similar in the decoupling regime, for two representative $\tan\beta$ values and from a scan over the $\mu$, $M_2$ parameters, and see where deviations are expected to be enhanced. Very low values,  $\tan\beta\simeq 3$, give tiny deviations that are not illustrated.

Note that the gluino mass is taken to be $m_{\tilde g}=3$ TeV which is above the present limit from the negative LHC searches, such as to maximize deviations since, as can be seen from eq. (\ref{Delb}), $\Delta^{\rm QCD}_b$ increases with $m_{\tilde g}$ as long as $m_{\tilde g} \lesssim m_{\tilde b_{1,2}}$. Note also that in our numerical analysis, we rather use exact (one-loop) expressions for the particularly sensitive QCD corrections  $\Delta^{\rm QCD}_b$, that differ from the large $\tan\beta$, $\mu\gg M_2$ approximations in eq.~(\ref{Delb}), the latter being thus not valid for small $\tan\beta$, $\mu$  also considered in our analysis. Indeed, the exact expressions tend to increase those corrections, by up to $30-40\%$ for $\tan\beta\simeq 3$ for which $\Delta_b$ is however very small, $ \sim 10^{-3}$.

As can be seen from Fig.~\ref{Fig:deltab1}, the deviations in the coupling $g_{Ab\bar{b}} \approx g_{H b\bar b}$ remain very moderate unless $\tan\beta$ is extremely large.  For instance,  for the value $\tan\beta\sim 30$, they can reach the level of $10\%$ for values of $|\mu|\sim 2$ TeV and even $15\%$ for $\mu$ values close to $3$ TeV. Nevertheless, the deviation is by far dominated by the QCD contribution.  Note that the dependence is very symmetrical for $\pm \mu$ and, thus, we only illustrate the effects only for positive $\mu$ values in the figure. 

Finally, for completeness, we also compare in Fig.~\ref{Fig:deltab1} the other relevant deviations in the coupling $g_{hb\bar{b}}$ of the lightest $h$ state and  the $A$ coupling top top quarks $g_{At\bar{t}} \approx g_{Ht\bar t}$.  Both couplings remain extremely moderate, illustrating the decoupling  of the corresponding contributions in the case of the SM-like $h$ boson and the absence of enhancement of the contribution at high $\tb$ values in the  case of the $A,H$ states.


\section{Constraints on the Gaugino-Higgsino Sector}
\label{sec:3}

In this section, we investigate the impact on the hMSSM by a light gaugino and higgsino sector, which is allowed by the present LHC data.\ In section\,\ref{sec:3.1}, we study the LHC search limits
on the charginos and neutralinos.\ Then, in section\,\ref{sec:3.2},
we analyze the LHC constraints on the Gaugino-Higgsino parameter space.\ For these analyses, 
we will use the SuSpect package\,\cite{Suspect}
to generate the supersymmetric particle spectra, and use the
packages SDECAY\,\cite{SDECAY} and SUSY-HIT\,\cite{SUSY-HIT}
to evaluate the decays of the heavier neutralinos and charginos into the lighter ones plus Higgs and gauge bosons.

\subsection{LHC Searches for Charginos and Neutralinos}
\label{sec:3.1}

Several searches for charginos and neutralinos have been performed by the ATLAS and CMS collaborations in various production channels and final state topologies. Following the spirit of out extended hMSSM framework, in which we assume the sfermions and in particular the squarks to be rather heavy and inaccessible at the LHC, we will ignore all channels in which the charginos and neutralinos originate from cascade decays of squarks. In addition, to simplify the discussion, we will assume that gluinos are also heavy and out of the LHC reach so that there are no gluino cascade decays neither. Thus, the main channel for the direct production of charginos and neutralinos is the Drell--Yan process:
\beq
pp \to q \bar q \to \chi_i^\pm \chi_i^\mp,\hs \chi_i^\pm \chi_j^0,\hs \chi_i^0 \chi_j^0\,.
\eeq
Of particular interest is the final state containing 
the lightest chargino and the next-to-lightest neutralino,
\beq
q\bar q' \to W^* \!\to \chi_1^\pm \chi_2^0 \,,
\eeq
which are produced only through the $s$-channel $W$-boson exchange.\ Another interesting channel should be the pair production of the lightest chargino through photon and $Z$-boson exchange and the next-to-lightest neutralino via $Z$-boson exchange only
\beq
q\bar q\!\to\! \gamma^*/Z^* \to \chi_1^\pm \chi_1^\mp \ , \ \ \ \ 
q\bar q\!\to\! Z^* \to \chi_2^0 \chi_2^0 .
\eeq

The production cross sections are known up to next-to-leading order (at least) in QCD \cite{pp-chi-NLO} and can be evaluated using, for instance, the program Prospino\,\cite{pp-chi-prospino}. 

\vspace*{1mm} 

For the first of the processes above, the most interesting decay products in the final state include the trileptons and missing energy from the channels ($\ell=e,\mu$) 
\beq
\chi_2^0 \to \chi_1^0 Z^{(*)}  \to \chi_2^0\ell\ell~~{\rm and}~~
\chi_1^\pm  \to \chi_1^0 W^{(*)}  \to \chi_1^0 \ell \nu \, ,  
\eeq
but we can also look for the possibility of the lightest Higgs boson in the final state,  
\beq
\chi_2^0 \to \chi_1^0 h \,. 
\eeq
In the case of chargino pair production, a powerful search channel is the one leading to two charged leptons and missing energy from the decay mode
\beq
\chi_2^\pm  \to  W^*\chi_1^0  \to \ell^\pm \nu\chi_1^0 \,. 
\eeq
If the charginos and neutralinos are very heavy and the phase space is favorable, they could also decay in channels in which the heavier MSSM Higgs bosons are 
present\,\cite{Haber-Gunion3,chi-decay-old,chi-decays}:
\beq
\chi_i \to \chi_j H_k \,, 
\eeq
where $\chi_i$ generically stand for the heavier neutralinos or charginos, $\chi_j$ for the lighter ones and $H_k$ to the MSSM Higgs bosons as already defined in section 2.2 with $k=1,2,3$ and $4$, corresponding respectively to the $H,h, A$ and $H^\pm$ states. 

\vspace*{1mm} 

The partial decay widths of heavier charginos and neutralinos $\chi_i$, decaying into lighter ones $\chi_j$ and gauge $V\! =\! W,Z$ or Higgs $H_k$ bosons are 
given by\,\cite{Haber-Gunion3,chi-decays}:
%
\beqs 
\begin{align} 
\Gamma(\chi_i \ito \chi_j V) =&\, \frac{\alpha}{8 c_W^2} m_{\chi_i}
\lambda^{\frac{1}{2}}(\mu_{\chi_j},\mu_V) \!\left\{ -12 \sqrt{\mu_{\chi_j}} g_{\chi_i \chi_j V}^L g_{ \chi_i \chi_j V}^R \right. \hspace*{3.5cm} \\
&  + \left. \!\left[ (g_{ \chi_i \chi_j V}^L)^2 \!+\! (g_{ \chi_i \chi_j V}^R)^2 \right]\!\! 
(1\!+\! \mu_{\chi_j}\!-\!\mu_V) 
\!+\!(1\!-\!\mu_{\chi_j} \!+\!\mu_V)(1\!-\!\mu_{\chi_j}\!\!-\!\mu_V) 
\mu_V^{-1} \right\} \!,  \nonumber \\
\hspace*{-10mm}
\Gamma(\chi_i \ito \chi_j H_k) =&\, 
\frac{\alpha}{8} \, m_{\chi_i}  
\lambda^{\frac{1}{2}}(\mu_{\chi_j},\mu_{H_k}) 
\!\left\{\! \left[ (g^L_{ \chi_i 
\chi_j H_k})^2 \!+\!(g^R_{ \chi_i \chi_j H_k})^2 \right]\! 
( 1\!+\! \mu_{\chi_j} \!\!-\! \mu_{H_k}) 
\right. 
\nonumber \\ 
&  \left. +\, 4 \sqrt{\mu_{\chi_j}} \, g^L_{ \chi_i  \chi_j H_k} g^R_{ \chi_i \chi_j H_k}  \right\} \!, 
\end{align}
\eeqs
%
with  the usual two--body phase space function defined as  
$\lambda(x,y)=1+x^2+y^2-2(xy+x+y)$ 
and given in terms of the reduced masses $\mu_X=m_X^2/m^2_{\chi_i}$. The couplings among charginos, neutralinos and the Higgs or massive gauge bosons have been presented in eq.\,(\ref{cp:inos}) and eq.\,(\ref{cp:V-gauginos}), respectively.\ 
The magnitude of these decays strongly depends not only on the phase-space, i.e. on the relative masses of the various chargino/neutralino states and the Higgs and gauge bosons, but also on the texture of the charginos and neutralinos and, hence, on the values of the $M_{1,2}$ and $\mu$ parameters. 

\vspace*{1mm}

To illustrate how the decay widths of, for instance,  the $\chi_2^0$ and $\chi_1^\pm$ states behave, let us ignore the masses of the decay products for simplicity and assume the decoupling limit with very heavy $H,A, H^\pm$ bosons that we ignore.  One can then express the partial decay widths of  these two, supposedly light, particles in units of $G_F M_W^2|\mu|/(8 \sqrt{2} \pi)$, as follows
\beq 
\Gamma( \chi_1^+ \!\!\to\!   \chi_1^0 W^+)  \! \approx \!  
 \Gamma(
\chi_2^0 \! \to \!  \chi_1^0 h ) \! \approx  \sin^2\!2\beta , \  \Gamma( \chi_2^0 \! \to \! \chi_1^0 Z ) \!\approx  
\cos^2 \!2\beta\,(M_2\!-\!M_1)^2\!/{4\mu^2}.~~~ 
\label{inodec-WZh} 
\eeq 
The first two decay widths are large at low $\tb$ values  when $\sin2\beta \!\approx\!1\,$,  while the last channel is important at high $\tb$ when $\cos2\beta \approx 1$ and when $M_{1, 2} \!\gg\! |\mu|\,$ to make the two neutralino states higgsino-like with a non-suppressed coupling to the $Z$ boson. 
\begin{figure}[!h]
\centering
\includegraphics[width=7.5cm]{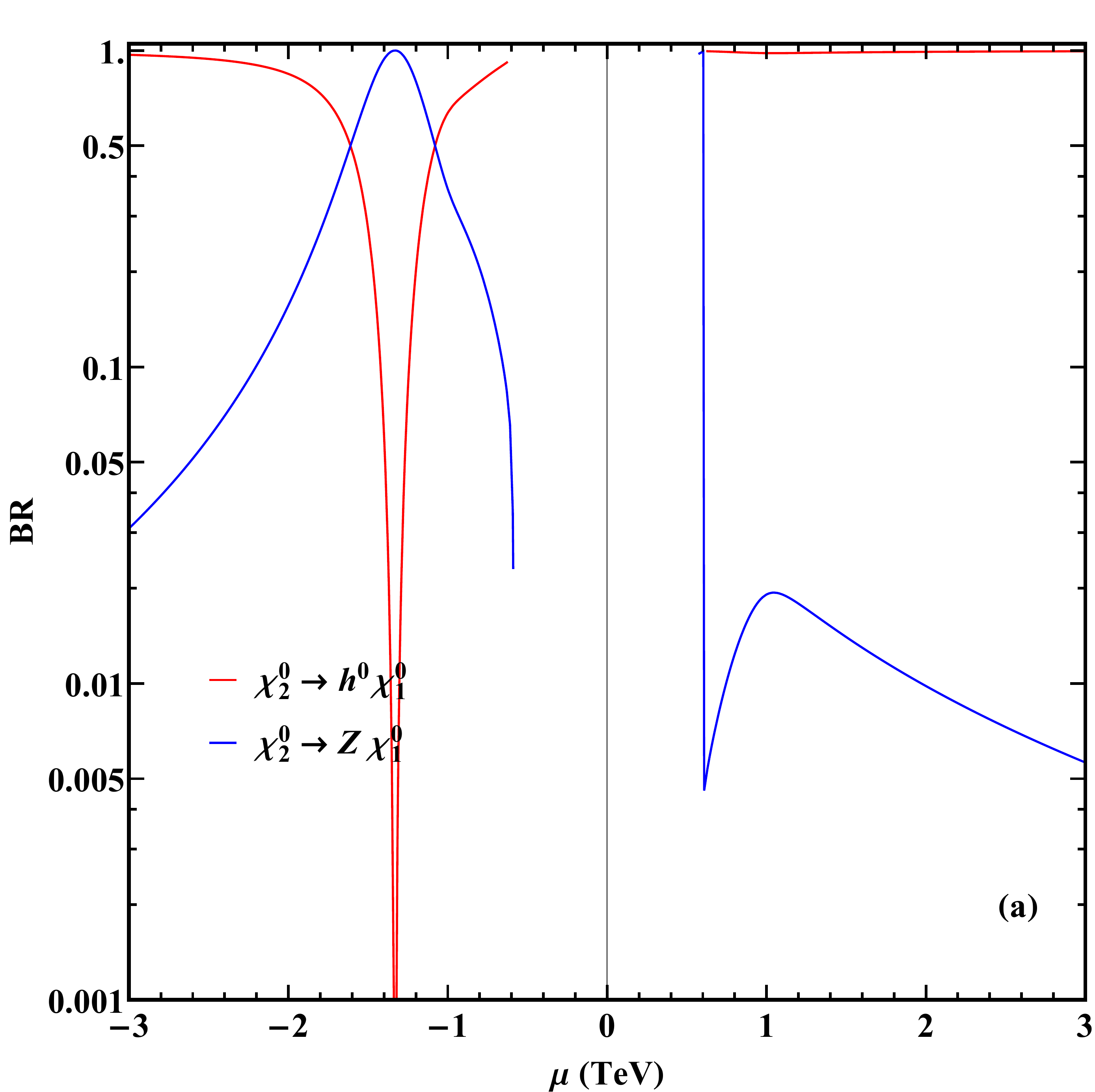}~
\includegraphics[width=7.5cm]{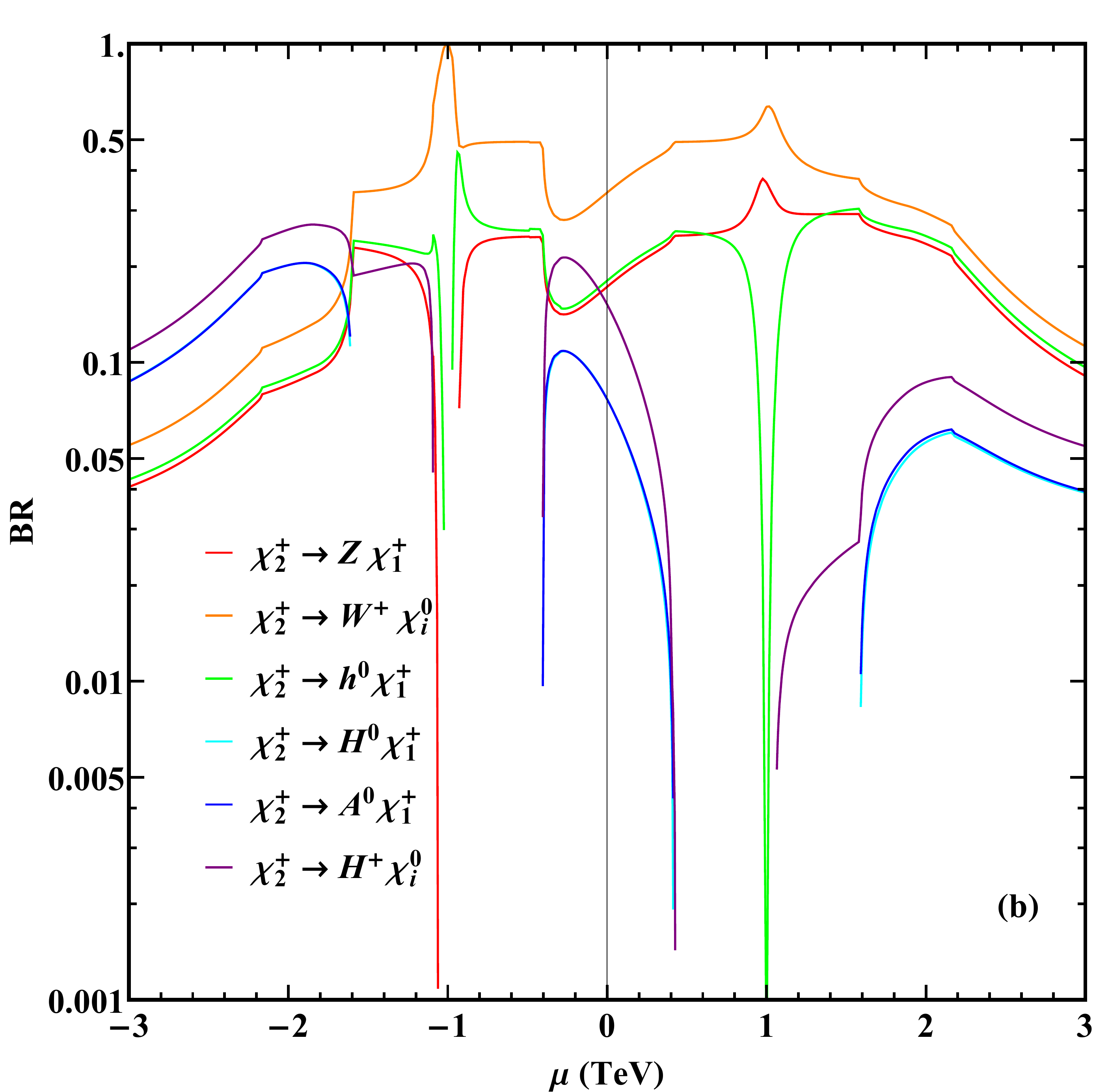}\\[3mm]
\includegraphics[width=7.5cm]{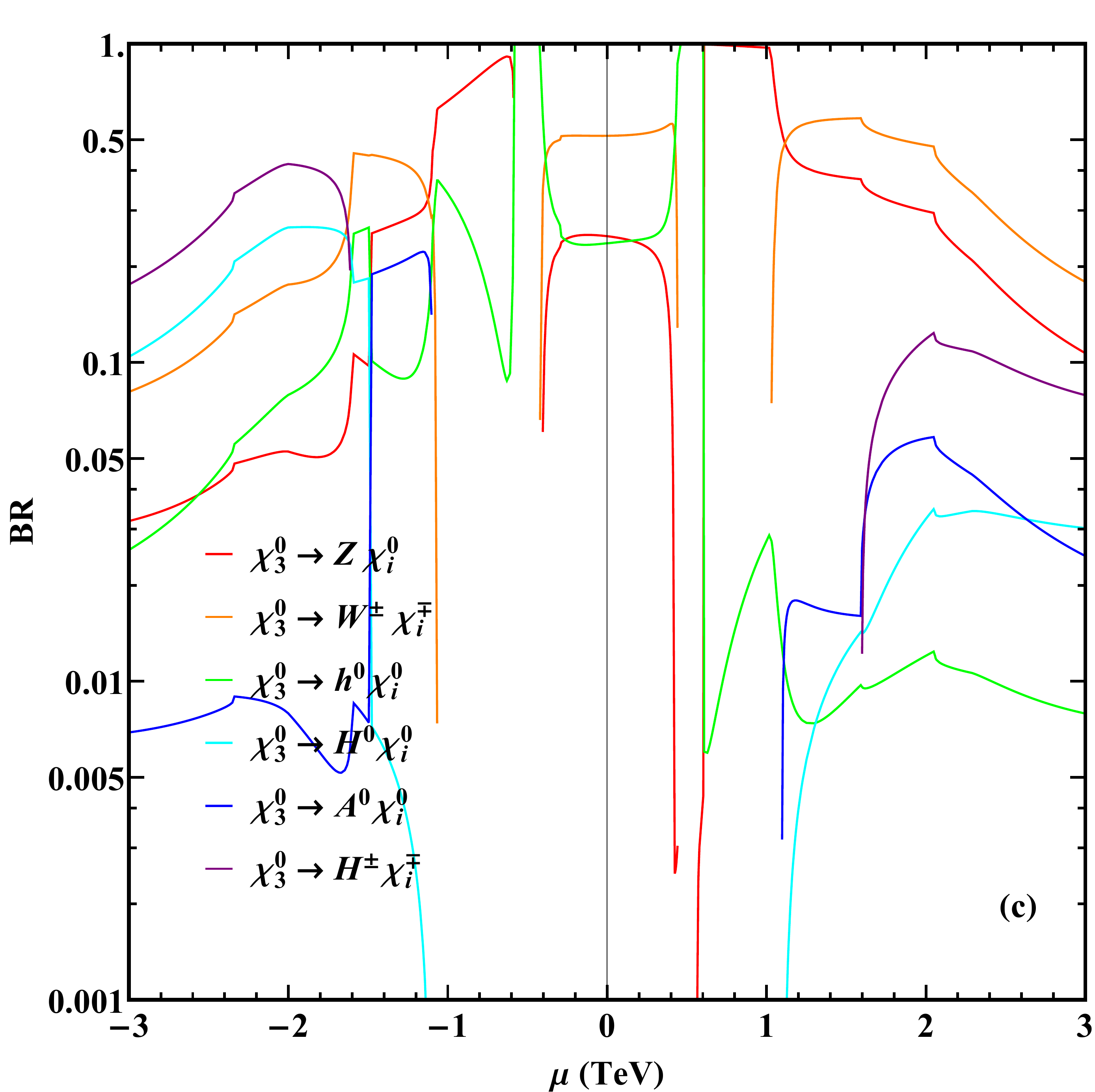}~
\includegraphics[width=7.5cm]{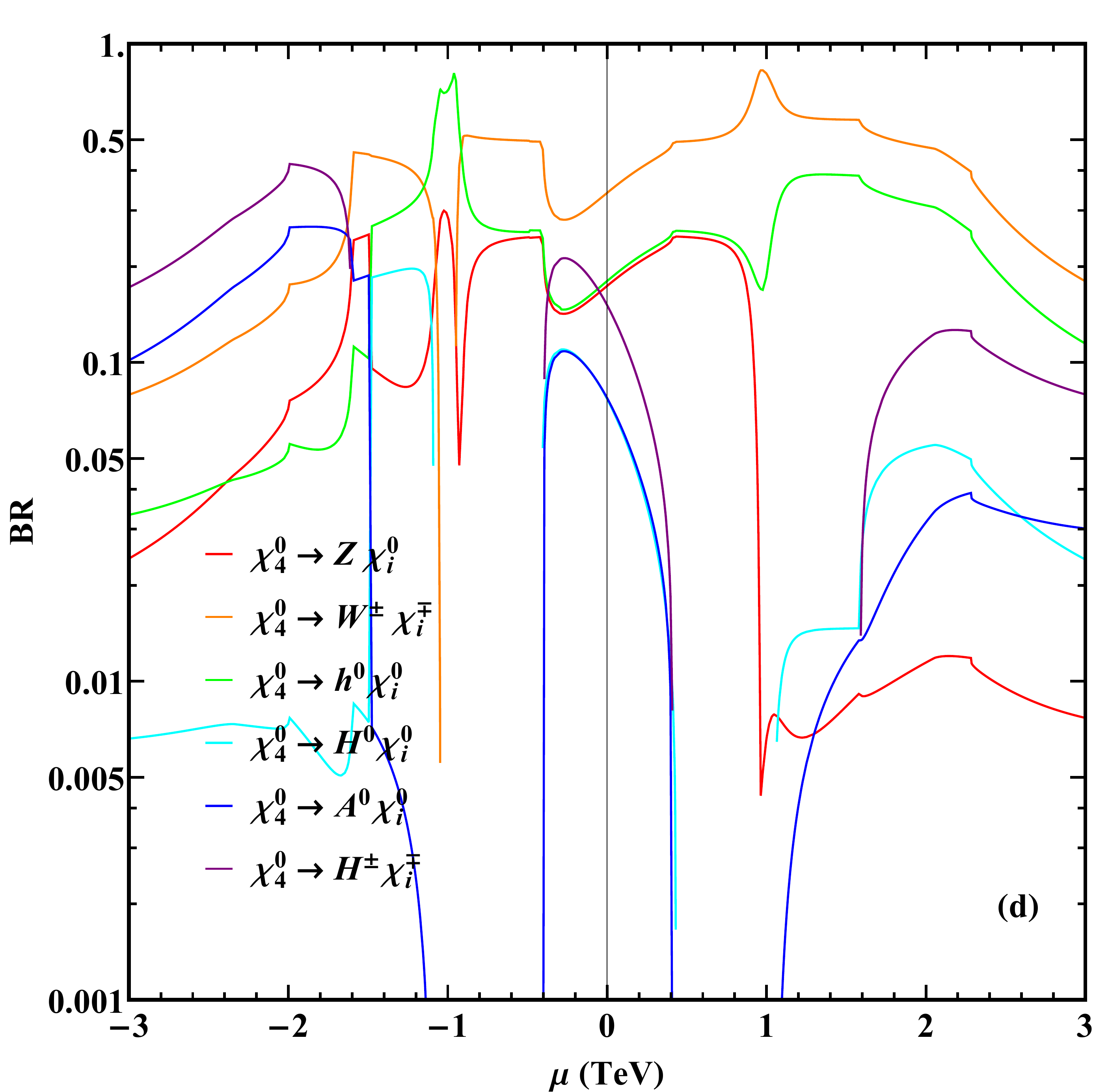}
\vspace*{-1mm}
\caption{Branching fractions of the various charginos and neutralinos decaying into gauge and Higgs bosons in the hMSSM, as functions of the $\mu$ parameter and for the set of input parameters $\tan\!\beta =3$, $M_A\!=\!600 $\,GeV and $2M_1\!=\!M_2\!=\!1 $\,TeV. }
\label{fig:sdecay}
\vspace*{-2mm}
\end{figure}

\vspace*{1mm} 

{We present in Fig.\,\ref{fig:sdecay} the branching fractions for the decays of the heavier neutralinos $\chi_{2}^0$, $\chi_{3}^0$, $\chi_{4}^0$ and chargino $\chi_{2}^\pm$ into the lighter ones plus Higgs and gauge bosons. They have been evaluated with the programs SDECAY \cite{SDECAY} and SUSY-HIT \cite{SUSY-HIT} in which we have generated the supersymmetric particle and Higgs spectra using the modified version of the SuSpect code \cite{Suspect} as discussed in the previous section.}  We have assumed that the sfermions, and also the gluinos ($m_{\tilde g} \gsim 3$ TeV),   are too heavy to have an impact on the numbers. The branching ratios are given as functions of the $\mu$ parameter which takes both signs and  we have set the other inputs to $\tan\!\beta =3$, $M_{\!A}^{}\!=600$\,GeV and $2M_1\!=\!M_2\!=\!1$\,TeV.

\vspace*{1mm}

For this parameter set, the next-to-lightest neutralino has two decay modes, $\chi_2^0 \!\to\! \chi_1^0\, h$ and $\chi_2^0 \!\to\!\chi_1^0 Z$.  The lightest chargino  can only decay into one channel,  $\chi_{1}^\pm \to W^\pm \chi_1^0$. Heavier charginos and neutralinos can decay not only into the light $h$ and  $Z,W$ bosons,  but also into the heavier $A,H, H^\pm$ states.  The branching fractions can vary significantly around some critical points, such as $|\mu|\!=\!M_1$ and $|\mu|\!=\!M_2$.  This is because these $\mu, M_1, M_2$ values determine how the gauge eigenstates form the physical mass eigenstates. 

\vspace*{1mm}

As can be seen from the figures, the decay pattern in this case is  rather involved as many possibilities could be allowed.  The mass difference between the parent and daughter particles should be large enough, firstly to avoid phase-space suppression for decays with $W,Z,h$ final states and secondly to open the possibility for cascade decays into the heavier MSSM $H/A/H^\pm$  bosons. The decays into Higgs bosons are particularly relevant 
if the lighter $\,\chi\,$ states are higgsino-like (gaugino-like)  and the heavier ones are gaugino-like (higgsino-like), which maximize the couplings as mentioned previously.\ 
{Those involving $H/A/H^\pm$ final states can be important, reaching sometimes the level of a few times $10\%$, but the decays into gauge bosons are in general dominant, in particular for large $\mu$ values.}  

\subsection{LHC Constraints on the Gaugino-Higgsino Parameters}
\label{sec:3.2}

Constraints on the gaugino-higgsino mass parameter space 
come mainly from LHC  searches of the lighter chargino and neutralinos in the  simple processes (without cascades) 
\beq pp\to\chi^0_2\chi^0_2,\chi^0_2\chi_1^\pm,\chi_1^\pm\chi_1^\mp \to \chi_1^0\chi_1^0 +XX\to XX+E_T^{\rm mis}\,,
\label{eq:ppxx}
\eeq
where $E_T^{\rm mis}$ is the transverse missing energy due to the escaping LSP neutralinos and the final states $X$ stand for the lightest Higgs and the massive gauge bosons,  $X\!=\!W^\pm,Z,h$. If the mass difference between the $\chi^0_2, \chi_1^\pm$ and the $\chi^0_1$ states is small, the $W$ and $Z$ boson could be off-shell  and would decay into (almost) massless quarks and leptons, off-shell $h$ bosons can be ignored as the total width $\Gamma_h^{\rm tot}=4.07$ MeV \cite{PDG} is too small.

\vspace*{1mm}

In most cases, only final state topologies with leptonic decays, that are subject to a significantly smaller QCD background than events with final state quarks, have been analyzed. The most famous signatures are the trilepton events, mainly from $ q\bar q \to \chi^0_2 \chi_1^\pm \to ZW + \chi^0_1 \chi_1^0 \to \ell^+ \ell^- \ell^{'\pm} + E_T^{\rm mis}$ (with $\ell, \ell'=e,\mu)$ which has a large cross section times branching ratio,  or the same sign dilepton events from processes such as the one quoted above but with one lepton which has not been detected. 

\vspace*{1mm}

Three or four leptons can also be obtained in the processes $q\bar q \to \chi^0_2 \chi_2^0 \to ZZ  \chi^0_1 \chi_1^0$, which nevertheless have a smaller cross section than the one above. Mono-$Z$ and mono-$h$ events from the process $q\bar q \to \chi^0_1 \chi_2^0$ which is more favored by phase-space and $\chi_2^0  \to \chi^0_1 Z$ or  $\chi_2^0 \to \chi_1^0 h$ with $Z \to \ell^+ \ell^-$ and $h \to b \bar b$ (and eventually $h \to \gamma\gamma $ as the much lower branching ratio could be compensated by the much smaller QCD background) are also considered but the rates are even smaller in general. 

\vspace*{1mm}

Finally, the process $q \bar q \to \chi^\mp_1 \chi_1^\pm \to WW \chi^0_1 \chi_1^0 \to \ell^\mp \ell^{'\pm} + E_T^{\rm mis}$, as it is also mediated by $s$-channel photon exchange,  has a significant cross section independently of the chargino texture, but the topology with opposite--sign leptons is subject to a larger background. 

\vspace*{1mm}

As already mentioned, the corresponding backgrounds, which mainly come from rare SM processes such as pair production of $W$ or $Z$ bosons or associated $W,Z$ bosons with top quarks or from events in which the leptons have been  misidentified or missed,  or result from decays of hadrons, are in general rather small.

\vspace*{1mm}

The experimental measurements on these channels have been performed by the ATLAS \cite{ATLAS:2018eui} and CMS \cite{CMS:2014kxj} collaborations and a very recent summary has been given in Ref.~\cite{Adam:2021rrw}. As already alluded to in the previous discussion,  the strongest constraints arise from the searches in the $\,pp\!\to\!\chi^0_2\, \chi_1^\pm\!\to\!\ell\ell \ell\nu +E_T^{\rm mis}$ topology. 

\vspace*{1mm}

In the case where the lightest chargino and the next-to-lightest neutralino are almost degenerate in mass with the LSP, the above searches are inefficient as the leptons or the other accompanying particles are too soft to be detected.\ In this case, one would have long-lived $\chi_1^\pm$ and $\chi_2^0$ states and the so-called ``disappearing track'' signature.\ 
This is mainly done in the pair production process of the lightest charginos $\,q \bar q \to \chi^\mp_1 \chi_1^\pm+g\,$ in which a high transverse momentum gluon is emitted in the initial state.\ Such signatures have also been searched for by the CMS \cite{CMS-compress} and ATLAS \cite{ATLAS-compress} collaborations, and constraints on the mass difference with the LSP $\chi_1^0$ neutralino or the lifetimes of the $\chi_1^\pm$ and $\chi_2^0$ states have been obtained.  

\vspace*{1mm}

In this case of ${\chi_1^0}$ being bino-like and 
$\chi_2^0/\chi_1^\pm$ being wino-like, 
the production rate of 
$\,p\!\to\!W^{\pm *}\!\to\!\chi_2^0\chi_1^\pm\,$
is not sensitive $\mu$ and $\tan\!\beta$
as shown in eq.\,\eqref{cp:V-gauginos} with $(i,j)\!=\!(2,1)$,
and $\chi_2^0/\chi_1^\pm$ can only decay to $\chi_1^0$ 
regardless of the parameter space.\ 
Almost all LHC experimental searches give bounds of 
$m_{\chi_1^0}^{}$ and $m_{\chi_2^0,\chi_1^\pm}^{}$ 
on this option.

\vspace*{1mm}

For other options such as $M_1^{}\!<\!|\mu|\!<\!M_2^{}\,$ 
or $\,|\mu|\!\ll\! M_{1,2}^{}$, the reactions 
in eq.\,(\ref{eq:ppxx}) are sensitive to $\mu$ and $\tan\beta$ because higgsino involved in the final states.\ 
The LHC experimental studies impose bounds on these options, 
but they used certain fixed $\mu$ and $\tan\beta$ values.\ 
Different experimental studies had different choices of fixed parameters, so we do not have universal bounds on these options.\ 
These experimental studies mostly gave the model-independent bounds on $\,\epsilon \times \sigma$\,, where $\,\epsilon\,$ is the detection efficiency which can be obtained only by detector-level simulations.\ This would require detector-level simulations for the whole parameter space.\ Thus it is valuable 
to further perform systematic experimental analysis 
in the near future.

\begin{figure}[!h]
	\vspace*{-1mm}
	\centering
	\includegraphics[width=7.5cm,height=8cm]{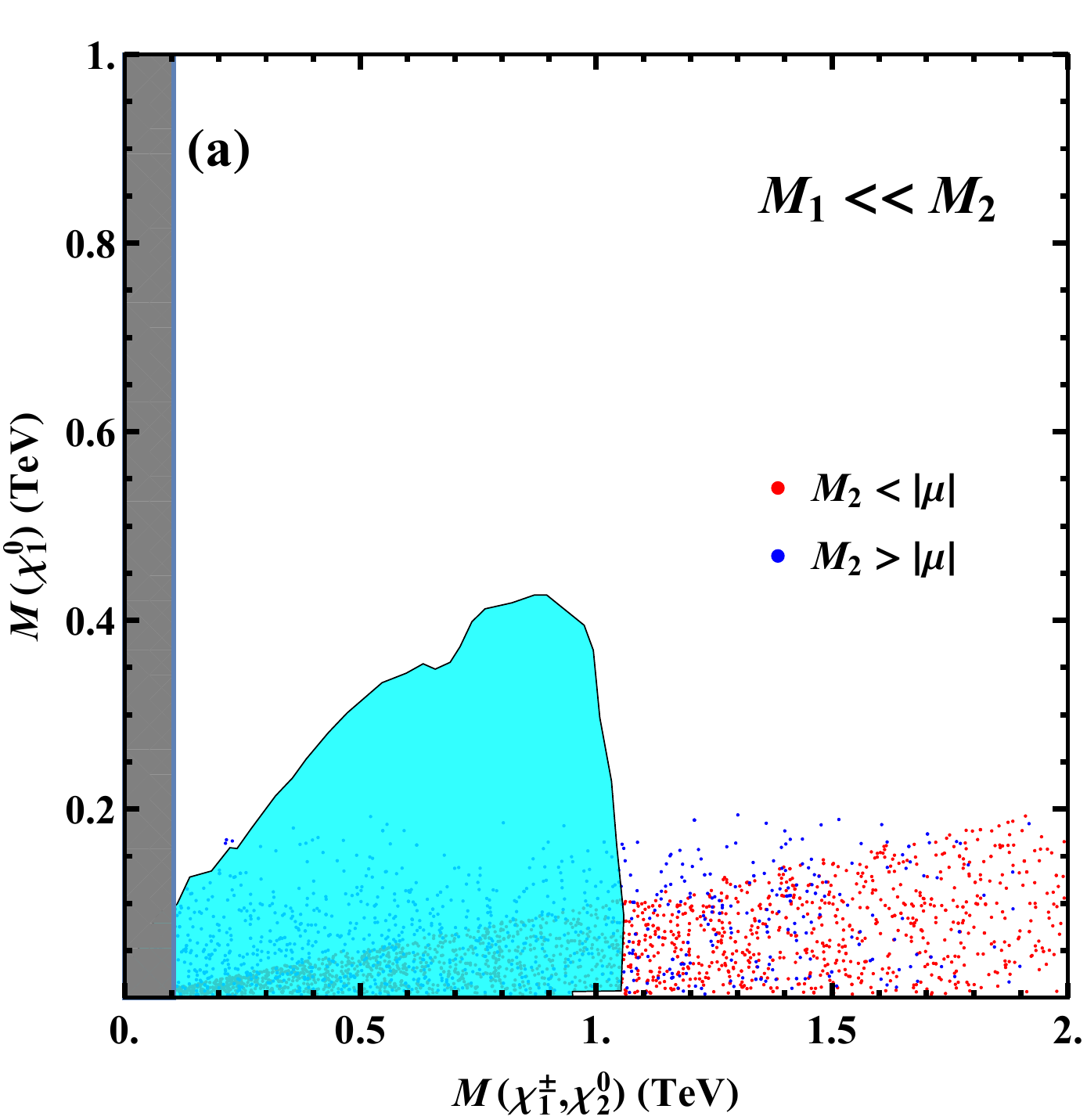}
	\includegraphics[width=7.5cm,height=8cm]{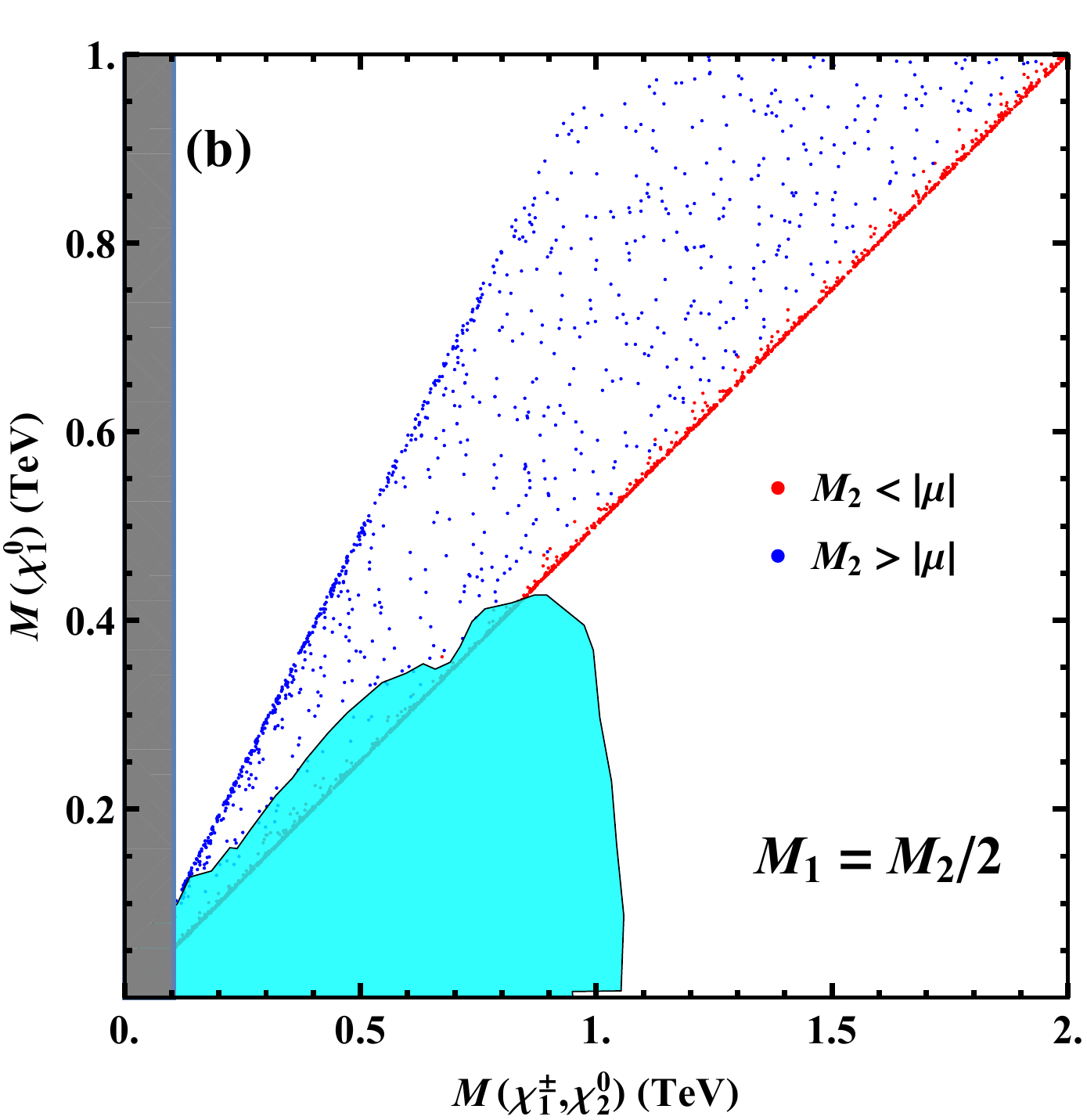}
	\\[2mm]
	\includegraphics[width=7.5cm,height=8cm]{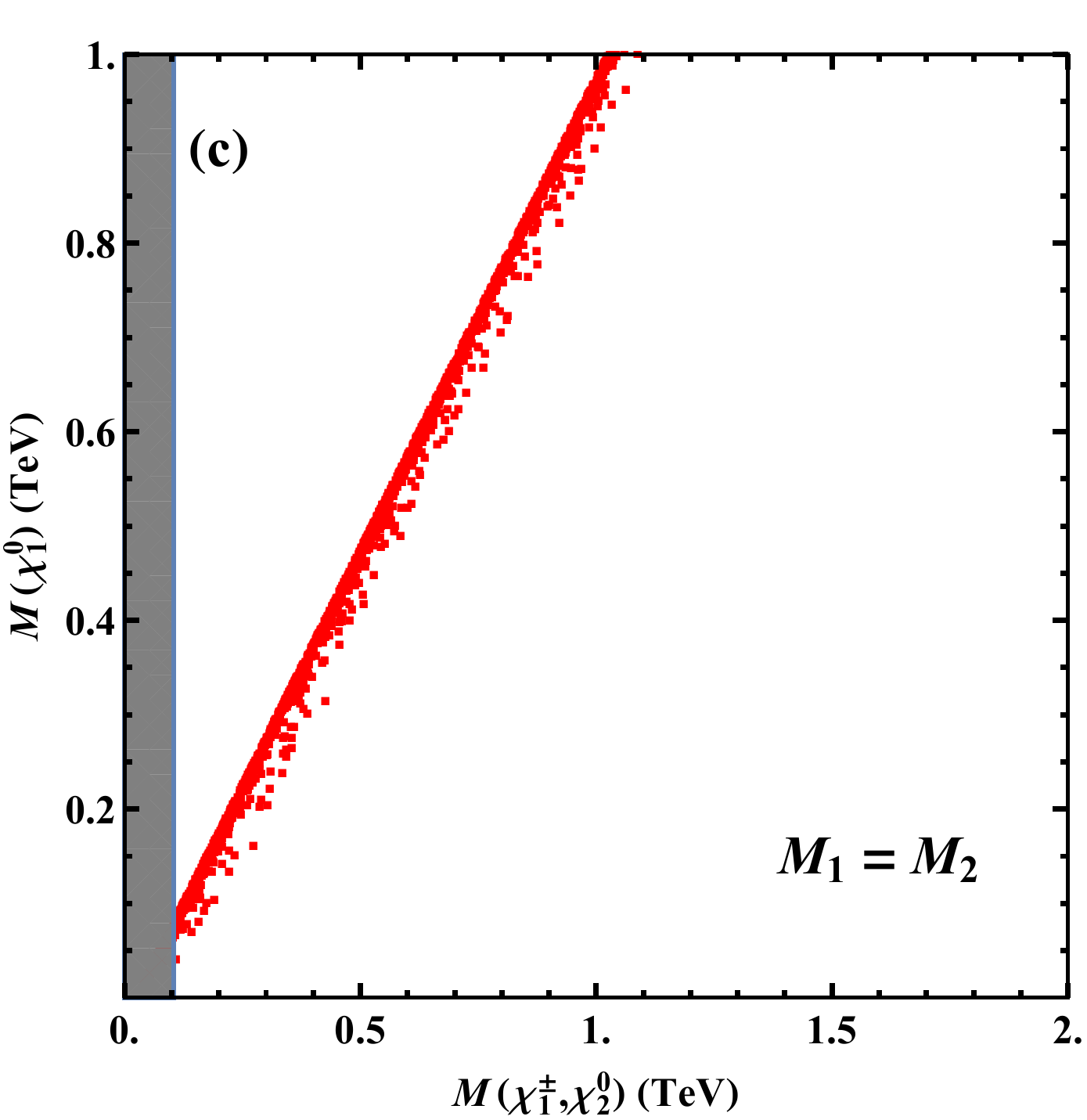}
	\includegraphics[width=7.5cm,height=8cm]{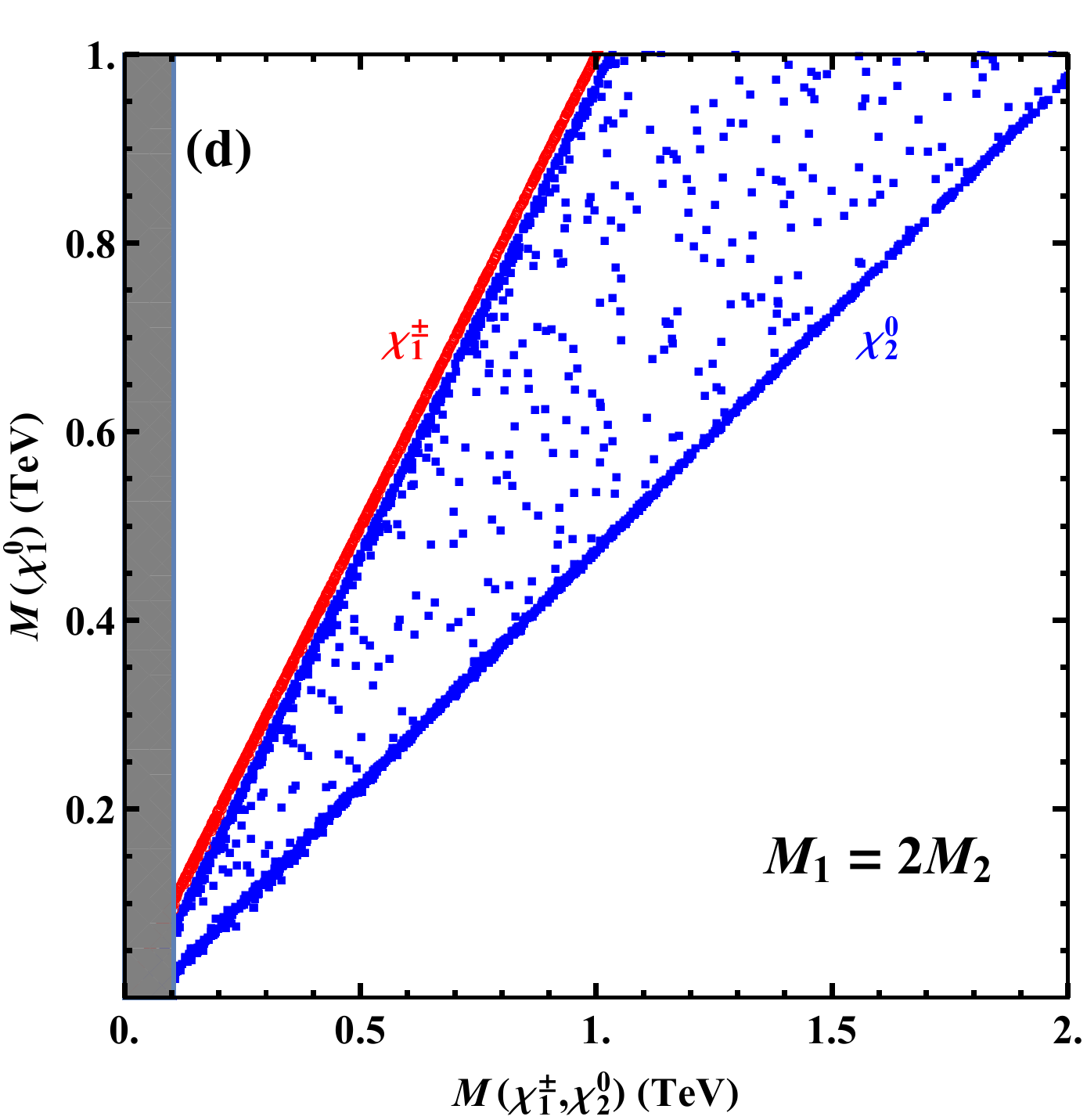}
	\vspace*{-2mm}
\caption{Allowed mass range of the lightest gaugino under the constraints from the gaugino and Higgs searches at the LHC, where the dots represent the hMSSM parameter space and the shaded blue regions are excluded by the existing LHC searches in the $(\tilde W\!,\,\tilde B)$ model ($M_1^{}\!<\!M_2^{}\!<\!|\mu|$) at the 95\%\,C.L.  
The analysis is presented for the four sample scenarios,
$\,M_1\!\ll\! M_2$ in panel\,(a),\
$\,M_1\!=\!M_2\,$ in  panel\,(b),\  
$\,M_1\!=\! \frac12 M_2\,$ in panel\,(c),\
and $\,M_1\!=2M_2\,$ in panel\,(d).
In the panels\,(a)--(b), the red dots correspond to the possibility  $\,M_2^{}\!<\!|\mu|\,$ while the blue dots correspond to the possibility $\,M_2^{}\!>\!|\mu|\,$. }
\label{fig:3}
\vspace*{-3mm}
\end{figure}
%


{In the following, we will give a concrete example
on how the constraints can be imposed on the viable parameter 
space in the plane of 
$[m_{\chi_1^\pm,\chi_2^0}^{},\, m_{\chi_1^0}^{}$] 
by using the current experimental searches and analyses.\ 
The latter have been done mostly for the wino-bino $(\tilde W,\,\tilde B)$ scenario as described above and the combination of all these experimental limits for this $(\tilde W,\tilde B)$ model are presented in Fig.\,\ref{fig:3}.\  
We have implemented the hMSSM in the package Suspect\,\cite{Suspect}  
as discussed in section\,\ref{sec:2.3}, generated the chargino, neutralino and Higgs spectra, and performed the following scan on the hMSSM parameter
space:
\begin{eqnarray}
\tan\beta\in(1,50),\quad M_A\in(0.5,2)\text{TeV},\quad M_2\in(0,2)\text{TeV},\quad \mu \in (-3,3)\;\text{TeV}\, .
\label{eq:scan}
\end{eqnarray}
For the remaining bino mass parameter $M_1$, we will study four benchmark scenarios in which it is connected to the wino mass parameter $M_2$ in the following way: 
\\[1mm]
(a).\ $M_1\!\ll\! M_2$ which will describe the pure bino-like possibility at large $|\mu|$ values. 
\\[1mm]
(b).\ $M_1\!=\! \frac12 M_2$ which features the GUT--like possibility with $M_2/M_1 = \alpha_2/\alpha_1$. 
\\[1mm]
(c).\ $M_1\!=\! M_2$ which leads to the scenario where the bino and wino are mass degenerate. 
\\[1mm]
(d).\ $M_1\!=\!2M_2$ in which the LSP can be wino-like and mass degenerate with $\chi_1^\pm$.
}

\vspace*{1mm}

{ 
As already stated, we set the other MSSM parameters such as the SUSY scale (which governs the sfermion masses) and the gluino mass parameter $M_3$ (which governs the gluino mass), 
to be large enough, 
so their effects on the current collider searches are negligible.\ 
Since the scenarios (c) and (d) have no intersection with the 
$(\tilde W,\tilde B)$ model,  we analyze their parameter space just for the comparison with the scenarios (a) and (b).
}

\vspace*{1mm}

We present the viable parameter space of the lightest gaugino mass in Fig.\,\ref{fig:3}. The shaded blue regions are excluded by LEP2 
searches of charginos which require $m_{\chi_1^\pm}>$ 104 GeV, 
and by the current direct searches at the LHC.\  
We see that sizable regions of the masses $m_{\chi_2^0},m_{\chi_1^\pm} \!\lesssim\! 1$\,TeV\, have been excluded by the LHC. For the case of $\,M_1\!\ll\! M_2$ ({we use the range $M_1\!<\!0.1 M_2$}) in Fig.~\ref{fig:3}(a), one can deduce that $m_{\chi_1^0} \! \lesssim\! m_{\chi_2^0}, m_{\chi_1^\pm}$.\ 
For the case of $M_1\!=\! \frac12 M_2$ in Fig.~\ref{fig:3}(b),  one infers the condition  $\frac12 m_{ \chi_2^0} , \frac12 m_{\chi_1^\pm}  \! < \! m_{\chi_1^0} \!<\!  m_{ \chi_2^0}, m_{\chi_1^\pm}$ while for the case of $M_1\!=\! M_2$ in Fig. \ref{fig:3}(c), one finds $m_{\chi_1^0}\! \lesssim \! m_{\chi_2^0}, m_{\chi_1^\pm}$.\ Then, for the case of $M_1^{}\!=\!2M_2^{}$ in Fig.\ref{fig:3}(d), one deduces  $\frac12 m_{\chi_2^0}  <\! m_{ \chi_1^0} \!\lesssim\! m_{ \chi_1^\pm}\!<\! m_{\chi_2^0}$.\ Finally, from the panels (a) and (b) of Fig.~\ref{fig:3}, one sees that for the $(\tilde W, \tilde B)$ model,  one of the two conditions must be satisfied: $\,150\,\text{GeV}\!<\! m_{\chi_1^0} \!<\! m_{\chi_2^0},m_{\chi_1^\pm}$, or  $m_{\chi_1^0}\!\ll\! 1\,\text{TeV}\!<\! m_{\chi_2^0},m_{\chi_1^\pm}$\,. 

\begin{figure}[t] 
\centering
\includegraphics[width=9.2cm]{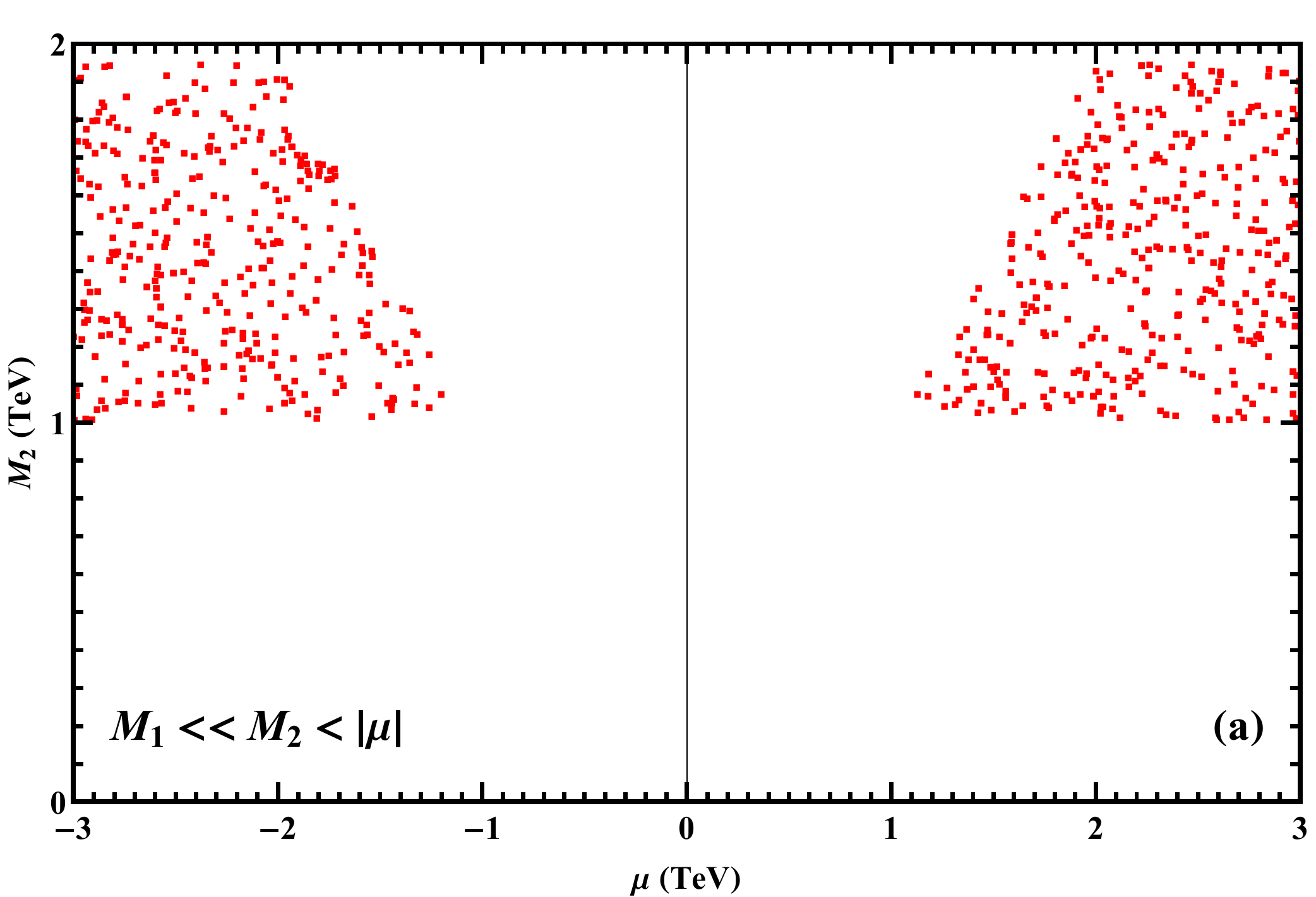}\\
\includegraphics[width=9.2cm]{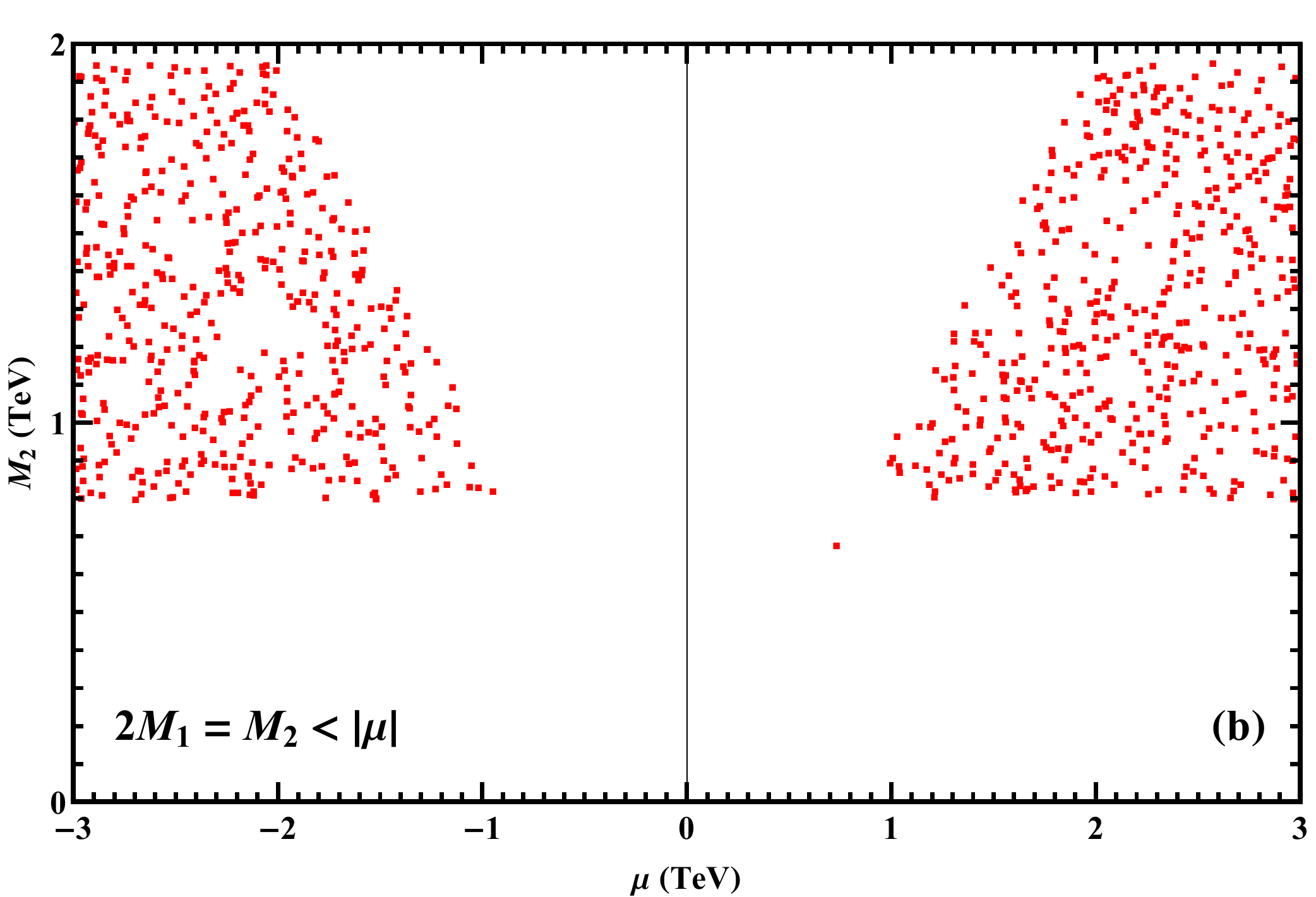}
\vspace*{-3mm}
\caption{{Allowed parameter space in the plane  $[\mu, M_2]$  
under the constraints (95\%\,C.L.) from gaugino searches  
 at the LHC.}\  
The analysis is presented for the two representative scenarios, $M_1\! \ll \! M_2 \!< \! |\mu|$ for panel\,(a) 
and $2M_1 \!=\!M_2\!< \! |\mu|$ for panel\,(b).}\label{fig:4}
	\vspace*{-3mm}
\end{figure}


Next, we further present in Fig.\,\ref{fig:4} the corresponding viable parameter space in the plane  $[\mu,\,M_2]$, where the constraints at the 95\% confidence level are derived from the negative searches of wino-like $\chi_2^0/\chi_1^\pm$, bino-like $\chi_1^0$ and Higgs bosons.\ The parameters $M_A$ and $\tan\beta$ which appear in the Higgs sector, are mainly constrained by the searches of the heavy Higgs states at the LHC and by the precision measurements of the lighter $h$ couplings to SM fermions and gauge bosons, which will be summarized in the next section.\    
This analysis has been performed  for the two representative scenarios  $M_1\!\ll\! M_2\!<\!|\mu|$ for panel (a) and $2M_1\!=\!M_2\!<\!|\mu|$ for  (b).


Then, using the constraints of Fig.\,\ref{fig:3},  one can further derive bounds on the $[\mu,\,M_2]$ parameter space as illustrated in Fig.\,\ref{fig:4} for the two possibilities $M_1\! \ll \! M_2 \! < \! |\mu|$ and $2M_1 \!=\!M_2\!  < \! |\mu|$. The two panels show that  the parameter regions with small $|\mu|$ and $M_2$ values have been excluded by the existing searches at the LHC.  To be more specific, we find that the regions of $|\mu |\!\lesssim\! 1.2$\,TeV and $\,M_2^{}\!\lesssim\! 1$\,TeV are excluded for the case $\,M_1\!\ll\! M_2\!<\!|\mu|$\, as in plot\,(a),\, while the regions of $\,|\mu |\!\lesssim\! 1$\,TeV and $\,M_2^{}\!\lesssim\! 0.8$\,TeV  are excluded for the case $2M_1\!=\!M_2\!< \! |\mu|$ as in plot (b). These bounds have already covered the bound of LEP2 searches of charginos ($M_2, |\mu| \gsim 100$\,GeV).

\vspace*{1mm}
\section{Collider Constraints on the Higgs sector}
\label{sec:4}
\vspace*{2mm}

In this section, we study the collider constraints on the
Higgs sector in the hMSSM formulation.\ 
In section\,\ref{sec:4.1} we analyze the productions and decays
of both the lightest Higgs boson $h$ and 
the heavier Higgs states $(H^0,\hs A^0,\hs H^\pm)$ 
of the hMSSM, including the SUSY corrections.\ 
In section\,\ref{sec:4.2}, we analyze the current constraints on 
the parameter space of the hMSSM Higgs sector as imposed by 
the ATLAS and CMS searches performed during the Run-II phase
of the LHC.\  
In section\,\ref{sec:4.3}, we study the decays of the 
heavier MSSM Higgs bosons into charginos and neutralinos
in the context of a light gaugino-higgsino sector, which 
can be significant in some of the hMSSM parameter space 
and thus can have important impact on the Higgs phenomenology 
of the hMSSM.

\vspace*{1mm}
\subsection{Higgs Production and Decay}
\label{sec:4.1}
\vspace*{1mm}

\subsubsection{Higgs Cross Sections and Decay Branching Fractions}
\label{sec:4.1.1}
\vspace*{1mm}

We first give a brief summary of the main Higgs production and decay modes in the MSSM \cite{Higgs-prod-decay} and start with the case of the lighter $h$ boson which is SM-like as soon as the mass of the pseudoscalar Higgs state is $\,M_A \!\gsim 300$\,GeV,\, which is indeed the case from the present LHC searches as will be described in the next subsection.  

\vspace*{1mm}

The SM-like $h$ boson mainly decays into $b \bar b$ pairs but the channels with $WW^*$ and $ZZ^*$ final states (before  allowing the gauge bosons to decay leptonically, $W \! \to \! \ell \nu$ and $Z\! \to  \! \ell \ell$ with $\ell\! =\! e,\mu$), as well as the $h\! \to \! \tau^+\tau^-$ channel, are also significant. The clean $h\to \gamma \gamma$ mode, induced by loops of top quarks and $W$ bosons in the SM, can be easily detected albeit  its small rates. The decays $h \to Z\gamma$  and $h\to \mu^+\mu^-$ will be accessible only at the high-luminosity (HL-LHC) LHC option \cite{HL-LHC}. 
The total $h$ decay width is rather small, $\Gamma_h = 4.07$ MeV, and any channel beyond the ones above will alter it significantly. This is particularly the case of invisible $h$ decays, which can be probed directly in a more efficient way.  We will use the program HDECAY \cite{HDECAY} to evaluate the corresponding branching fractions.  

\vspace*{1mm}

As for the Higgs production processes, they will be evaluated
using the programs of Refs.~\cite{Michael} which include all relevant higher order QCD corrections. The dominant process is gluon--fusion $gg\to H$  (ggF) and has rates that are at least an order of magnitude larger than the two subleading channels,  vector boson fusion  $qq\! \to\! Hqq$ and Higgs-strahlung  $q\bar q \!\to \!HV$ with $V\!=\!W,Z$. Associated $p p\!\to \!t\bar  tH$  production  has an even smaller rate.

\vspace*{1mm}

Turning to the heavier $H, A$ and $H^\pm$ bosons, they are almost degenerate in mass in the decoupling regime, when $M_A \gsim 300$--500 GeV, decouple from the $W/Z$ bosons and interact only with fermions with couplings that are enhanced by powers of $\tb$ for $b$-quarks and $\tau$-leptons and suppressed as $1/\tb$ for $t$-quarks. The production and decay rates strongly depend on $\tan\beta$ \cite{Higgs-prod-decay}.  
At high values, $\tan\beta \gsim 10$, the neutral $H,A$ states are mainly produced in $b\bar b$ and $gg$ (through the $b$-loop contributions) fusion  with large rates and decay almost exclusively into $b\bar b$ and $\tau^+\tau^-$, with branching ratios of respectively, 90\% and 10\%. The $H^\pm$ bosons can be produced in the $gb \to t H^-$ mode and would decay into $tb$ and $\tau \nu$ final states, again with branching fractions of 90\% and 10\%, respectively.

\vspace*{1mm}

In the low $\tb$ region of 
$\,\tb\!\lsim 3\hs$, and for Higgs masses above the $2\hs m_t$ threshold, the heavy neutral states will be produced essentially in the $gg \to H/A$ processes with the top quark loop providing the main contribution and will almost exclusively decay into $t\bar t$ final states\footnote{In the process $gg\to H/A \to t\bar t$, one has to take into account both $H$ and $A$ contributions and also their interference   with the $gg\to t\bar t$ QCD background \cite{tt-interference}. Nevertheless, as the experimental collaborations are only starting to consider this interference, we will ignore it in our analysis.}. The $H^\pm$ bosons will mainly decay into $tb$ states with a branching ratio of almost 100\%.

\vspace*{1mm}

For the intermediate $\tb$ region of 
$\hs 3 \lsim \!\tb\! \lsim 10\hs$, the main Higgs production mode will be $gg \to H/A$ with some small additional contributions from $b\bar b$ fusion; the rates are nevertheless smaller than usual as the coupling $g_t^{H/A}$ is suppressed while  $g_{b}^{H/A}$ is not yet enhanced.\ 
For the decays,  there will be a competition between the 
$H/A \ito t\bar t$ and $b\bar b$ modes.\ 
Any additional mode, such as decays into charginos and neutralinos, will impact the rates.

\subsubsection{Diphoton Decay Rates of Higgs Bosons}
\label{sec:4.1.2}
\vspace*{1mm}

A first effect of the gaugino-Higgsino spectrum on the Higgs sector could be seen in the couplings of the lightest $h$ boson which are rather precisely measured at the LHC. The measurement of the so-called $h$ signal strengths in a given channel, such as the $h \to XX$ decay, gives a direct constraint on the coupling $g_X^h$ or its reduced form when it is normalized to the coupling of the SM Higgs boson denoted by $H_{\rm SM}$, which are defined as  
\beq
\kappa_{XX}^{pp \to h}  = \frac{(g_X^h)^2}{(g_X^{H_{\rm SM} })^2 } = \frac{\sigma (pp \to h) \times {\rm BR} (h \to XX)}  { \sigma (pp \to H_{\rm SM}) \times  {\rm BR} (H_{\rm SM} \to XX)} \, . 
\label{s-strengh}
\eeq
 These $\mu_{XX}$ values depend not only on the angles $\alpha$ and $\beta$ outside the decoupling regime, when $\alpha \! \neq \!  \beta \! - \! \frac{\pi}{2}$, but also on the loop contributions of the new particles if they are not too heavy. For most of the couplings, these radiative corrections are small but there might be a notable exception with the $h\gamma\gamma$ coupling as the decay $h \to \gamma\gamma$ proceeds through loops.  Besides the standard ones from the top quark and the $W$ boson, there will be those due to SUSY particles which appear at the same order. In our case, there will be only two new contributions: the one due to the charged Higgs boson which, in any case, is rather small and should already be taken into account in the the context of the MSSM, but also the contribution due to charginos that we we discuss in the following.\footnote{We will ignore here the corresponding chargino (and eventually neutralino) contributions to the other loop induced decay mode, namely $h\to Z\gamma$ \cite{HZp}, which can be probed only with extremely high statistics and which will provide essentially the same information in the hMSSM context.}

\vspace*{1mm}

 These loops have been discussed in several instances and  we will closely follow the relatively recent analysis given in Ref.~\cite{Hgamma} that we will update. We use the program SUSY-HIT \cite{SUSY-HIT} to evaluate the fraction BR$(h \to \gamma\gamma)$ including the chargino contributions in the hMSSM context, that is, we use  the program  HDECAY \cite{HDECAY} where all these loop contributions are included at one-loop order (which should be largely sufficient for our purpose here) with the SUSY particle spectrum generated by the package Suspect\,\cite{Suspect}.  
 
\begin{figure}[t]
\centering
\includegraphics[height=6cm,width=7.8cm]{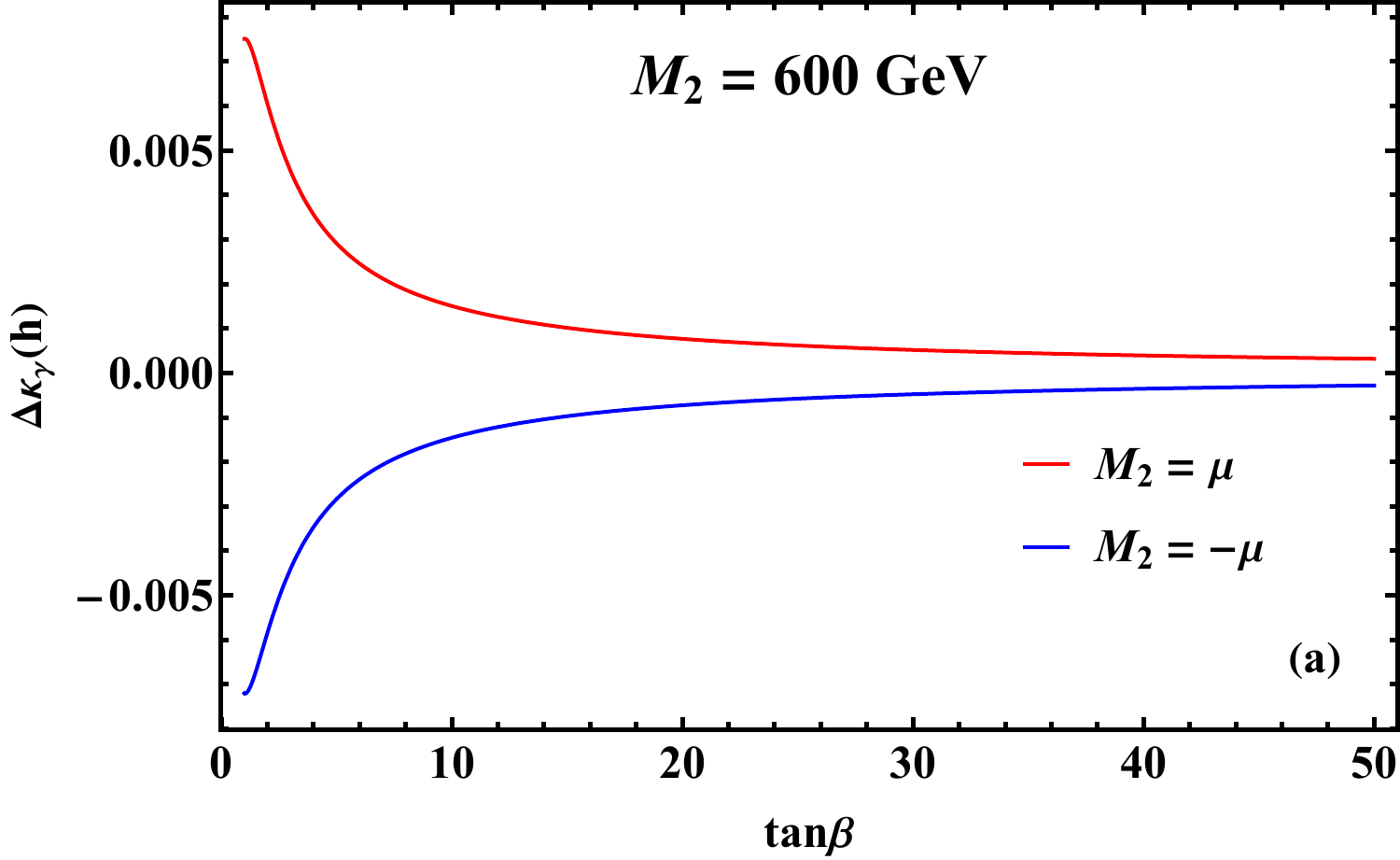}
\includegraphics[height=6cm,width=7.8cm]{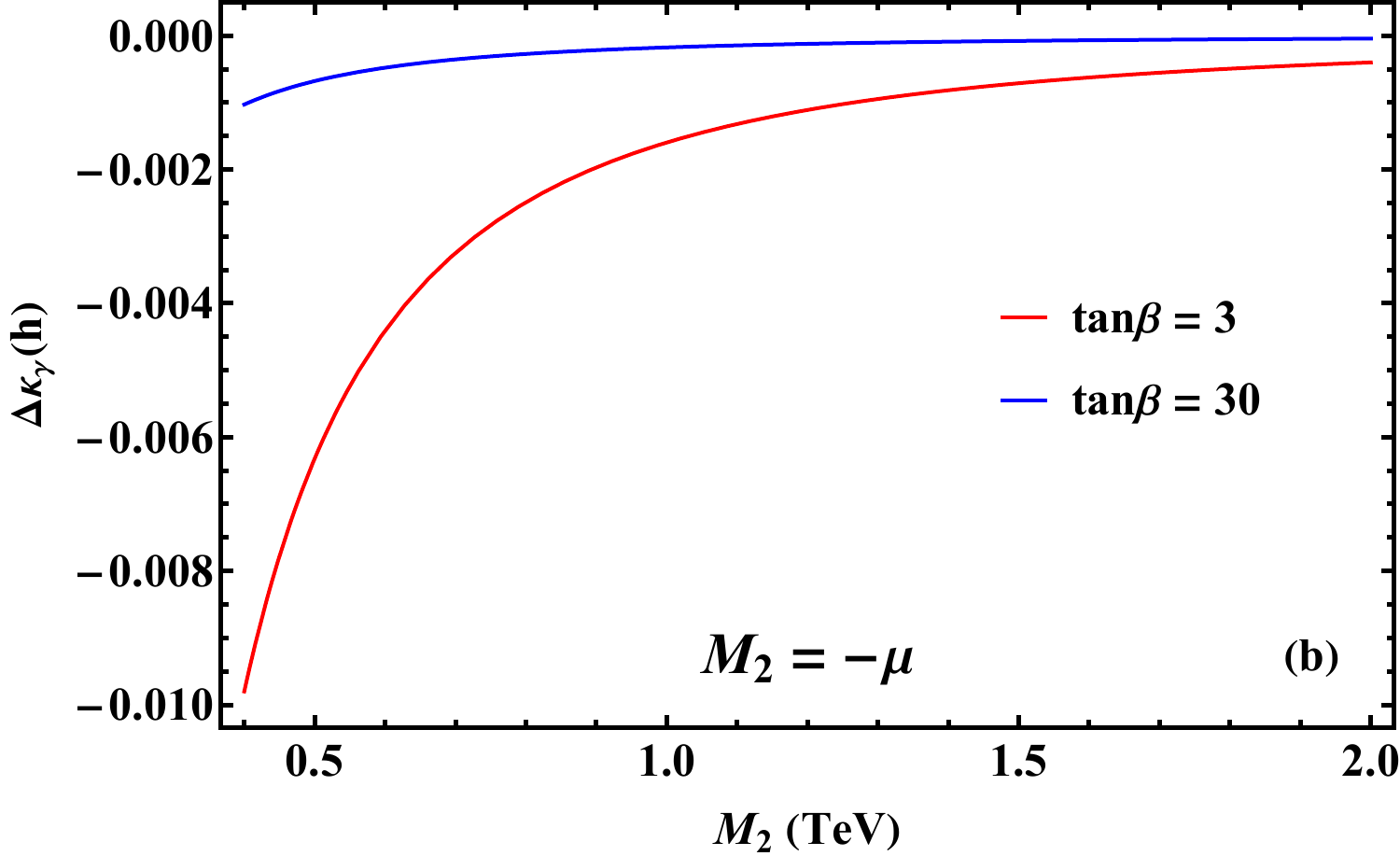}
\vspace*{-2mm}
\caption{Deviations of the $h\gamma\gamma$ coupling 
as a function of $\tan\!\beta$ in the upper panel (a) and of the gaugino mass parameter $M_2$ in the lower panel (b). For (a), the red (blue) curve corresponds to $M_2^{}=\mu\,$ ($M_2^{}=-\mu\,$)  $=600$\,GeV.
For (b), the red (blue) curve corresponds to $\tan\!\beta \!=3\,$ ($\tan\!\beta \!=30\,$) and we set a relation $M_2^{}=-\mu$\,.
}
\label{fig:ga}
\end{figure}

\begin{figure}[t] 
\centering
\includegraphics[height=6cm,width=7.5cm]{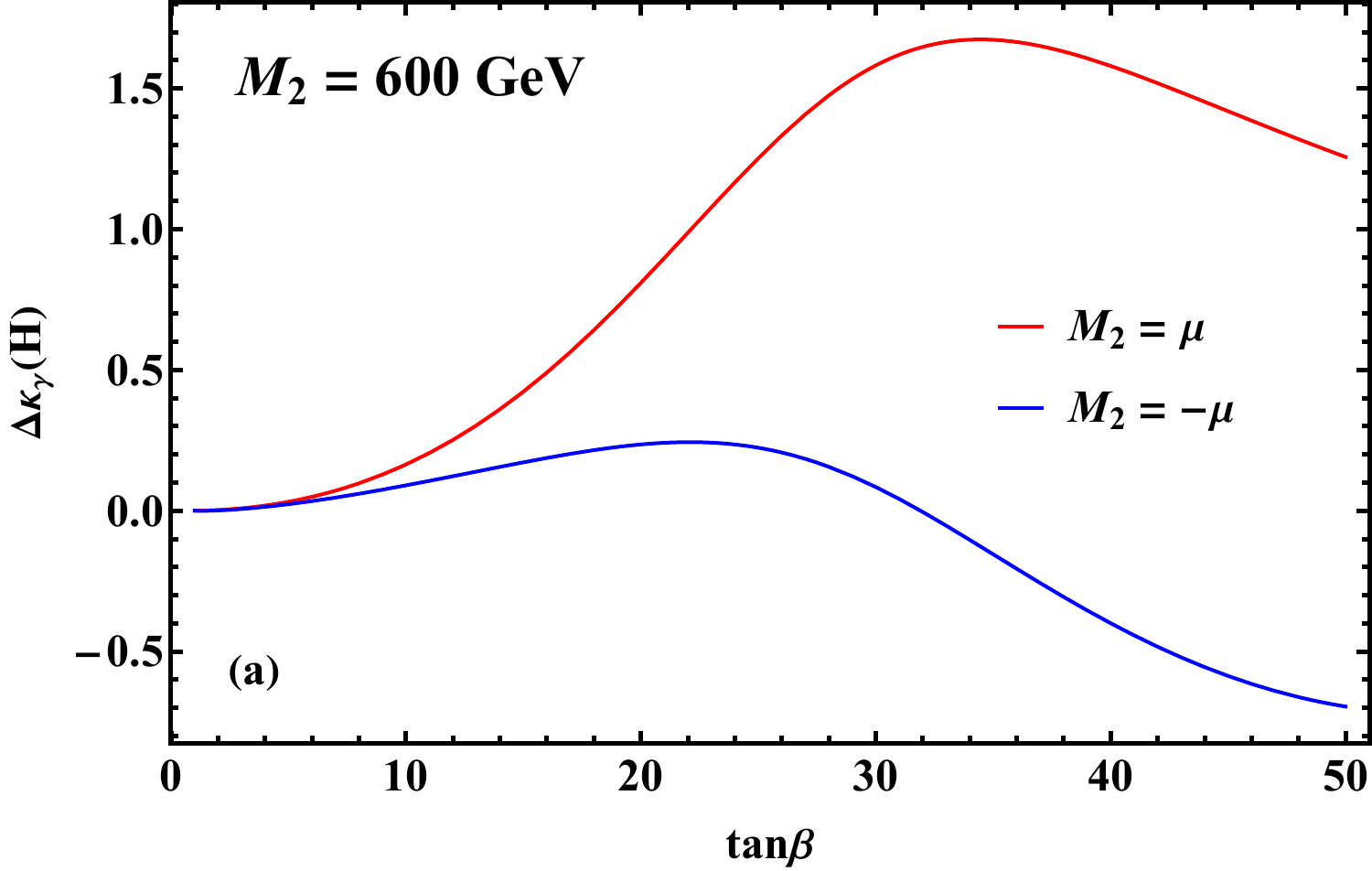}
\includegraphics[height=6cm,width=7.5cm]{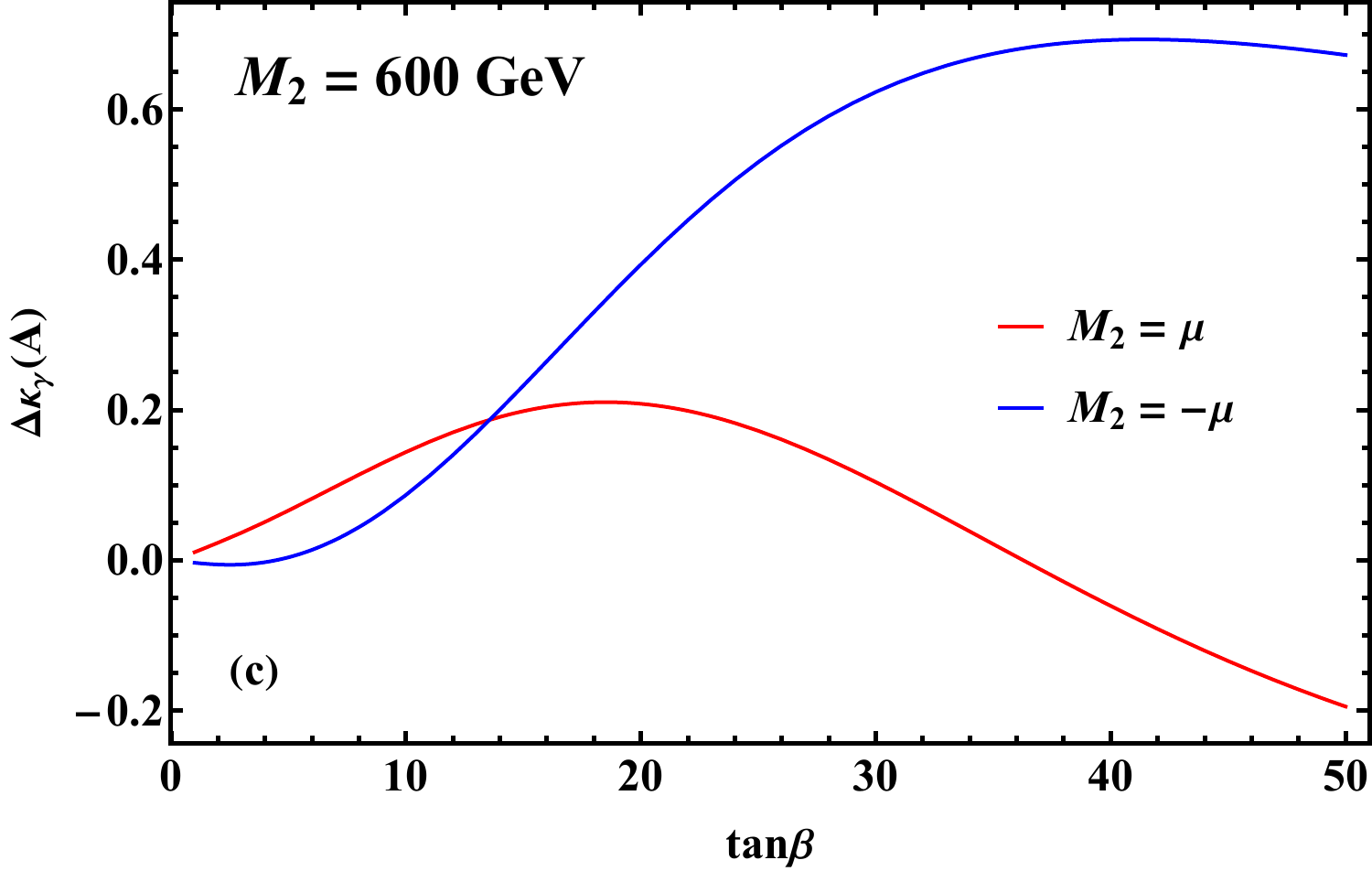}
\\[3mm]
\hspace*{1mm}
\includegraphics[width=7.5cm,height=6cm]{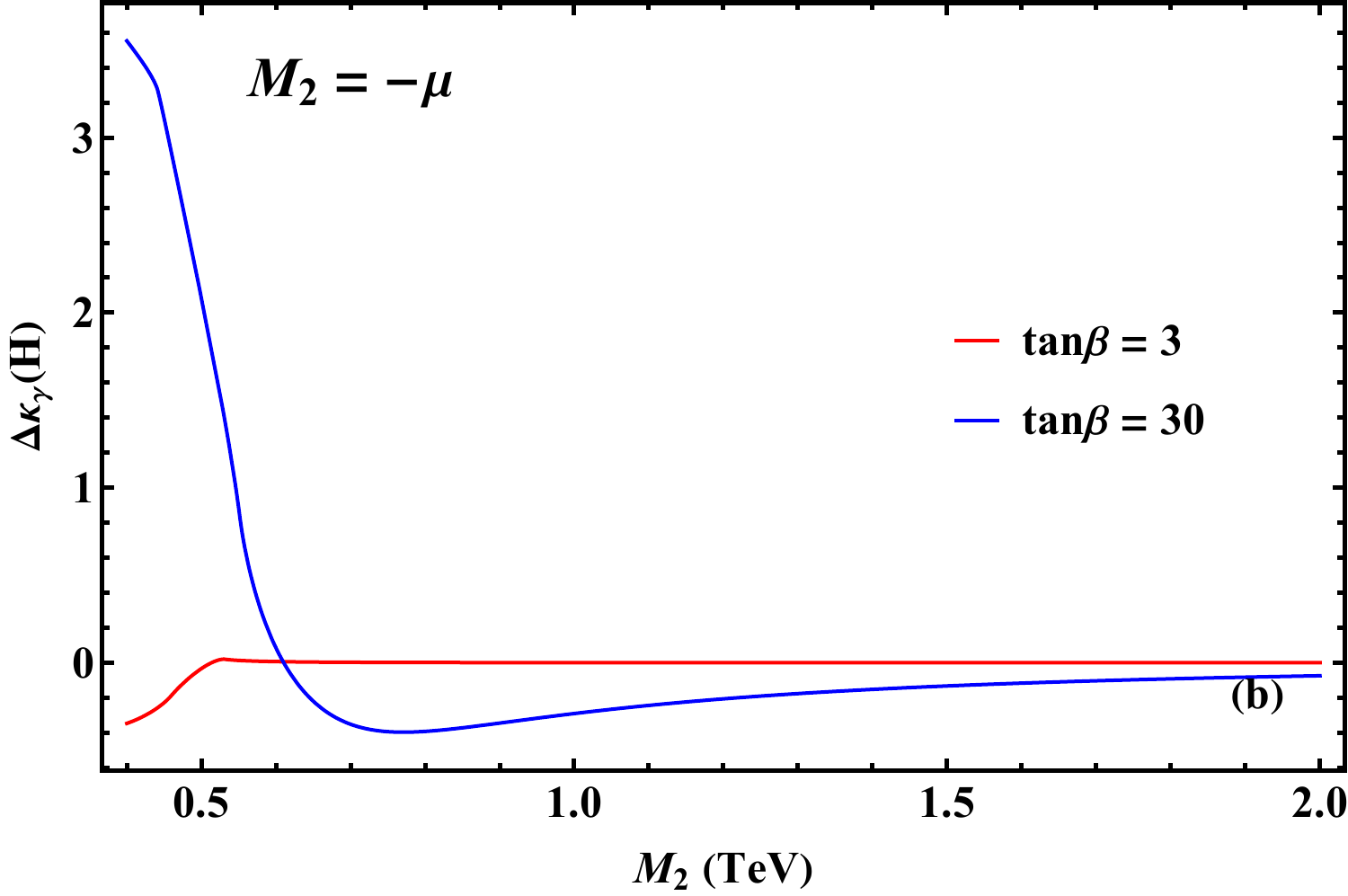}
\includegraphics[width=7.5cm,height=6cm]{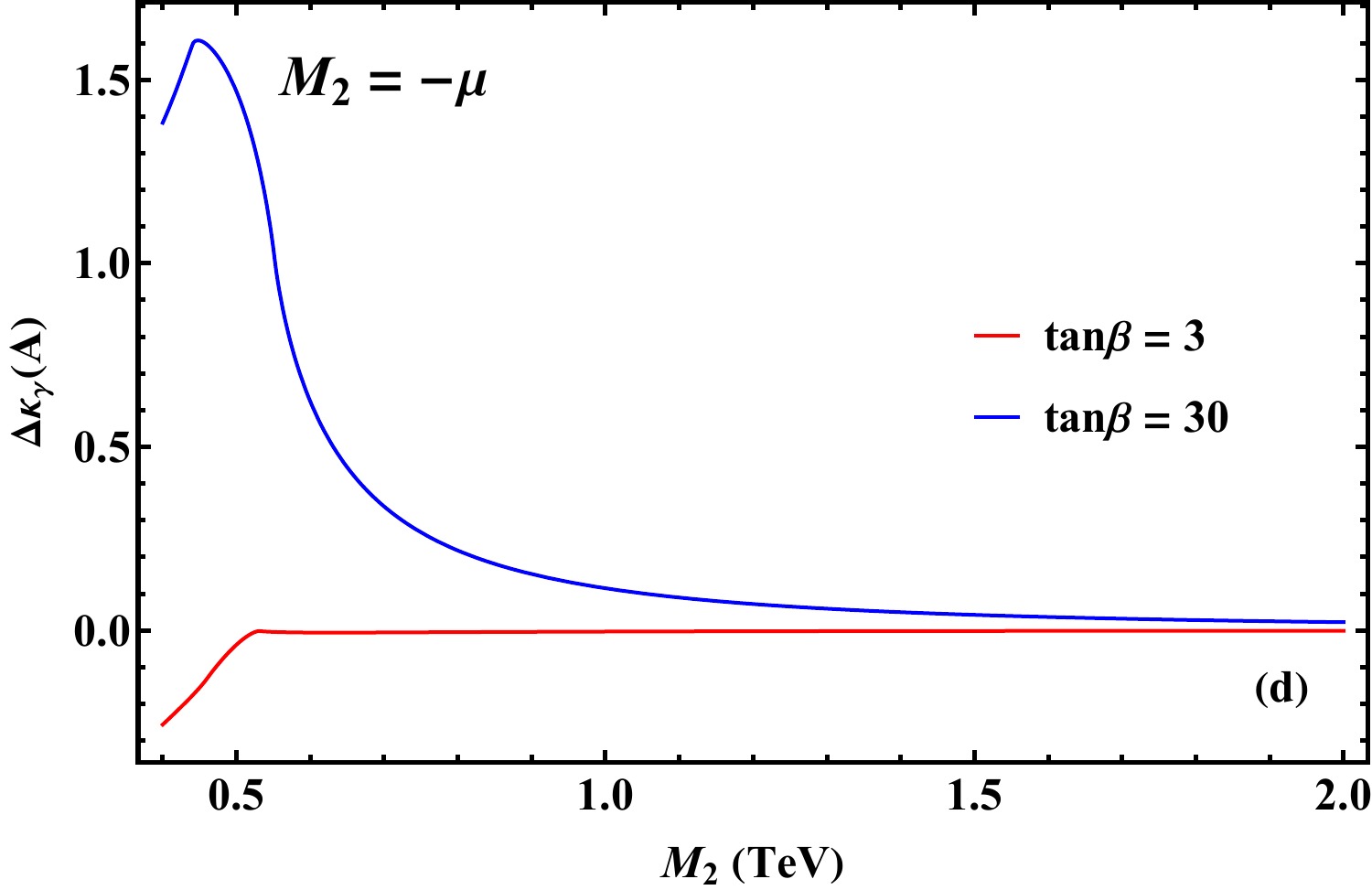}
\vspace*{-3mm}
\caption{Deviations of the $H\gamma\gamma$ and $A\gamma\gamma$ couplings as a function of $\tan\!\beta$ in the upper panel (a-c) and of the gaugino mass parameter $M_2$ in the lower panel (b-d). For (a-c), the red (blue) curve corresponds to $M_2^{}\!=\!\mu$ ($M_2\!=\!-\!\mu)=\!600$\,GeV while for (b-d ), the red (blue) curve corresponds to $\tan\!\beta \!=3$ ($\tan\!\beta \!=30$) and we set $M_2\!=\!-\mu$.
}
\label{fig:gaHA}
\end{figure}

\vspace*{1mm}

The deviation of  the $h\gamma\gamma$ coupling when including the SUSY-loop contributions relative to the case without them, $\Delta\kappa_\gamma^h$
will depend on the values of $(M_2,\,\mu)$ that enter the chargino sector in addition to $\tb$.\ In Fig.\,\ref{fig:ga}, we present the deviation $\,\Delta\kappa_\gamma^h\,$ for the chargino loop corrections to the $h$-diphoton coupling as a function of $\tan\!\beta$ in the upper plot and as a function of the gaugino mass $M_2$ in the lower plot. In both cases, we fix the pseudoscalar mass to $M_A=1$\,TeV, which makes that we are in the decoupling limit and the Higgs couplings are not affected by the angles $\alpha$ and $\beta$. We assume the equality $M_2= |\mu|$ which maximizes the $h$ couplings to the charginos as described in 
section\,\ref{sec:2.2}. In the upper panel but we take take both signs of $\mu$ and vary $\tan\beta$, while in the lower panel, we take $M_2^{}\!=-\mu\,$ only and study the chargino impact as a function of $M_2$ for the two values 
$\,\tb =3\,$ and $\,\tan\!\beta \hsm =30\,$.\  
From this figure, we see that the possible deviation $\Delta\kappa_\gamma^h$ has the same sign as the $\mu$ parameter.
Moreover, $|\Delta\kappa_\gamma^h|$ is smaller than 1\% 
and decreases with the increase of $\tan\beta$ 
[as in Fig.\,\ref{fig:ga}(a)] and the increase of $M_2$ 
[as in Fig.\,\ref{fig:ga}(b)]. 

\vspace*{1mm} 

For comparison, we also present the chargino loop corrections to the $H\gamma\gamma$ and $A\gamma\gamma$ couplings in Fig.\ref{fig:gaHA} for $M_A\!=\!1\,$ TeV. Because the masses of charginos can be smaller than $H$ and $A$, $\Delta^H \kappa_\gamma^{}$ and $\Delta^A \kappa_\gamma^{}$ can be larger than 1. There are significant differences near $M_A\!=\!2M_2 \!\approx \! 2m_{\chi_i^\pm}$ thresholds and the sign of $\mu$ has an opposite impact on $\Delta\kappa_\gamma^{H}$ and $\Delta\kappa_\gamma^{A}$.

\vspace*{1mm}
\subsubsection{SUSY Corrections to Higgs Production and Decays}
\label{sec:4.1.3}
\vspace*{1mm}

As discussed in section 2.4, SUSY particles contribute directly to the Higgs couplings to fermions $g_f^\Phi$ and, hence, impact the MSSM Higgs production and decays rates in a way that cannot be absorbed into corrections to the angle $\alpha$. In particular, at high $\tan\beta$,  the direct correction $\Delta_b$ to the Higgs--$b\bar b$ vertices, the leading part of which is given in eq.~(\ref{Delb}), can be significant when $\mu$ is also large. The SUSY--QCD part of the correction, $\Delta_b^{\rm QCD}$, being proportional to $\tan\beta \mu m_{\tilde g} / {\rm max} (m _{\tilde g}^2, M_S^2)$ and the electroweak one 
$\Delta_b^{\rm EW} \!\propto\! \tan\!\beta\hs \mu /M_S^2 
\!\times\! (\mu, M_1, M_2)\hs$,\, can be made small by setting the SUSY scale $M_S$, and hence the squark masses since $\,M_S^2 \!=\! m_{\tilde t_1}^{} m_{\tilde t_2}^{} 
\!\approx\! m_{\tilde q}^2\hsx$,\,  
to very large values as we are assuming in the hMSSM considered here.\ This would also be the case of the stop--chargino and sbottom--neutralino contributions to the electroweak correction $\Delta_b^{\rm EW}$.

\vspace*{1mm}

\begin{figure}[t]
\begin{center}
\includegraphics[height=10cm]{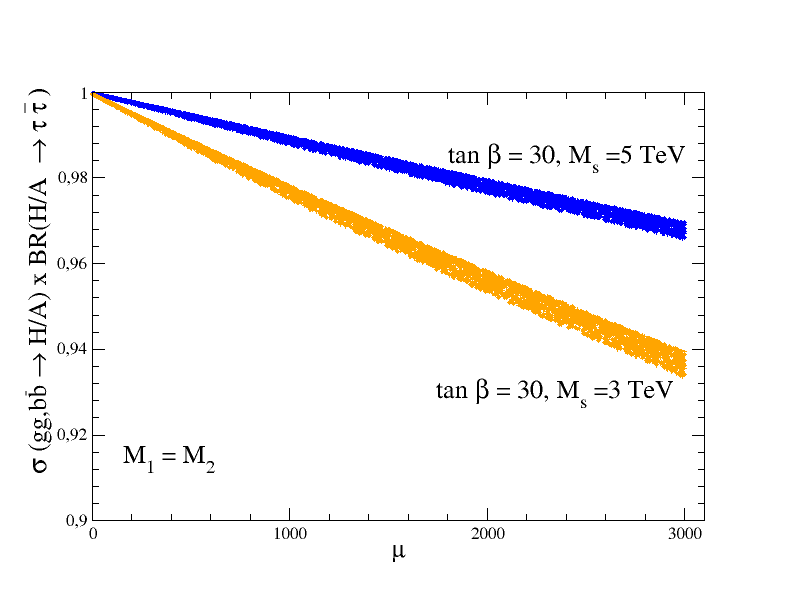}
\vspace*{-10mm}
\caption{Impact of the $\Delta_b$ correction on the production times decay rates $pp\! \to\! H/A\! \to\! \tau^+ \tau^-$  and $pp\! \to\! tb H^\pm\! \to\! tb  \tau \nu$ for the large $\tan\!\beta$  
from a scan on the $(M_2,\hs \mu)$ parameter space.}
\label{fig:sigBRHA}
\vspace*{-6mm}
\end{center}
\end{figure}

Nevertheless, even if squarks and gluinos are light enough to have sizeable contributions to $\Delta_b$, the latter will have  only a  limited impact in the main detection channels of the heavy MSSM Higgs states when the full production times decay rates in the dominant processes are taken into account. Indeed, for $H/A$, the main processes considered at the LHC are $gg, b\bar b \to H/A \to \tau^+ \tau^-$  and while the production cross sections are modified as
\beq
\sigma (gg,b\bar b \to H/A)  \propto (1+ \Delta_b)^{-2} \, , 
\label{Delta_b_sig}
\eeq
one would have the following modification f
on the decay branching fractions: 
\beq
\hspace*{-5mm}
{\rm BR}(H/A \ito \tau\tau) =  \Gamma(H/A \ito \tau\tau)/[(1+ \Delta_b)^{-2}\Gamma(H/A \ito b\bar b)+
\Gamma(H/A \ito \tau\tau)] \, ,  
\label{Delta_b_br}
\eeq
assuming, as it is generally the case, that the corresponding $\Delta_\tau$ correction is small enough to be negligible.\  
The $\Delta_b$ correction  will then largely cancel out in the product of the cross section and branching fraction:
\beq
\sigma (gg,b\bar b \to H/A) \times {\rm BR}(H/A \to \tau\tau)  \simeq  1-  \Delta_b/5 \, .  
\label{Delta_b_sbr}
\eeq
Hence, only when the $\Delta_b$ correction is very large (say, of order unity), its impact on the 
\,$pp \ito H/A \ito \tau\tau$\, 
rate would become of the order of the theoretical uncertainty of the process (stemming from the scale and the PDF uncertainties),  
which is estimated to be about 20\% \cite{Higgs-prod-decay}. 
Thus the $\Delta_b$ effect is insignificant.\ 
Such a large $\Delta_b$ correction does not occur in our hMSSM scenario as we assume the squarks to be heavy enough.\ This is illustrated in Fig.\,\ref{fig:sigBRHA}, from a scan over the parameters $M_2$ and $\mu$ and for a representative large 
$\hs\tb \hsm =\hsm 30\hs$ value and not too heavy squark masses, $M_S\!=\!(3\!-\!5)$\,TeV. Note that it would reach about $10\%$ only for values $\tb \!=\!50$, $M_S\!=\!3$\,TeV, and large 
$|\mu|\!\simeq\! 3$\,TeV.\ 
Hence, the limits set by ATLAS and CMS on the heavier MSSM Higgs bosons, which are dominantly derived from the channels above, should not be affected by  these SUSY direct corrections.

\vspace*{1mm}

The discussion above holds partly in the case of the charged Higgs for which the main production 
process is $bg \to H^\pm t $.  At high $\tan\beta$, its cross section is also modified as in eq.~(\ref{Delta_b_sig})  and in one the main detection channels, $H^\pm \to \tau \nu$, it has the same branching ratio as in  eq.~(\ref{Delta_b_br}) since BR$(H/A \to \tau\tau,  b\bar b)$ should be replaced by BR$(H^\pm \to  \tau \nu, tb)$. The product of the two, $\sigma ( bg \to H^\pm) \times {\rm BR}( 
H^\pm \to \tau \nu)$,  will then also behave as in  eq.~(\ref{Delta_b_sbr}) and, hence, the correction will largely cancel out in the cross section times branching ratio. However, another important channel for the charged Higgs boson at high $\tan\beta$ (as well as low) values, is the $H^\pm \to tb$ channel which could also be strongly affected by the $\Delta_b$ correction. The cross section times decay branching ratio will behave in this case as 
\beq
\sigma (gb \ito H^\pm) \!\times\! {\rm BR}(H^\pm \ito tb)  
\,\simeq\,  1\!-\! {11} \Delta_b/5 \, ,  
\label{Delta_b_H+}
\eeq
and will be thus more significantly affected if the $\Delta_b$ correction is large. In this case, one will need to take into account the value of the relevant SUSY parameters (in particular the gluino and squarks masses which enter the dominant SUSY-QCD corrections) in order to fix this corrections. But again, one can choose a benchmark scenario in which all these masses are  large enough not to affect the Higgs vertices.

\vspace*{1mm}

In any case, for the charged Higgs boson,  the $H^\pm \to tb$ channel is up to now not as sensitive as the $pp \to H/A \to \tau^+\tau^-$ process and will not change the LHC  present sensitivity limits of the ATLAS and CMS experiments. Note that we have ignored possible effects in the $H^\pm tb$ vertex stemming from corrections due to the top/stop sector at low $\tan\beta$. The $\Delta_t$ corrections, as in the case of the $\Delta_\tau$ corrections are rather small.  

\vspace*{1mm}

Let us finally make an important remark on the case of the lightest $h$ boson couplings that are now measured rather precisely at the LHC. In principle, the $hbb$ coupling also receives $\Delta_b$ corrections, but as discussed recently in Ref.~\cite{last-SUSY-paper} for instance, close to the decoupling limit $\hs M_A \!\gg\! M_Z\,$ the deviation from 
its SM--like value will be given by
\beq 
\Delta g_d^h  \,\simeq\, \Delta_b 
\frac{\,1\!+\! \tan^2\!\beta}{\tan\!\beta\,} 
\!\times\! \frac{M_Z^2}{\,2M_A^2\,} \sin\! 4\beta 
\stackrel{\tb \gg 1}{\longrightarrow}  - \Delta_b \!\times\!  \frac{M_Z^2}{\,8M_A^2\,} \,, 
\eeq   
which is very strongly suppressed for 
$\,M_A \!\gsim\hsm 2 M_Z$\,.\ 
Hence, these direct corrections should be very small and will not affect the signal strengths of the $h$ boson  measured at the LHC to which we turn our attention now. 

\vspace*{1mm}
\subsection{Constraints on the Parameter Space of the Higgs Sector}
\label{sec:4.2}
\vspace*{1mm}

\begin{figure}[!h]
\begin{center}
\includegraphics[width=10.5cm]{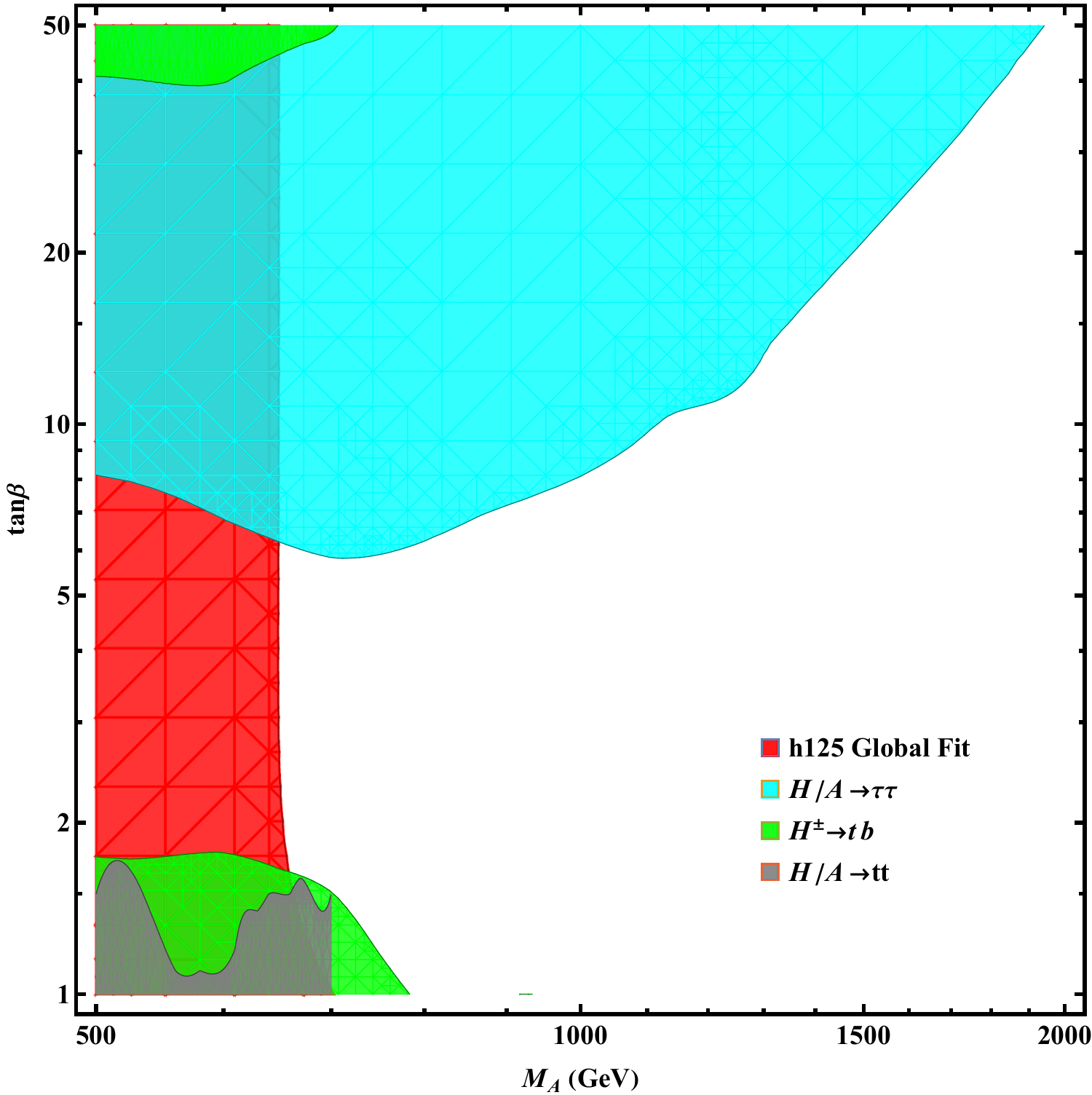}
		\vspace*{-1mm}
\caption{Constraints on the hMSSM viable parameter space
in the $[M_A^{},\,\tan\beta]$ plane at 95\%\,C.L.,\ 
using the LHC measurements of the signal strengths of
the SM-like Higgs boson $h$ (the red area)  
as well as the direct searches of the heavier neutral Higgs bosons in the reaction $pp\!\to\! H/A \!\to\! \tau^+ \tau^-$  (the blue area), the charged Higgs boson in the reaction $pp\!\to\! H^+ \!\to\! t\bar b$ (the green area) and the neutral Higgs bosons in the channel $pp\!\to\! H/A \!\to\! t\bar t$ 
(the grey area).}
\label{fig:1}
\vspace*{-3mm}
\end{center}
\end{figure}

In this subsection, we study the present constraints on 
the parameter space of the hMSSM Higgs sector as imposed by 
the ATLAS and CMS searches performed at the Run-II of the LHC.\  
These constraints arise from two sources.\ 
The first one is due to the measurements of 
various couplings of the observed Higgs boson of 
mass 125\,GeV \cite{ATLAS:2020qdt}\cite{CMS:2020gsy} which, in our context, corresponds to the lightest $h$ state.\  
Another source of constraints arise from the direct searches 
of the heavier neutral and charged Higgs bosons in 
various channels \cite{ATLAS:2020zms}\cite{ATLAS:2020jqj}\cite{CMS:2022rbd}\cite{CMS:2019pzc}.\ 
Both constraints already exclude a significant portion of the $[M_A,\,\tb ]$ parameter space.\ We recast the two sets of constraints above in our hMSSM context 
and we start by summarizing our results in Fig.\,\ref{fig:1} 
that we will describe below. 

\vspace*{1mm}

For the direct or indirect searches, the main precision measurements of the SM-like Higgs couplings come from the bosonic decays 
$h \ito ZZ,WW$ and $\gamma\gamma$ with the $h$ boson dominantly produced from gluon fusion, $gg\ito h\,$.\ The measurements of the fermionic couplings in the decays $h\ito \tau^+\tau^-$ and $b\bar b$ are less precise and the contribution of the other production channels such as vector boson fusion $\,pp\ito h qq\,$ and associated production $pp \ito hV$ add only little.\ 
The measured signal strengths in these main production and decay channels, $\mu^{\rm ggF}_{\gamma\gamma, ZZ, WW}$ are as defined in eq.\,(\ref{s-strengh}). The combination of measurements of the SM-like Higgs production and decay channels using the full set of 139 or 137\,fb$^{-1}$ data collected at the LHC with $\sqrt{s}=13$ TeV by ATLAS\,\cite{ATLAS:2020qdt} and CMS \,\cite{CMS:2020gsy} 
gives:
\beqs 
\begin{eqnarray}
\hspace*{-8mm}
\text{ATLAS}:&&\mu_{\gamma \gamma}^{ggF}\!=1.03^{+0.11}_{-0.11},\,~~ \mu_{ZZ}^{ggF}\!=0.94^{+0.11}_{-0.10},\,~~
\mu_{WW}^{ggF}\!=1.08^{+0.19}_{-0.18},\noindent 
\\[1mm]
\hspace*{-8mm}
\text{CMS}:&&\mu_{\gamma \gamma}^{ggF}
\!=1.09^{+0.15}_{-0.14},\,~~ 
\mu_{ZZ}^{ggF}\!=0.98^{+0.12}_{-0.11},\,~~
\mu_{WW}^{ggF}\!=1.28^{+0.20}_{-0.19}\, , 
\end{eqnarray}
\eeqs
where the errors correspond to the total theoretical plus experimental uncertainties which have been added in quadrature. As can be seen, up to the level of about 10\%, all signal strengths are around unity, meaning that the Higgs particle has SM-like couplings. In the MSSM, as the $h$ couplings should be modified by the angles $\alpha$ and $\beta$ outside the decoupling regime, one concludes that the value of the pseudoscalar $A$ boson should be large, $M_A \gg M_Z$, in order to be close to this regime. To quantify the implications on these measurements  on the $[M_A,\,\tan\beta]$ parameter space, we perform the following  $\chi^2$ fit\footnote{The fit has also  been performed by ATLAS (see e.g. \cite{ATLAS:2021ayy}) and CMS and in a more accurate way. We nevertheless need to do our own fit as we will extend it later to include some additional SUSY effects.}  
\begin{eqnarray}
\chi^2=\sum {\,(\mu_{\rm hMSSM}-\mu_{\rm EXP})^2\,}/
{\sigma_{\rm EXP}^2},
\end{eqnarray}
where the quantities $\mu_{\rm EXP}, \sigma_{\rm EXP}$ are the signal strength and uncertainty from the observations given above, and the quantity $\mu_{\rm hMSSM}$ is the corresponding signal strength in the hMSSM which is a function of $M_A$ and $\tan\beta$. As shown by the red area in Fig.~\ref{fig:1},  the hMSSM fit has ruled out the mass range $M_A \lsim 600$ GeV at the 2$\sigma$ level. The excluded  area does almost not depend on the value of $\tan\beta$. The reason is that $\cos(\beta-\alpha)$ which measures the departure from the decoupling limit reads at high $M_A$ values \cite{last-SUSY-paper}
\beq
\cos (\beta-\alpha) \stackrel{M_A \gg M_Z}{\longrightarrow} \frac12 \frac{M_Z^2}{M_A^2} \sin 4\beta \longrightarrow
\frac12 \frac{M_Z^2}{M_A^2} \times \bigg\{ 
\begin{array}{l} - 4/\tb~~      {\rm for}~\tb \gg 1 \\ 
1- \tan^2\beta~   {\rm for}~\tb \; \sim \, 1 
\end{array}
\ \to 0 \, , 
\eeq
is suppressed not only by ${M_Z^2}/{M_A^2}$ but also by $\tb$, both at high and low $\tan\beta$ values.

\vspace*{1mm}

Moving to the constraints from direct LHC searches, we will take into account those of heavy Higgs bosons performed by ATLAS \cite{ATLAS:2020zms,ATLAS:2020jqj} and CMS \cite{CMS:2022rbd,CMS:2019pzc} again using the full integrated luminosity of 139 fb$^{-1}$ at $\sqrt s = 13$ TeV.  The search that provides by far the strongest constraint is the one performed in the channel $pp \to gg, b\bar b \to H/A \to \tau^+\tau^-$ \cite{ATLAS:2020zms,CMS:2022rbd} and it excludes at the 95\% CL the large part of the $[M_A,\tan\beta]$ plane depicted by the blue area in Fig.\,\ref{fig:1}. Values of $M_A$ below the TeV range are excluded for $\tan\beta \gsim 10$. 

\vspace*{1mm}

The search of the charged Higgs boson in the channel $pp \to tbH^+$ with $H^\pm\!\to tb$ \cite{ATLAS:2020jqj}  is much less constraining than the previous one at high $\tan\beta$, since it is sensitive only for $\tb$ close to our upper limit $\tan\beta \approx 50$ and Higgs mass values $M_{H^\pm} \approx M_A$ which are already excluded by the $h$ signal strength measurements. However, the search is also sensitive to very small $\tb$ values, 
$\tb = 1\!-\!2$ for $M_{H^\pm} \lsim 800$\,GeV.   

Note that in this low $\tan\beta$ area, the search channel $pp \to gg \to H/A \to t\bar t$ should be in principle very efficient, in particular when interference effects with the QCD background are taken into account. However, the one performed by the CMS collaboration with only a luminosity of about 36 fb$^{-1}$ \cite{CMS:2019pzc} is, for the time being, weaker than the one from the $H^\pm \to tb $ mode discussed above. The corresponding ATLAS search, performed with the same integrated luminosity \cite{ATLAS:2018rvc},  did not include the interference effects and hence, has not been interpreted in the Higgs resonance context. We also find that the bounds imposed by other heavy Higgs search channels \cite{ATLAS:2021ayy}, including $H\to WW,ZZ, hh$ and $A \to hZ$ for instance, are weaker and are already covered by the exclusion regions shown in Fig.\,\ref{fig:1}.

\vspace*{1mm}

In a next step, we perform the previous $\chi^2$ fit of the $h$ coupling measurements in the $[M_A,\tan\beta]$ parameter space, but this time, taking account the possible effect of the parameters $\mu$ and $M_2$ for which we perform the scan in eq.\,(\ref{eq:scan}) with  $M_1 \ll M_2$ and $M_2=2 M_1$ in, respectively, the left and the right panels. Again,  the red dots represent the hMSSM parameter space which obeys all the experimental constraints that we considered here. The blue curves in each panel represent the constraints from the   $\chi^2$ fit of the SM-like $h$ coupling measurements 
with the two parameters $M_A,\tan\!\beta$ only. Comparing the allowed region (marked by red dots) by the  four-parameter fit of ($M_A^{},\,\tan\!\beta,\,M_2,\,\mu$) 
with the bound (blue curve) by the two-parameter fit of 
($M_A^{},\,\tan\beta$), we see that the lower bound on $M_A^{}$ in four-parameter fit is relaxed modestly.\

\vspace*{1mm}

{The reason follows from the fact that the decay channel $h\!\to\!\gamma\gamma$ is very important to the $\chi^2$ fit of the $h$ measurements. Since low $M_2$ and $\mu$ values can affect the $h\gamma\gamma$ coupling,  with two more degrees of freedom ($M_2,\,\mu$), 
there is more available parameter space for $(M_A,\tan\beta)$.\ 
The red dots satisfy the constraint of heavy Higgs search without considering the impact of charginos and neutralinos.\ We will discuss the bounds including the decays of heavy Higgs bosons to charginos and neutralinos in the next subsection.}  

\begin{figure}[t]
	\centering
	\includegraphics[width=7.8cm]{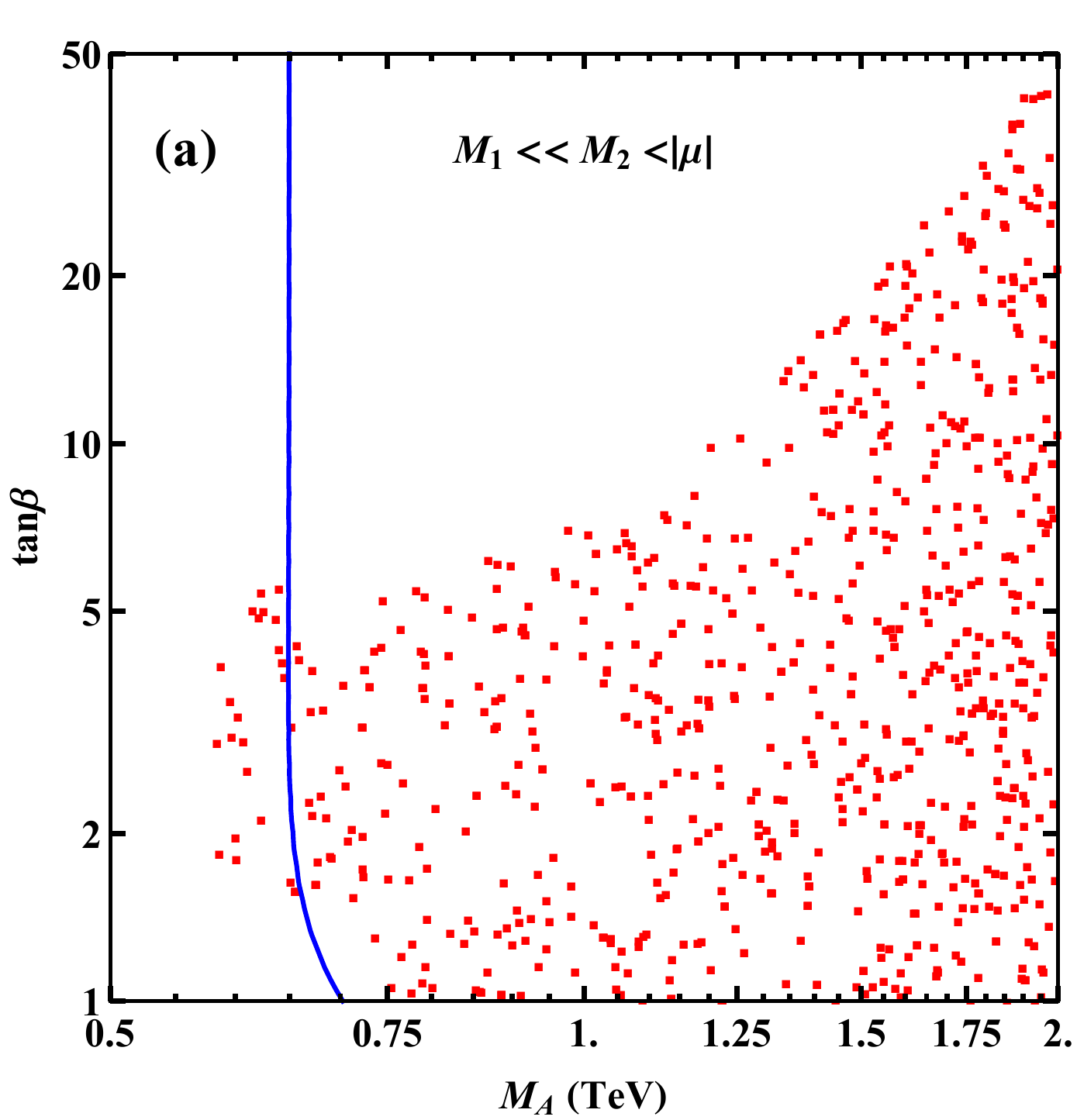}
	\includegraphics[width=7.8cm]{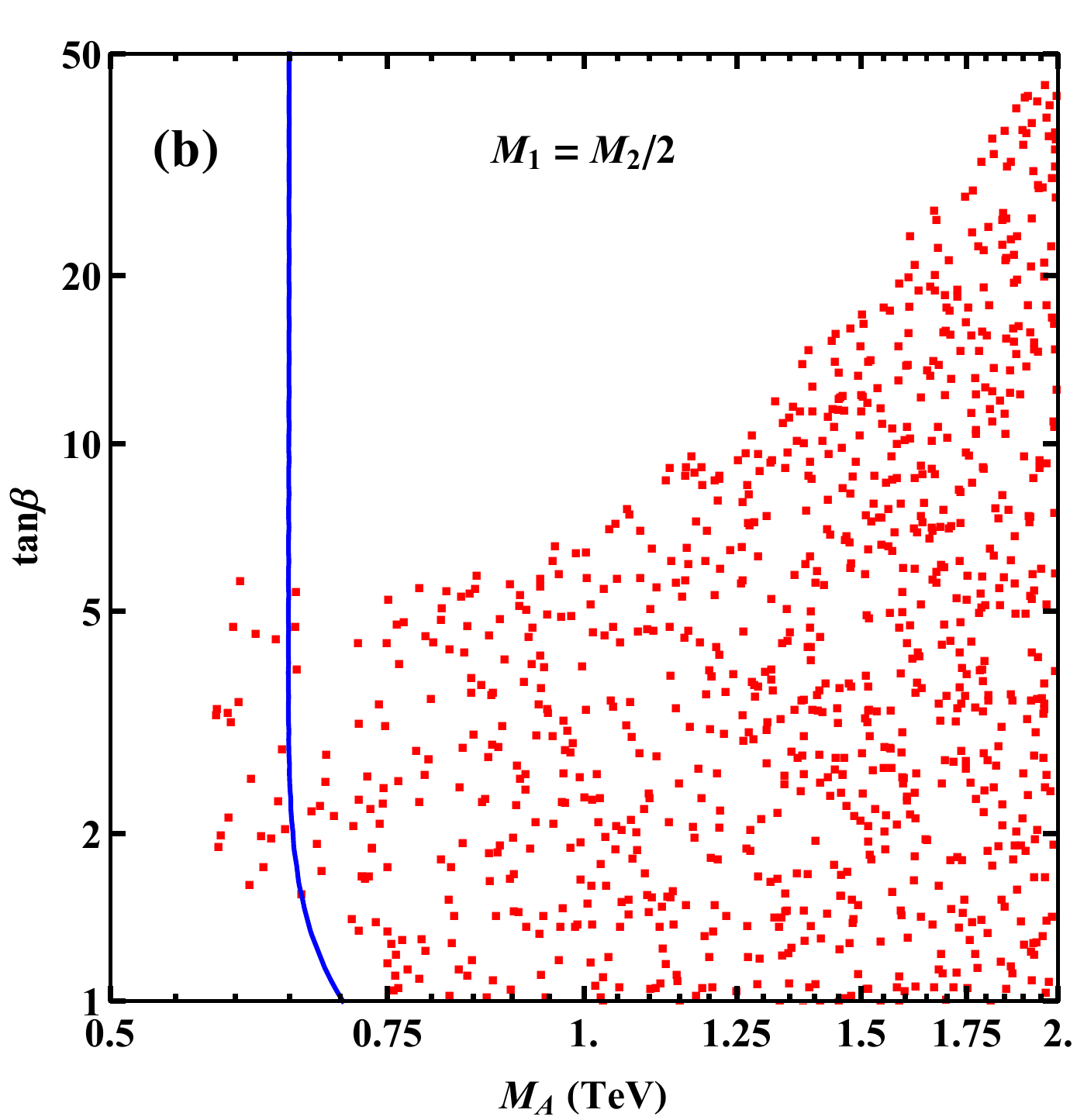}
\vspace*{-3mm}
\caption{Allowed parameter space in the $(M_A, \tan\beta)$ plane at the 95\%\,C.L., obtained by fitting the $h$ coupling measurements and direct search of $H/A$ at the LHC and including the effects of the parameters $M_2$ and $\mu$ when setting $M_1 \!\ll\! M_2$ (left panel) and $M_2=2 M_1$ (right panel).\ The regions on the left of the blue curves in each plot are excluded by the LHC $h$ coupling measurements with the parameters 
$(M_A, \tan\!\beta)$ only.}
\label{fig:2}
\end{figure}

\vspace*{1mm}
\subsection{Higgs Decays into Charginos and Neutralinos}
\label{sec:4.3}
\vspace*{1mm}

A very important feature in the context of a light gaugino-higgsino spectrum is that it allows for the decays of at least the heavier MSSM Higgs bosons into charginos and neutralinos.\ 
These decays can be significant in some of the hMSSM parameter space and thus can make a large impact on the phenomenology of the Higgs sector, as these will affect the LHC Higgs searches, but also the chargino and neutralino sector, as they provide a new window 
for the detection of these particles.\ 
We analyze this aspect in this subsection.

\begin{figure}[!h]
\centering
\includegraphics[height=7cm,width=7.5cm]{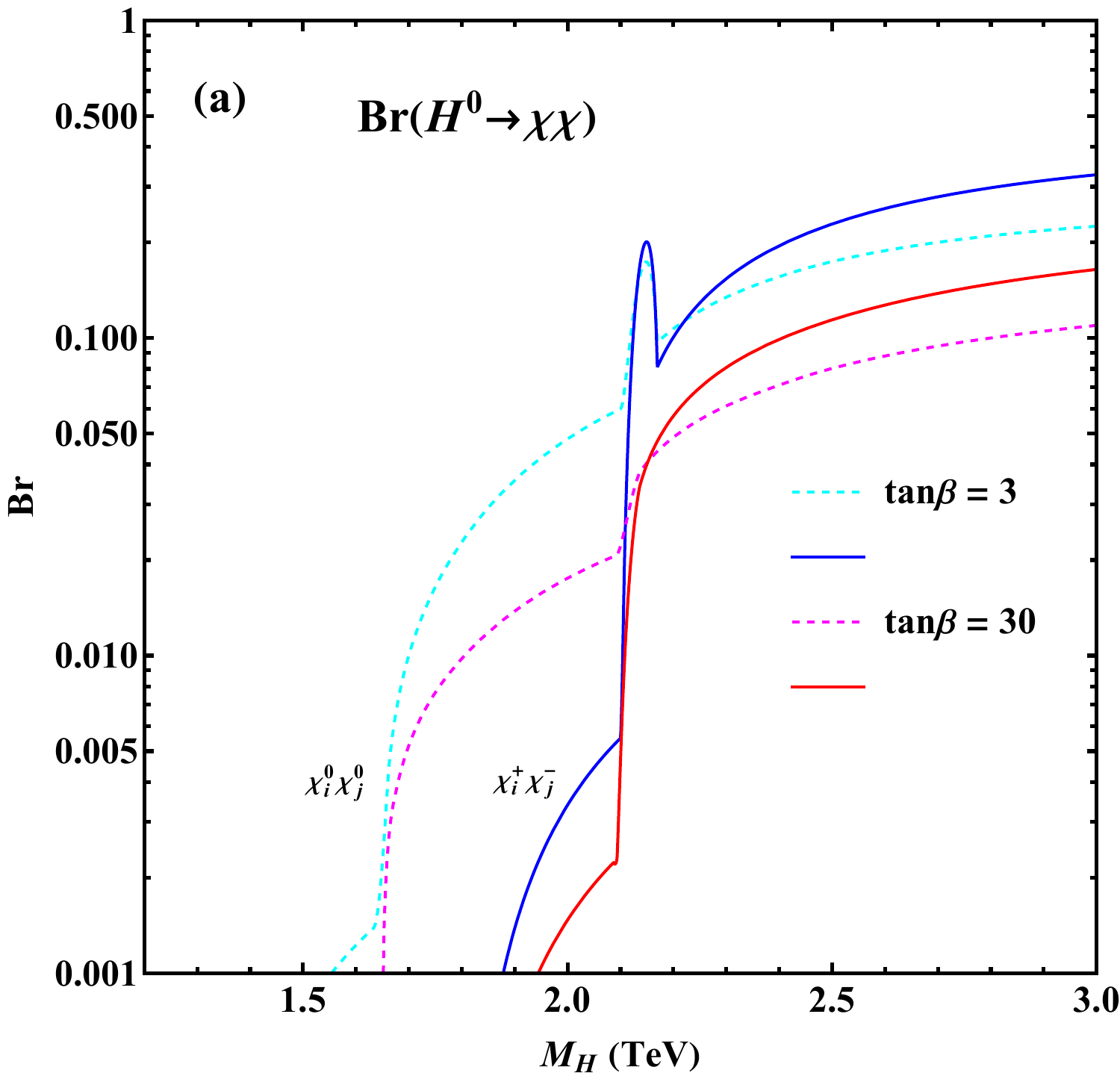}
\includegraphics[height=7cm,width=7.5cm]{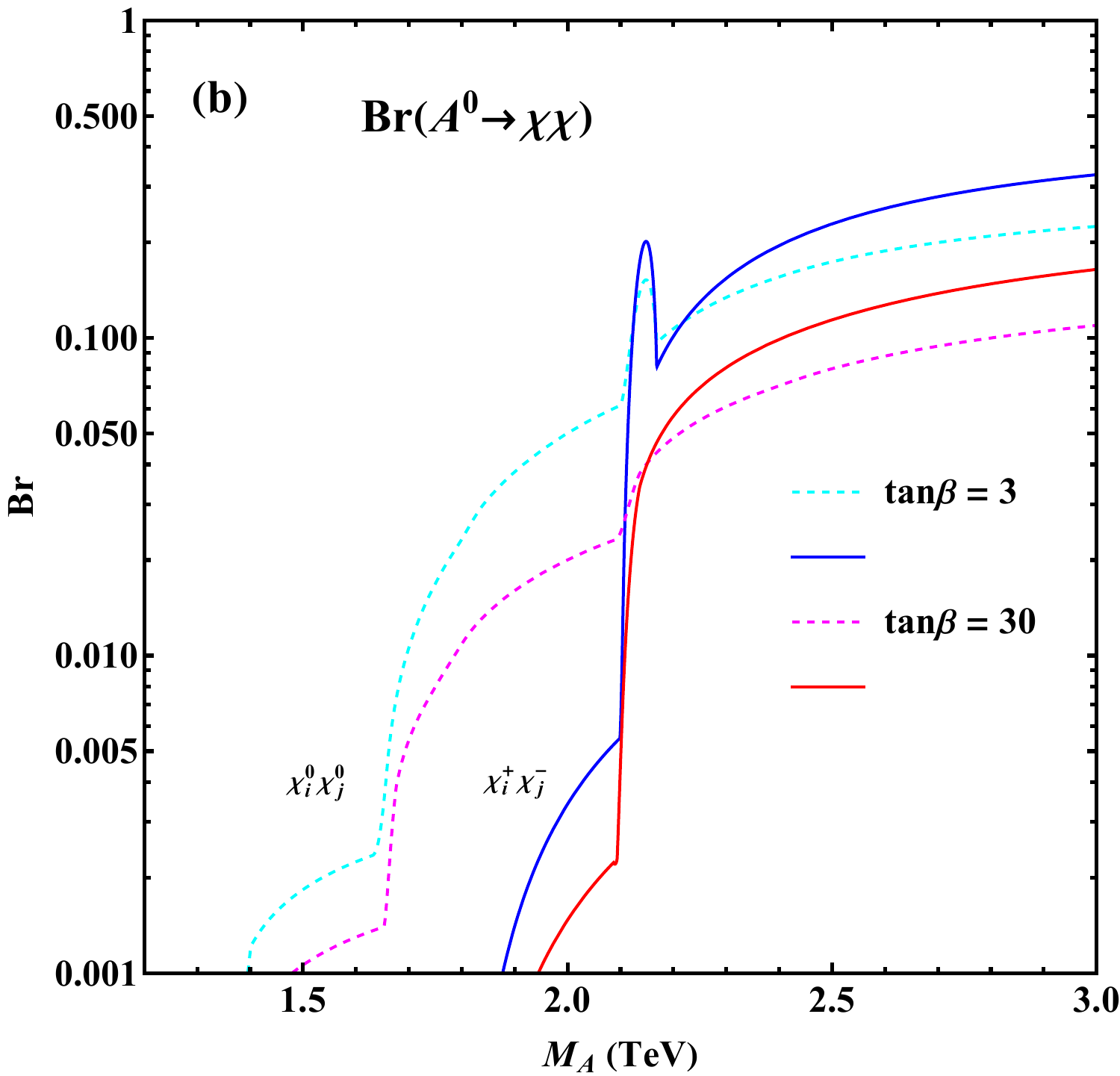}
\\[2mm]
\includegraphics[height=7cm,width=7.5cm]{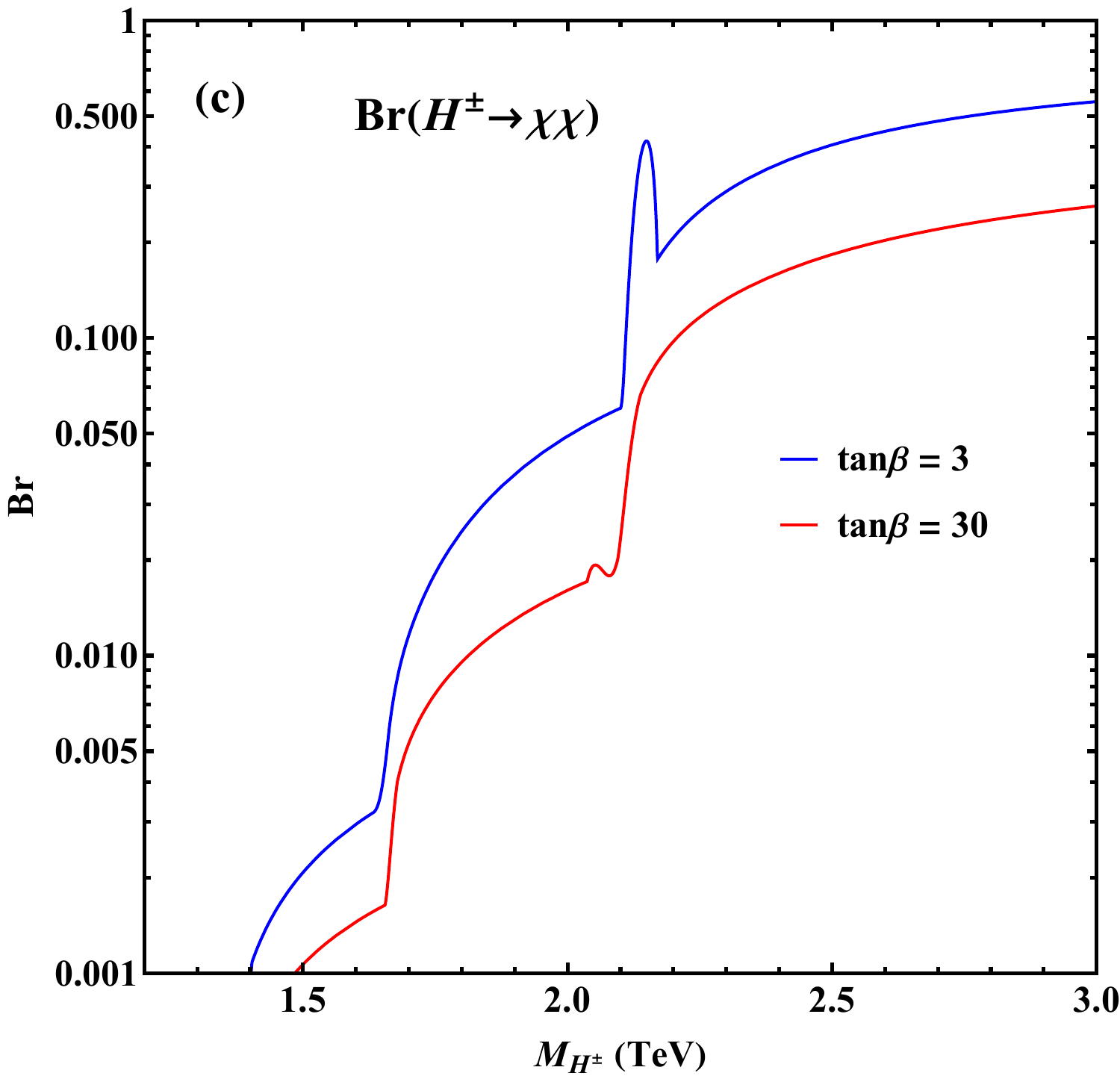}
\includegraphics[height=7cm,width=7.5cm]{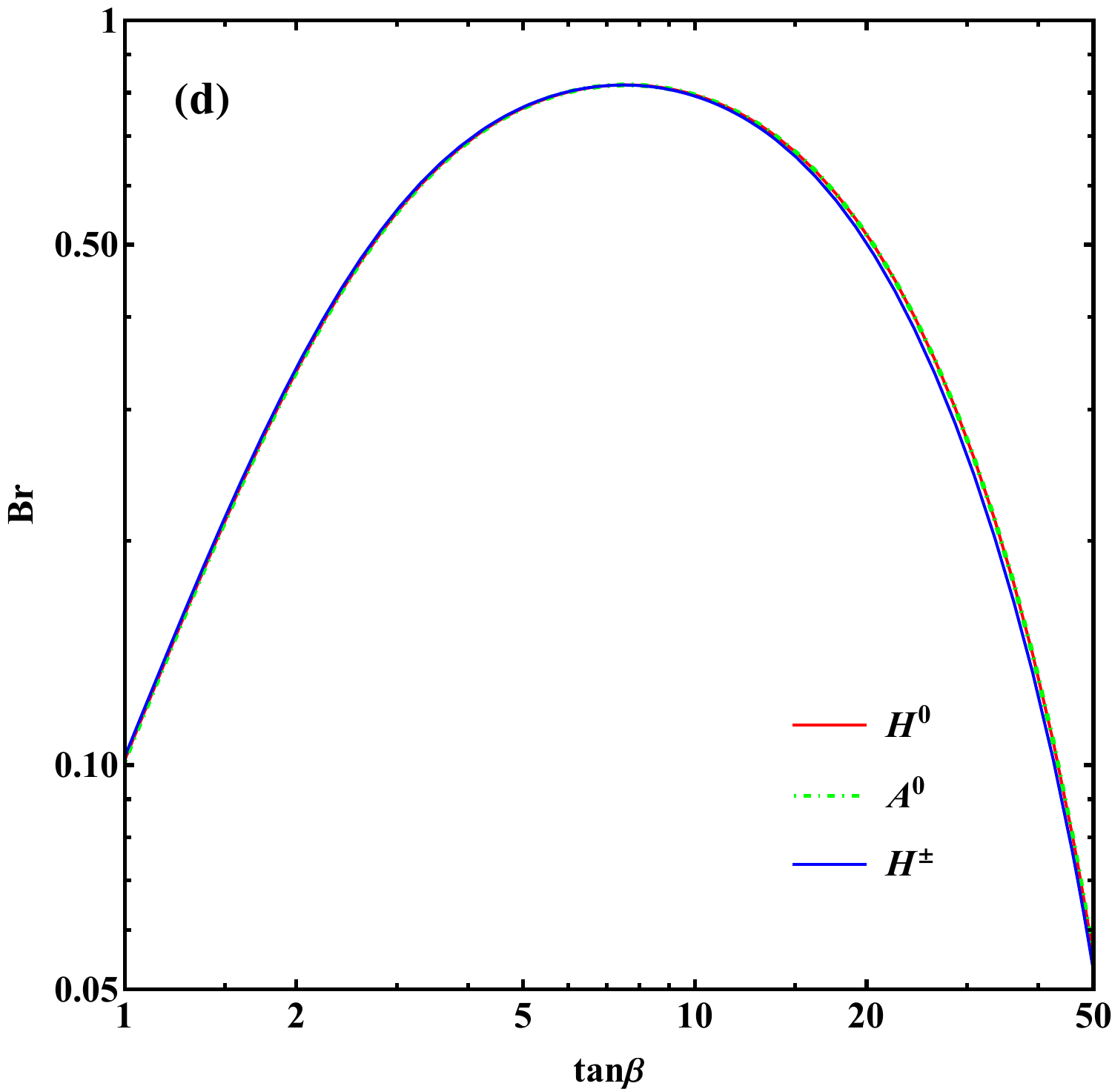}
\vspace*{-3mm}
\caption{Branching fractions of the heavy Higgs decays into charginos and neutralinos as functions of their masses for the inputs $\,\tan\!\beta\!=\!3,\,30$,  $M_2\! =\!2M_1\!=0.9$\,TeV and $\mu = -1.2$\,TeV. In the panel\,(d), we further set the input $M_A^{}\!=\!3$\,TeV,\, such that all the SUSY-decay channels of the heavy Higgs bosons are open and sum up all the decay modes.}
\label{fig:BR}
\vspace*{-5mm}
\end{figure}

Denoting as in section\,\ref{sec:2.2} 
the MSSM Higgs bosons as $H_k^{}$ [with $k\!=\!1,2,3,4$   
for ($H,\,h,\,A,\,H^\pm$)] and the neutralinos and charginos collectively by $\chi_i^{}\hs$, the partial widths of the decays  $\,H_k^{}\!\!\to\!\!\chi_i^{}\chi_j^{}\,$ into light gauginos and higgsinos can be written 
as\,\cite{Griest:1987qv}\cite{Gunion:1988yc}\cite{Djouadi:1992pu}\cite{Djouadi:1996pj}:
\\[-6mm]
\begin{eqnarray}
\hspace*{-1.5mm}
\Gamma (H_k\! \to \!\chi_i \chi_j) \!= \!
\frac{G_\mu M_W^2 s_W^2}{2 \sqrt{2} \pi} \frac{ M_{H_k} \lambda_{ij}^{\frac12 } }
{1\!+ \! \delta_{ij}}\! 
\left( \left[ (g_{ijk}^L)^2\! + \! (g_{jik}^R)^2 \right]\! (1\!-  \!
\kappa_i^2 \! - \!  \kappa_j^2)\! -\!4 \epsilon_i \epsilon_j g_{ijk}^L g_{jik}^R \kappa_i \kappa_j  \right)\!,~~\,
\end{eqnarray}
where use the abbreviation $\,\kappa_i^{}\!= m_{\chi_i^{}}^{}/ M_{H_k}^{}$ and where   $\,\delta_{ij}^{}=0\,$ unless the final state consists of two identical (Majorana) neutralinos in which case  $\delta_{ii}^{}\!=1$. $\,\epsilon_i^{} \!=\pm 1\,$ 
stands for the sign of the $i$-th eigenvalue of the neutralino mass matrix, but for charginos, one has $\epsilon_i^{}\!=\!1$;  $\lambda_{ij}^{} \!=\! 1\!+\!\kappa_i^4\!+\kappa_j^4\! -2(\kappa_i^2 \kappa_j^2 \!+\! \kappa_i^2 \! +\! \kappa_j^2)$ is the usual phase space factor. The Higgs couplings to charginos and neutralinos have been already given in eqs.~\eqref{cp:inos}-\eqref{ed-coefficients}.

\vspace*{1mm}

In the gaugino (higgsino) limit for the lightest $\chi$ states $|\mu| \!\gg\! M_{1,2}^{}$ (or $|\mu| \!\ll\! M_{1,2}^{}$),  the neutral Higgs  decays into identical neutralinos and charginos $A/H\!\to\!\chi_i^{}\chi_i^{}$, together with the charged Higgs decays $H^\pm\to \chi_{1,2}^0 \chi_1^\pm, \chi_{3,4}^0 \chi_2^\pm$,  will be strongly suppressed by the couplings even if  phase-space favored.  The Higgs decays to the mixed heavy and light $\chi$ states will in turn be favored by the larger couplings.  For example, in the gaugino limit and if one ignores the phase-space suppression by taking  $M_{H_k}\!\gg\! |\mu| \!\gg\! M_2$, the partial widths of the heavy Higgs decays into mixed $\chi$ states,  in units of the factor    
$\,G_F M_W^2 M_{H_k}/(4\sqrt{2}\pi)$, will be simply
given by the simple expressions 
\bea
\Gamma(H/A \! \to \! \chi_i^0 \chi_j^0) \propto 
\frac12 \,\xi_i^{}\, (1 \pm\sin 2\beta ),~~~~ 
\Gamma(H/A \to \chi_1^\pm \chi_2^\mp) \propto 1   ,
\eea
with $\xi_1^{}\!=\!\tan^2 \theta_W$ and 
$\xi_2^{}\!=\!1$,\, so that the decays of one of the neutral Higgs bosons are not suppressed when $\tb$ is either large or close to one.  The charged Higgs boson decays $H^\pm \!\to\!\chi_i^\pm \chi_j^0$ do not depend on $\tb$ in this limit and the decay width is simply either 1 or $\tan^2\!\theta_W$  in the unit above. We present in Fig.\,\ref{fig:BR} 
the branching fractions of the three heavy Higgs bosons 
decaying into the neutral and charged $\chi$ states.\ 
We see that the turning points of the heavy Higgs masses around 
$\,M_2^{}\!+|\mu|\!=\!2.1$\,TeV are due to the mass relation  
$m_{\chi_2^0},m_{\chi_1^\pm} \approx M_2$ and $m_{\chi_3^0},m_{\chi_4^0},m_{\chi_2^\pm} \approx |\mu|$\,.

\vspace*{1mm}

For large Higgs masses 
$\,M_A\! \approx\! M_H\! \approx\! M_{H^\pm} \!\! \gg \!\! M_Z\,$,\, 
when all decay channels are kinematically accessible, the branching fractions can be significant  and sometimes can even become dominant, also for the low and large values of $\tb$ which, respectively, enhance the top and bottom decay modes.   However, the maximal Higgs decay rates into these states are  obtained at moderate $\tb$ when all channels are kinematically accessible. In this case, as a consequence of the unitarity of diagonalizing the $\chi$ mixing matrices,  the sum of the partial widths does not depend on any supersymmetric  parameter when phase space is neglected.  For instance, one gets the following expressions for the total branching fraction  by summing up all the possible decay modes
\begin{eqnarray}
\label{BR:tanb}
\hspace*{-7mm}
{\rm BR}\big(\! \Phi \to \!\sum_{i,j} \chi_i^{} \chi_j^{}\big) 
= \frac{ \left( 1\!+\hsm\frac{1}{3}\tan^2\!\theta_W\right)\! M_W^2}
{\,\left( 1\!+\!\frac{1}{3}\tan^2\!\theta_W\hsm\right)\! M_W^2 
\hsm + \overline{m}_t^2 \cot^2\!\beta \hsm +\hsm 
(\overline{m}_b^2 \!+\! \frac13 m^2_\tau) \tan^2\!\beta~}   
\,,
\end{eqnarray}
where, besides the decays into the superparticles, only the leading channels $t\bar{t}$, $b\bar{b}$ and $\tau \tau$ for the neutral  $H/A$ and the dominant modes $t b$ and $\tau \nu$  for the charged $H^\pm$ bosons are included in the total decay widths, which is indeed the case in the decoupling limit where other decays become negligible, and the mass effects have been neglected.

\vspace*{1mm}

In Fig.\,\ref{fig:BR}, we illustrate the branching fractions 
for the heavy Higgs decays into charginos and neutralinos.\
In the panels (a)-(c), these branching fractions are plotted for
neutral and charged heavy Higgs bosons as functions of their masses. In all these cases, we choose the sample inputs of 
$\,\tan\!\beta =3,\,30\,$ and the SUSY parameters 
 $M_2\! =2M_1\!=0.9$\,TeV and 
$\mu = -1.2$\,TeV. One sees that the pattern for the decay branching fractions of the heavy $H,\,A,\,H^\pm$ bosons is quite similar.\ 
For the  mass values 
 $\,M_{H_k}^{}\!\!\lesssim\! M_2^{}\!+\!|\mu|\!=\!2.1$\,TeV, the branching fractions are generally below the 10\% level, and as the Higgs masses become larger, they  can increase up to the level of $(40\!-\!50)\%$.

\vspace*{1mm}

In the last pane\,(d), we also present the Higgs decay branching fractions into all $\chi$ states as function  of $\,\tan\!\beta\,$, where a sample input  $M_A^{}\!=\!3$\,TeV is taken,  such that all the chargino and neutralino decay modes of the heavy Higgses are open. It shows that the decay branching fractions of the three heavy Higgs bosons nearly coincide and can reach values around $40-80\%$
for $\tb$ values in the range $\,\tan\!\beta \in\! [2.5,\,25]$.

\vspace*{1mm}

As mentioned earlier, for very large $\,\tan\!\beta\,$ values, 
the partial decay widths of the 
$H/A\!\!\to\!b\bar b, \tau^+\tau^-$ and  
$H^+ \!\!\to\! t\bar b,\tau^+\nu\,$ decays are so strongly enhanced, 
that they leave little room for the SUSY-decay channels.\ 
At low $\,\tan\!\beta\,$ values, the decay rates of  
$\,H/A \!\to\! t\bar t$ when kinematically accessible  
and  $\,H^+ \!\to\! t\bar b$ are large and can be dominating.  
Thus, the heavy Higgs decays into the neutralinos and charginos
could play a significant role mainly for the intermediate values 
of $\,\tan\!\beta\,$ and possibly for
$M_{H} = M_{A} \!\lsim 350$\,GeV.  
However, two requirements should be fulfilled also in this
case. First of all, to make some SUSY decay modes 
$H_k \!\to\!\chi\chi$ kinematically possible,
certain $\chi$ states should be light, 
$M_{H_k}^{} \!\gsim 2m_{\chi}^{}$ (with $H_k \!=\!H\!,A$).\ 
Secondly, the
$H_k \chi\chi$ couplings should be large enough,  
 meaning that the $\chi$ states should be gaugino-higgsino mixtures as previously discussed.

\begin{figure}[t] 
	\centering
	\includegraphics[width=7.5cm]{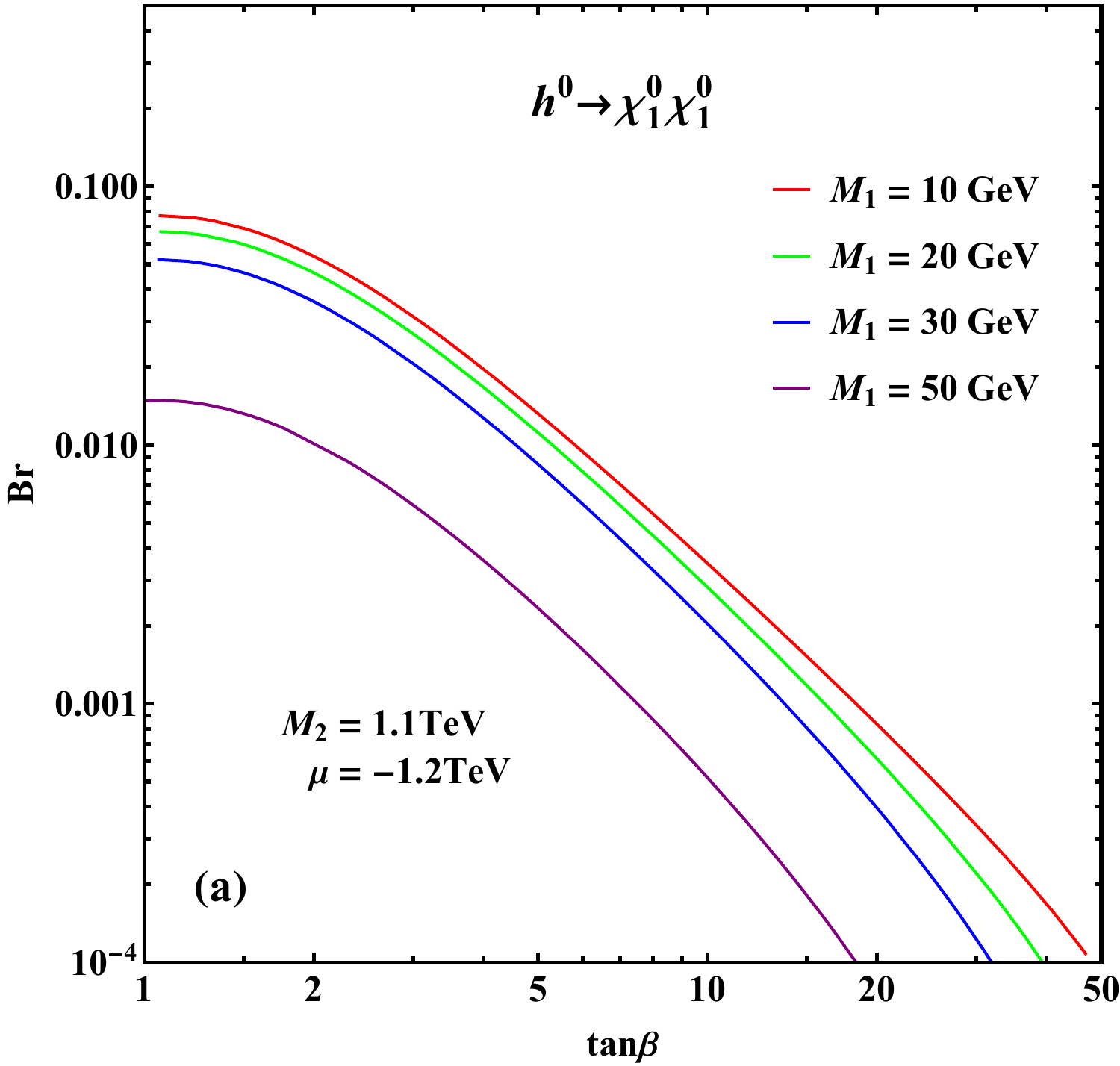}
	\includegraphics[width=7.5cm]{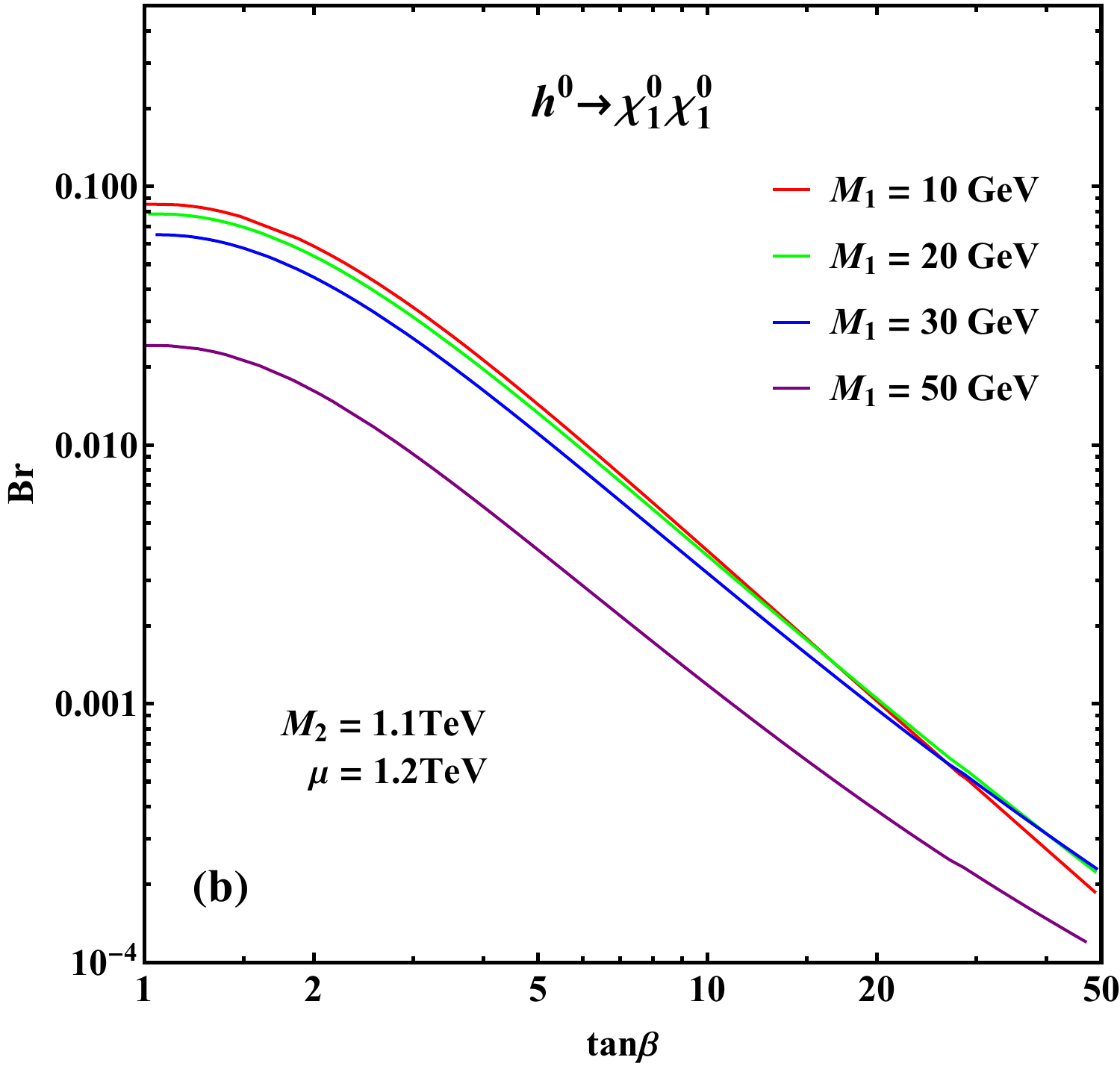}
\vspace*{-3mm}
\caption{The invisible decay branching fraction  
BR($\,h\!\!\to\!\!\chi_1^0\chi_1^0)$  as a function of $\tan\beta$, where the (red,\ green,\,blue,\,purple)
curves correspond to a gaugino mass parameter 
$M_1^{}\!=\!(10,\,20,\,30,\,50)$ GeV, respectively.\  
We choose the sample inputs $M_2\!\!=\!\!1.1$\,TeV, 
$\,\mu \!=\! -1.2$\,TeV for the plot\,(a) and $\,\mu \!=\! 1.2$\,TeV
for the plot\,(b). 
}
\vspace*{-3mm}
\label{fig:BRh}
\end{figure}

\vspace*{1mm}

Finally, let us make a few comments on the SUSY-decay channels of the light CP--even and SM-like $h$ boson.  The experimental bound $m_{\chi_1^\pm}^{} \!\gsim\!104$\,GeV from LEP2 searches  does not allow the $h$ boson to decay into any chargino pair or 
neutralino pair except for the invisible decays into a pair of
the LSP neutralinos, $h \!\to\!\chi_1^0\chi_1^0\,$.  This is especially valid in the case where one equates the gaugino masses at the GUT scale, leading to the relation  $M_1^{} \!\sim\!\frac{1}{2} M_2^{}$ at the electroweak scale, is relaxed. This would  lead to the possibility of very light LSP neutralinos while the LEP2 bound on $m_{\chi_1^\pm}$ still holds. However, as $\chi_1^0$ should be primarily bino-like in this case,  $M_1^{}\!\ll\! M_2^{}, |\mu|$,  the $h\chi_1^0\chi_1^0$ coupling is suppressed and leads to small invisible branching fractions.  Nevertheless, as it competes with modes that have small partial widths, the rate can still reach a few percent level.  Hence,  it can be probed by future measurements of the $h$ signal strengths or in the direct searches for invisible decays  at HL-LHC or at future $e^+e^-$ and $pp$ colliders.

\vspace*{1mm}

Nevertheless, even with such a small $h$ branching fraction, one can arrange that the LSP has the required cosmological density, since it will annihilate efficiently through the exchange of the $h$ Higgs boson,  as will be discussed in the following astrophysics section.

\vspace*{1mm}

In Fig.\,\ref{fig:BRh}, we present the branching fractions of 
the lightest $h$ boson when decaying into the LSP neutralinos,
 $\,h\!\to\!\chi_1^0\chi_1^0\,$,  
as a function of $\tan\!\beta$. 
Here, the (red,\ green,\,blue,\,purple) curves correspond to 
the bino mass parameter taking the values
$M_1\!=\!(10,\,20,\, 30,\,50)$\,GeV, respectively.  
We choose the other input parameters as 
$M_2=1.1$\,TeV, $\,\mu \!=\! -1.2$\,TeV for the panel\,(a),   
and $M_2\!=\!1.1$\,TeV, $\,\mu \!=\! 1.2$\,TeV for the panel\,(b).

\vspace*{1mm}

\begin{figure}[!h]
	\centering
	\includegraphics[width=10cm]{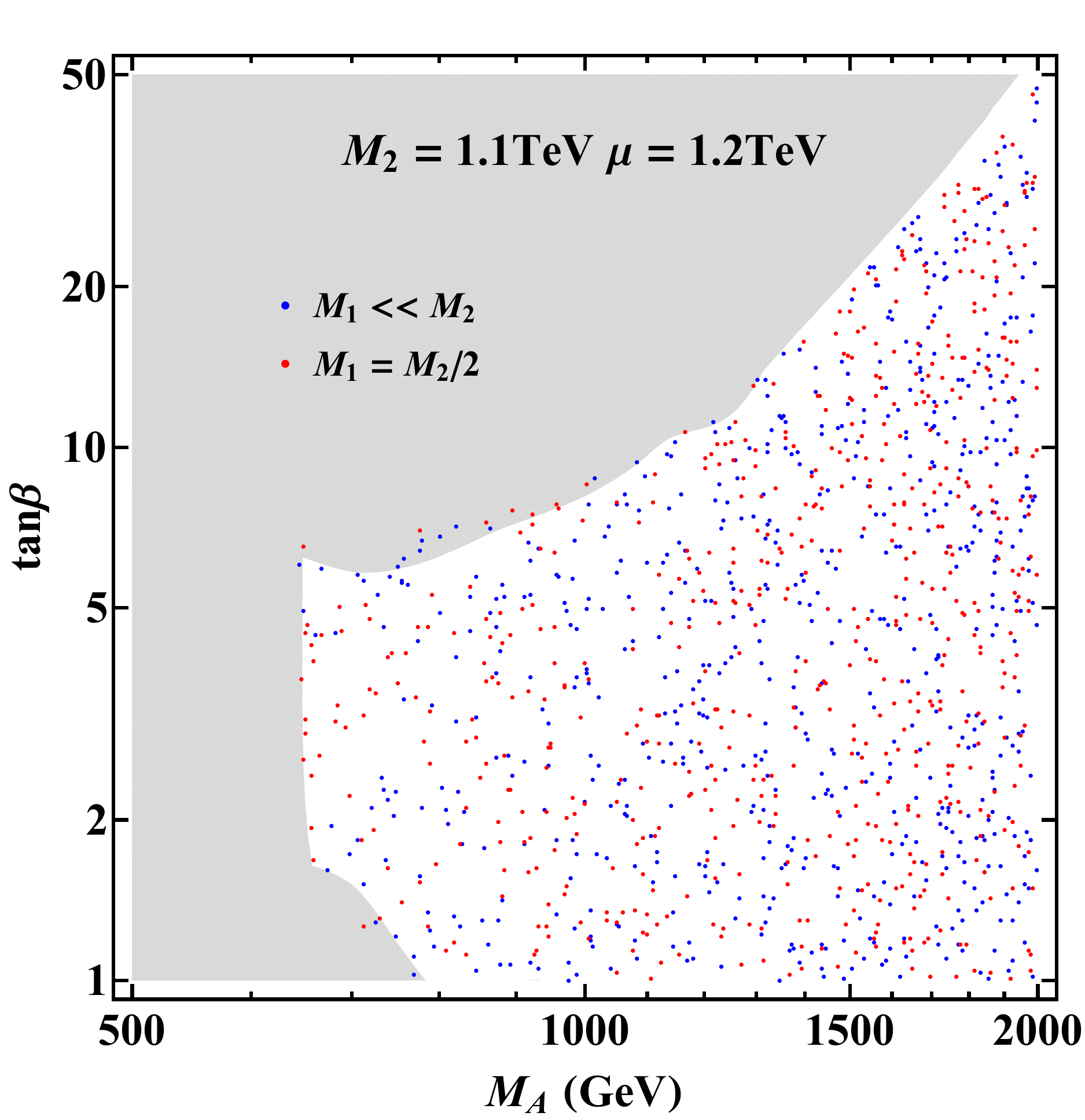}
\vspace*{-3mm}
\caption{Allowed parameter space in the $[M_A, \tan\beta]$ plane at the 95\%\,CL, obtained by fitting the $h$ coupling measurements and direct search of heavy Higgs bosons at the LHC, 
where we set $M_2=1.1$\,TeV, $\mu=1.2$\,TeV, $M_1 \ll M_2$ 
(shown by blue dots), and $M_2=2 M_1$ (shown by red dots).\ 
The grey area presents the exclusion region under the condition that SUSY particles have no contribution to Higgs production and decays.}
\label{fig:matbh}
\end{figure}

We note that for the small bino mass  
$\hs M_1^{}\!=\!(10\hsm -\hsm 20)$\,GeV, 
 the decay branching fraction of $\,h\!\to\!\chi_1^0\chi_1^0\,$
is around $(2\!-\!8)\%$  for 
$\hs\tan\!\beta\!\lesssim\! 5\,$.\ 
On the other hand, for larger mass values 
$M_1^{}\!\!=\!50$\,GeV, the decay branching fraction lies around the 
percent level only for  $\tan\!\beta\!\lesssim\! 2\,$.\ 
We find that the branching fraction 
for $\,\mu \!=\! 1.2$\,TeV is generally larger than that 
for $\,\mu \!=\! -1.2$\,TeV.

\vspace*{1mm}

Fig.\,\ref{fig:matbh} presents the parameter space allowed by the combined $h$ coupling measurements and the direct Higgs searches, including contributions of SUSY particles to the decays as discussed in this section.\ 
Here, we set the input parameters $M_2\!=\!1.1$\,TeV and 
$\mu \!=\!1.2$\,TeV for illustration.\ 
For comparison, we also present the exclusion region without the contribution of SUSY particles (as given by Fig.\,\ref{fig:1}), 
and this is shown as the grey region of Fig.\,\ref{fig:matbh}.\
The presence of SUSY particles tends to reduce the branching fractions 
of heavy Higgs decays to the SM particles and thus allows 
for a larger parameter space.\ 
Fig.\,\ref{fig:matbh} considers the case of 
$\,M_2\!<\!\mu\,$ and has the inputs of $(M_2,\,\mu )$ 
lie in the allowed parameter region of Fig.\,\ref{fig:4} 
(represented by the red points).



\vspace*{1mm}
\section{Astrophysical Constraints of the hMSSM}
\label{sec:5}
\vspace*{1.5mm}

The previous discussion about the relation between the invisible  decay branching ratio of the SM-like Higgs boson and the relic density of the DM particle allows us to make a smooth  transition towards the astrophysical and astroparticle aspects of a neutralino LSP in the context of the hMSSM. As is well known, in an MSSM in which $R$--parity \cite{R-parity} is conserved, this particle was for a long time considered to be the best candidate for a thermal DM particle \cite{LSP-DM,bulk}.  Under this assumption, the requirement of a cosmological relic density compatible with the latest measurement of the PLANCK satellite \cite{Planck:2018vyg} 
\beq
\Omega_{\chi_1^0} h^2 \simeq 0.12 \pm 0.01, 
\label{eq:omegah2}
\eeq
as well as compatibility with direct and indirect detection experiments, provide very strong constraints on the MSSM Higgs sector that are complementary to the  collider constraints discussed before. We will briefly review below the various DM constraints, focusing on relic density and direct detection, being, in general, the most stringent. Whenever appropriate, will account for indirect detection while illustrating our numerical results.

First, concerning the cosmological relic density of the LSP neutralino that should be compatible with the measured value given in  eq.~(\ref{eq:omegah2}), it is determined, up to non standard assumptions about the cosmological history of the early universe, by the freeze-out paradigm. According to the latter, the DM abundance is directly related to the thermally averaged cross section $\langle \sigma v \rangle$ of the annihilation of DM pairs into lighter particles, $\Omega_{\chi_1^0} \propto 1/\langle \sigma v \rangle$. In the MSSM, this cross section crucially depends on the composition of the lightest neutralino and on the supersymmetric particle spectrum since co-annihilation processes, occurring when the next–to–lightest supersymmetric particle (NLSP) is almost degenerate in mass with the DM one, play an important role.  

In a general MSSM with possibly light sfermions, two main configurations in the model parameter space that lead to an appropriate  relic density, have been discussed. A first one, called the ``bulk region" \cite{bulk}, is when the main DM annihilation process is into leptonic final states. This is mediated by $t$–channel slepton exchange and, in particular, the exchange of the $\tilde \tau$ state which is expected to be the lightest slepton. The other configuration is when the neutralino $\chi_1^0$ has a mass very close to that of a sfermion $\tilde f$, leading to efficient LSP-$\tilde f$ and $\tilde f \tilde f $ co-annihilations.  The possibilities which were advocated most are co-annihilations with again $\tau$-sleptons \cite{coanih-stau} but also top--squarks \cite{coanih-stop}. Both sfermions can be made the lightest sfermions by enforcing a strong mixing between the left- and right-handed states. 

However, in our hMSSM context in which the sfermions are assumed to be very heavy, these two possibilities become absent. DM annihilations will thus occur only through the $s$-channel exchange of the $Z$ and neutral bosons,  \`a-la Higgs/$Z$-boson portal scenarios \cite{Hportal}, or via gauge interactions. The size of such LSP  annihilation cross sections and thus, the DM relic abundance, will then primarily depend on the gaugino-higgsino composition of the neutralino DM. This composition also controls the mass difference between the LSP and the other lighter neutralinos and charginos, which is important for co-annihilation. 

The correct $\chi_1^0$ relic density  can be therefore obtained only in the following situations.\smallskip

$i)$ \underline{A mostly bino-like LSP neutralino}. This scenario resembles the Higgs portal simplified models, with the DM mostly annihilating into SM fermions and gauge bosons via the $s$-channel exchange of neutral bosons \cite{Hportal}. As the bino annihilation cross sections via the exchange of the CP--even Higgs bosons, as well as the exchange of the $Z$ boson, are $p$-wave suppressed, only annihilation processes into SM fermions via the exchange of the pseudoscalar $A$ state, are in general relevant as they lead to an $s$-wave dominated cross section. These processes feature a potentially nice correlation with the LHC Higgs searches illustrated in the previous sections. Nevertheless, for a light DM with a mass $m_{\chi_1^0} \lsim  100$ GeV, $Z$ and $h$ mediated annihilation channels become relevant but the corresponding cross sections typically lie below the thermally favored value, of order $ \langle \sigma v \rangle \approx 10^{-26} {\mbox{cm}}^3 {\mbox{s}}^{-1}$, unless ``pole" enhancements that occur for  $m_{\chi_1^0}\simeq \frac12 M_{Z}$ or $m_{\chi_1^0}\simeq \frac12 M_{h}$ are present \cite{Hpole}.\smallskip
    
$ii)$ \underline{A higgsino or wino-like LSP neutralino}. In this case, the pairs of $\chi_1^0$ DM states will annihilate via gauge interactions into pairs of $W$ and $Z$ bosons, with a relic density mainly set by the value of the LSP mass. The cosmologically favored value is achieved for  $m_{\chi_1^0} \approx  1 \,\mbox{TeV}$ and $m_{\chi_1^0} \simeq 3\,\mbox{TeV}$ for, respectively, purely higgsino-like and purely wino-like LSPs \cite{limit-LSP} and the DM would be under-abundant for lighter masses. In these cases, the correlations with searches of heavy Higgs bosons at the LHC would be very weak or even absent. Hence, dedicated DM searches will be the primary probes for this type of scenario.\smallskip

$iii)$ \underline{The ``well tempered" bino-higgsino and bino-wino regimes} \cite{tempered}. Here, the correct relic density is achieved, away from resonances and for DM masses of the order of few hundred GeV, by having a suitable admixture between a bino-like LSP, with very suppressed interactions, and a higgsino and/or wino component, which is more efficiently interacting. As it will be clarified in the discussion below, bino-higgsino DM has enhanced interactions with CP-even Higgs bosons in contrast to the mixed bino-wino lightest neutralinos. The former is thus disfavored by DM direct detection, while the  later regime can instead evade these constraints provided that some fine tuning of parameters occurs.\smallskip 

This brings us to the issue of direct detection. In general, the neutralino DM features  both spin--independent (SI), mediated by the exchange of the neutral CP-even Higgs bosons and generating a DM--nucleon scattering cross section on the proton of the form 
\begin{equation}
\label{eq:sigmaSI}
    \sigma_{\chi_1^0 p}^{\rm SI}=\frac{\mu_{\chi_1^0 p}^2}{\pi}\frac{m_p^2}{v^2}{\left \vert \sum_q f_q^p \left(\frac{g_{ \chi_1^0 \chi_1^0 h} g_{q}^h}{M_h^2}+\frac{g_{\chi_1^0 \chi_1^0 H} g_{q}^H}{M_H^2}\right) \right \vert}^2 , 
\end{equation}
and spin--dependent (SD) interactions, mediated by the $Z$ bosons and  leading to a scattering cross-section on protons which reads 
\begin{equation}
    \label{eq:sigmaSD}
    \sigma_{\chi_1^0 p}^{\rm SD}=3\frac{\mu_{\chi_1^0 p}^2}{\pi M_Z^4}{\left[g_{\chi_1^0 \chi_1^0 Z }\left(g_u^A \Delta_u^p+g_d^A\left(\Delta_d^p+\Delta_s^p\right)\right)\right]}^2 . 
\end{equation}
In these equations, $m_p$ is the proton mass while $\mu_{\chi_1^0 p} = 
m_{\chi_1^0} m_p/( m_{\chi_1^0 } + m_p)$ is the DM–proton reduced mass. 
For the coefficient $f_q^p$ and $\Delta_q^p$ giving the contributions of the various light and heavy quarks, we adopt the numerical values given in Refs.~\cite{f-values}. 

For both types of interactions, we will impose the most recent constraints provided by the LUX-ZEPLIN (LZ) collaboration \cite{LZ:2022ufs}, which recently superseded the limits by XENON1T \cite{XENON:2018voc,XENON:2019rxp}. Analogous constraints have been also obtained by the PANDA-X experiment and add little so that we do not include them. 

In the next subsections, we show how these DM constraints complement the collider ones in the different configurations for the  various regimes of the bino and wino mass parameters that we have identified in the previous sections. In all cases, our numerical results will be based on the implementation of the modified version of the program SuSpect for the hMSSM discussed in section 2 in the program Micromegas \cite{MicroMegas} for DM.

\subsection{$\mathbf{M_1 \simeq \frac12 M_2}$}

\begin{figure}
    \centering
    \subfloat{\includegraphics[width=0.5\linewidth]{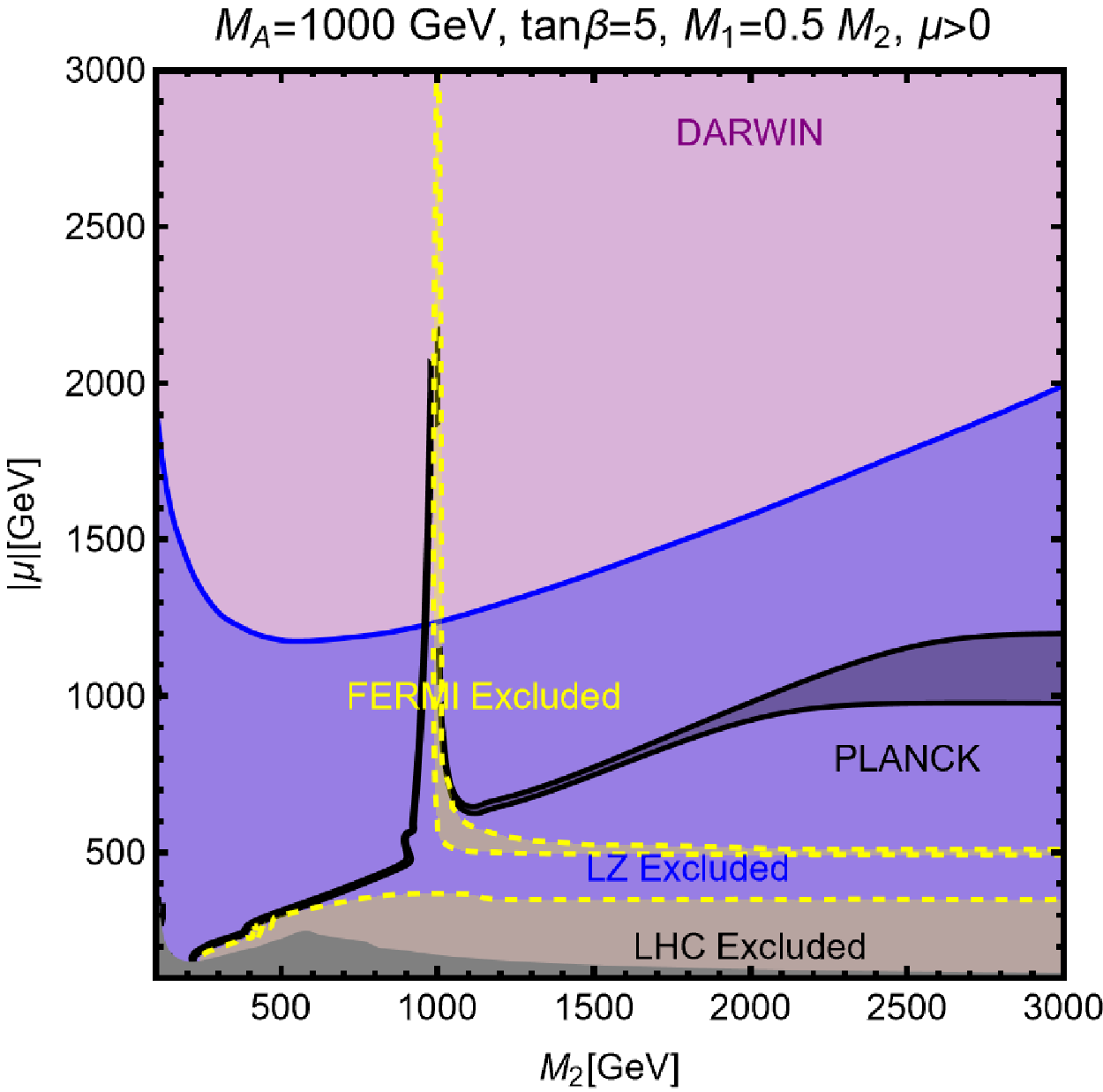}}
    \subfloat{\includegraphics[width=0.5\linewidth]{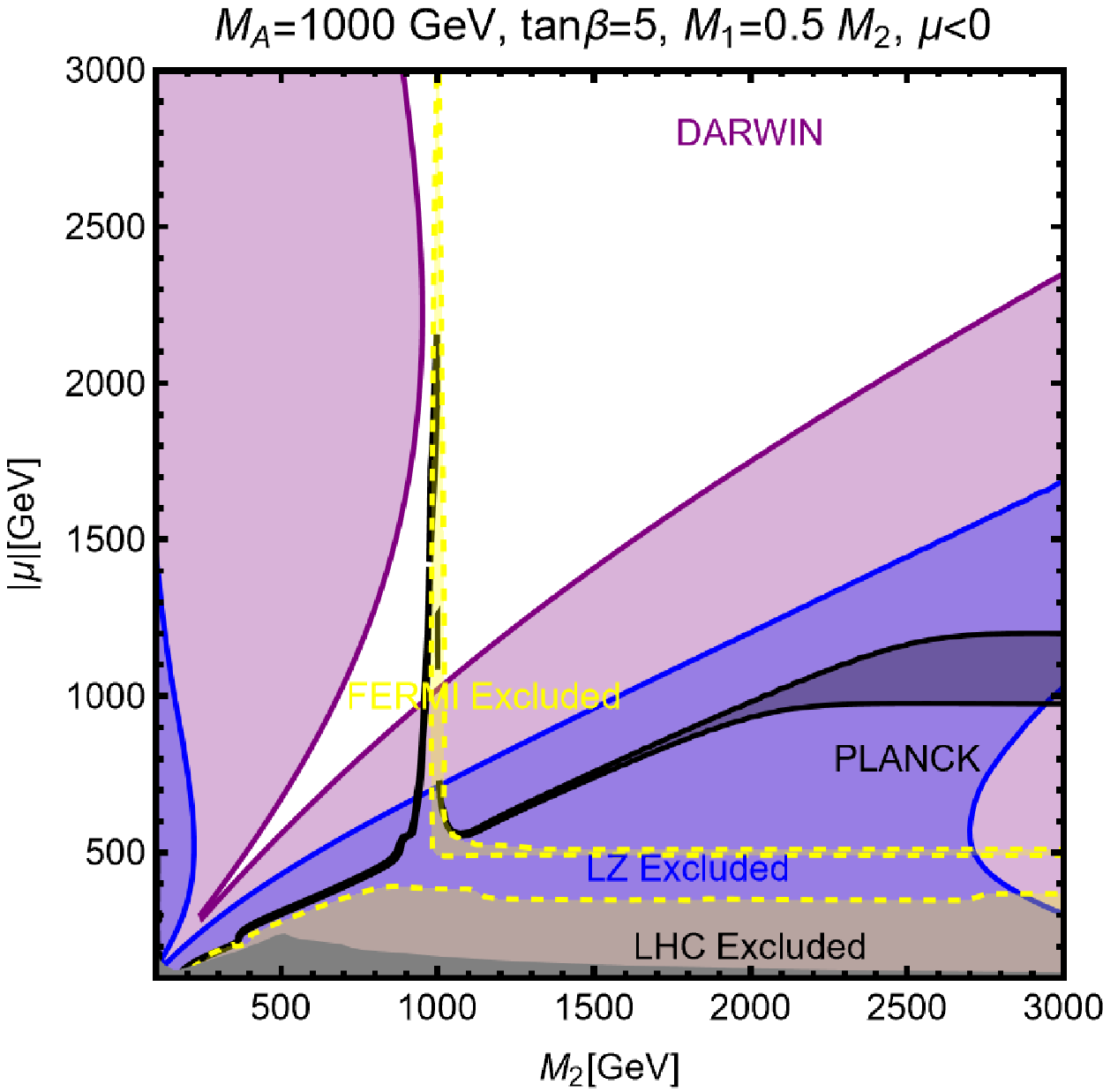}}\\[3mm]
    \subfloat{\includegraphics[width=0.5\linewidth]{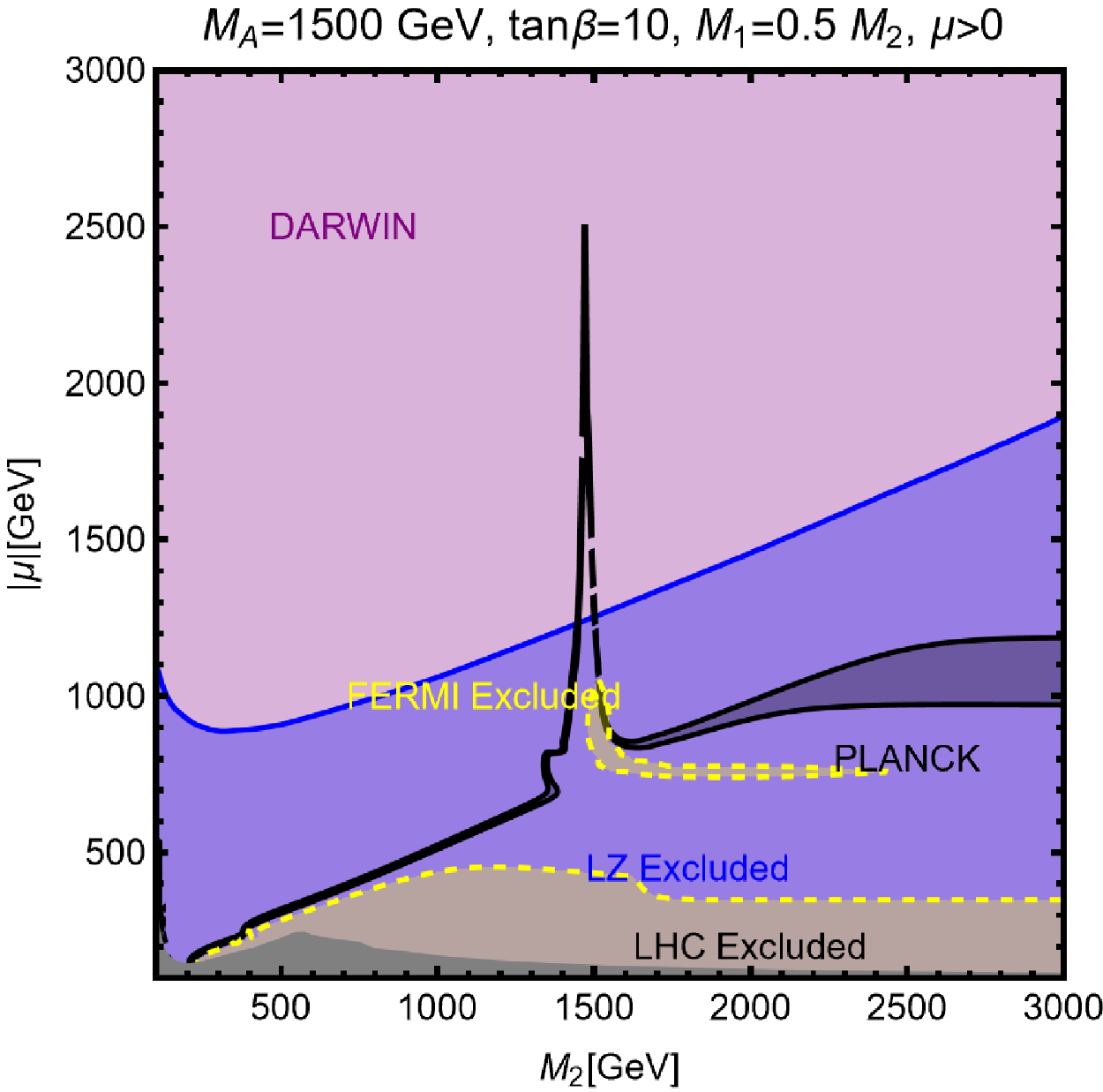}}
    \subfloat{\includegraphics[width=0.5\linewidth]{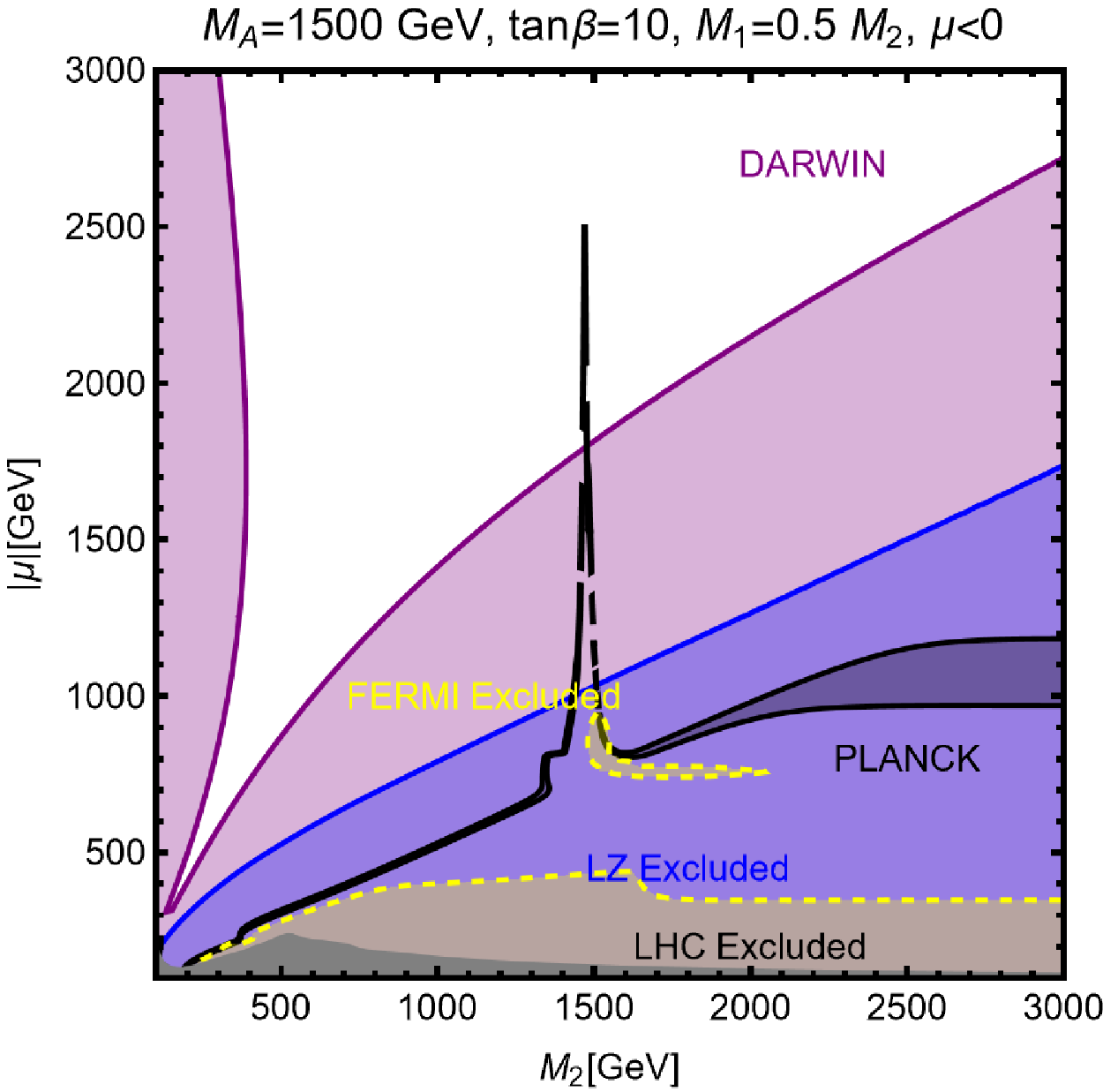}}
    \caption{Summary of DM constraints in the $[M_2,|\mu|]$ plane for the $M_2=2 M_1$ scenario. The upper panels are for the $(M_A,\tan\beta)$ assignments $(1\,\mbox{TeV,\, 5})$,  while the lower panels ones consider $(1.5\,\mbox{TeV},10)$. For each benchmark we have considered both signs for the $\mu$ parameter. In all panels, the black isocontours correspond to regions in which the correct DM relic density is reproduced, the colored contours the present and future sensitivities regions of, respectively, the LZ and DARWIN direct detection experiments and the regions marked in gray are those ruled out by LHC searches of charginos/neutralinos.}
    \label{fig:scans_gut}
\end{figure}

Let us first consider the case in  which the soft SUSY-breaking wino and bino mass parameters are linked by the GUT relation $M_2 \simeq 2 M_1$. In such a case, the neutralino DM can be either bino-like or higgsino-like, or a mixture of these two components. Representative examples of the combined DM/LHC constraints in such a scenario are provided in Fig.~\ref{fig:scans_gut}. 

The four panels of the figure show the constraints on the bidimensional $[M_2, |\mu|]$ parameter space for two benchmark scenarios with $(M_A,\tan\beta)$ Higgs sector parameters of $(1\,\mbox{TeV,\, 5})$ for the upper panels and $(1.5\,\mbox{TeV}, 10)$ for the lower ones. For each of these two scenarios, we considered the two possibilities $\mu>0$ and $\mu <0$. We have then applied on this plane the constraints on the relic density from the PLANCK satellite, the present and future sensitivities of the direct detection experiments and finally, the constraints on this parameter space from LHC searches of charginos and neutralinos discussed previously. 

Given the precision in the experimental determination of $\Omega_{\chi_1^0}h^2$, the correct DM relic density, assuming the WIMP paradigm and a standard cosmological evolution for the early universe, is achieved only in the narrow contours of the $[M_2,|\mu|]$ plane marked in black. For the shown benchmarks  to be viable, at least one portion of the relic density isocontours should lie outside both the blue regions, corresponding to the exclusion from LZ, the yellow regions, corresponding to the exclusion from searches of continuous spectra of $\gamma$-rays from DM annihilations as performed by the FERMI collaboration \cite{fermi},and the gray regions, excluded by neutralino/chargino searches at the LHC. The areas of the $[M_2,|\mu|]$ plane will be, in turn, ruled out if the next future results 
from the DARWIN (purple) \cite{DARWIN} experiment will confirm the absence of DM signals.  

The outcome shown in Fig.~\ref{fig:scans_gut} can be understood as follows. In the regions corresponding to $m_{\chi_1^0}\! <\! \frac12 M_{H,A}$, the correct DM relic density can achieved  only for an LSP with a substantial bino-higgsino mixture, a possibility that is ruled out by direct detection limits. For $m_{\chi_1^0}\! \simeq\! \frac12 M_{H,A}$, the vicinity of the Higgs poles enhances the DM annihilation rate, in particular in the pseudoscalar Higgs case. The correct relic density could be then obtained for a bino-like DM that evades, at least partially, the direct detection constraints. The increased sensitivity of upcoming detectors will, however, strongly erode this region in the absence of a signal. Finally, the relic density curve reaches a plateau at around $\mu \simeq 1.1\,\mbox{TeV}$. In this region the relic density is saturated by the annihilation processes into gauge boson pairs of a higgsino-like DM state. For $|\mu|$ and $M_{1,2}$ values of the same order, this region appears to be excluded by DM direct detection. Direct detection bounds can be relaxed by considering the possibility $M_{1,2}\gg |\mu|$. This scenario would be not very interesting for what concerns LHC SUSY particle searches.

Comparing the left (for positive $\mu$) and right (for negative $\mu$) panels,  one notices that in the latter case, there are regions in which direct detection constraints become sensitively weaker. Indeed, in such regions, the DM scattering cross section is suppressed as a consequence of ``blind spots'', i.e. assignments of the model parameters corresponding to a cancellation of the coupling of the DM with the light $h$ boson or destructive interference between the contributions associated with the exchange of the two $h$ and $H$ states \cite{blind-spots}. An approximate analytic expression to have these ``blind spot" regions reads
\begin{equation}
    2 \left(m_{\chi_1^0}+\mu \sin 2 \beta \right)/{M_h^2} \simeq -\mu \tan\beta /{M_H^2} \, . 
\end{equation}
In the limit in which the $H$ state is very heavy, $M_H \approx M_A \gg M_Z$, the condition above reduces to $m_{\chi_1^0}+\mu \sin 2\beta \simeq 0$, which requires a negative $\mu$ value to be satisfied.
Concerning DM indirect detection, limits affect the regions of parameter space with mostly higgsino like DM. This is due to the fact that in the latter case the DM annihilation cross-section is s-wave dominated.

\subsection{$\mathbf{M_1 \simeq M_2}$}

\begin{figure}[!ht]
    \centering 
    \subfloat{\includegraphics[width=0.5\linewidth]{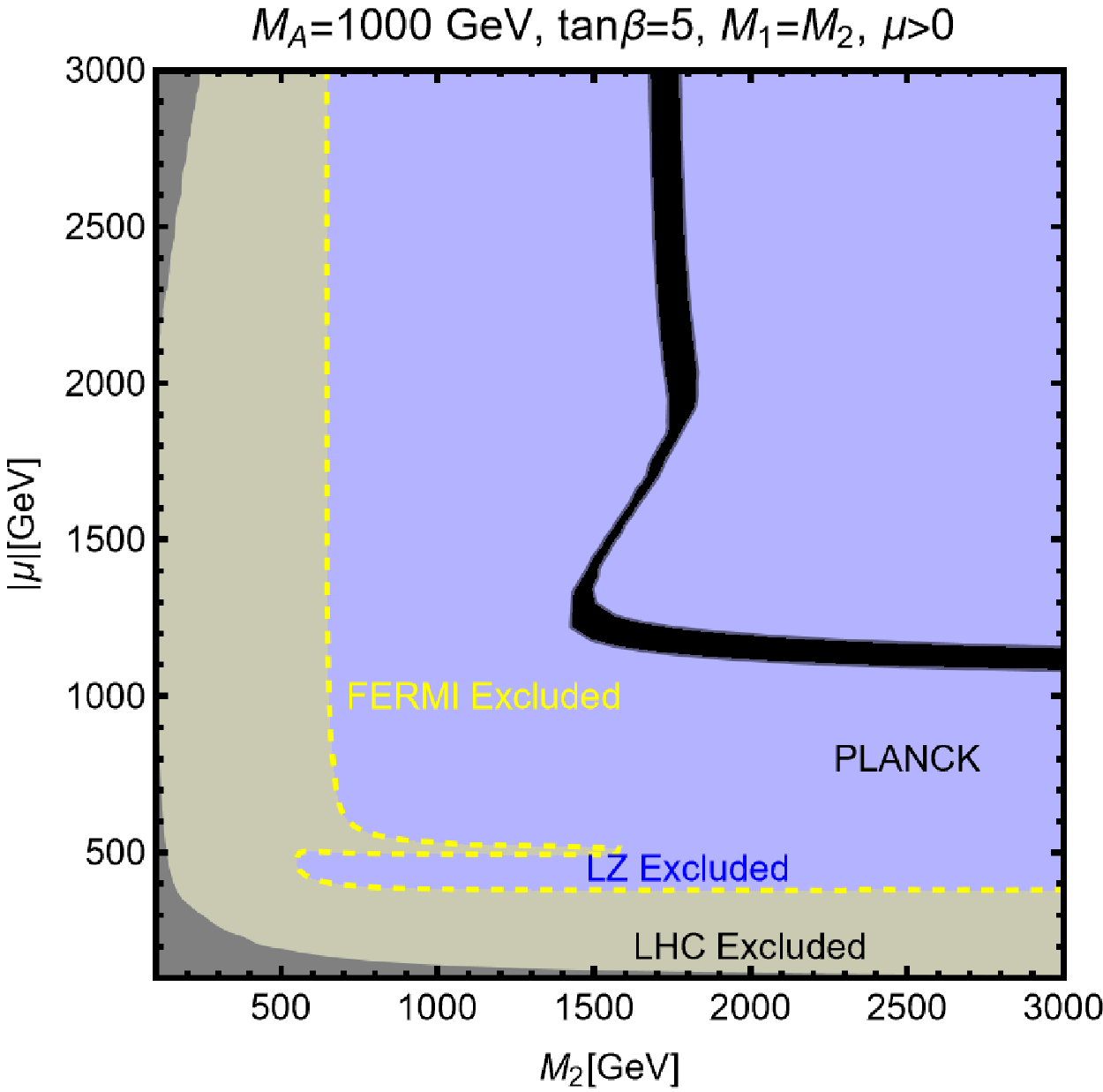}}
    \subfloat{\includegraphics[width=0.5\linewidth]{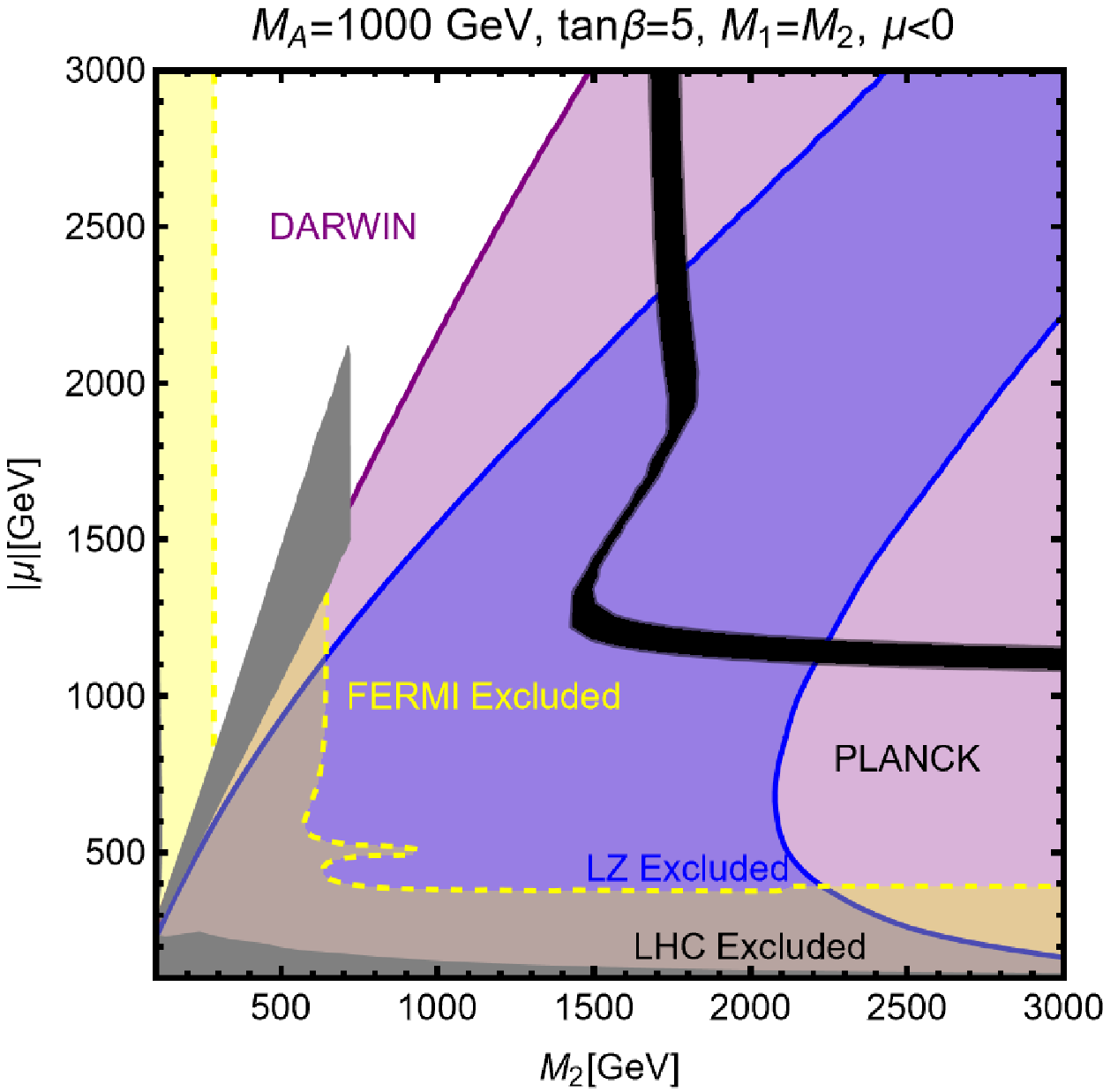}}\\[2mm]
    \subfloat{\includegraphics[width=0.5\linewidth]{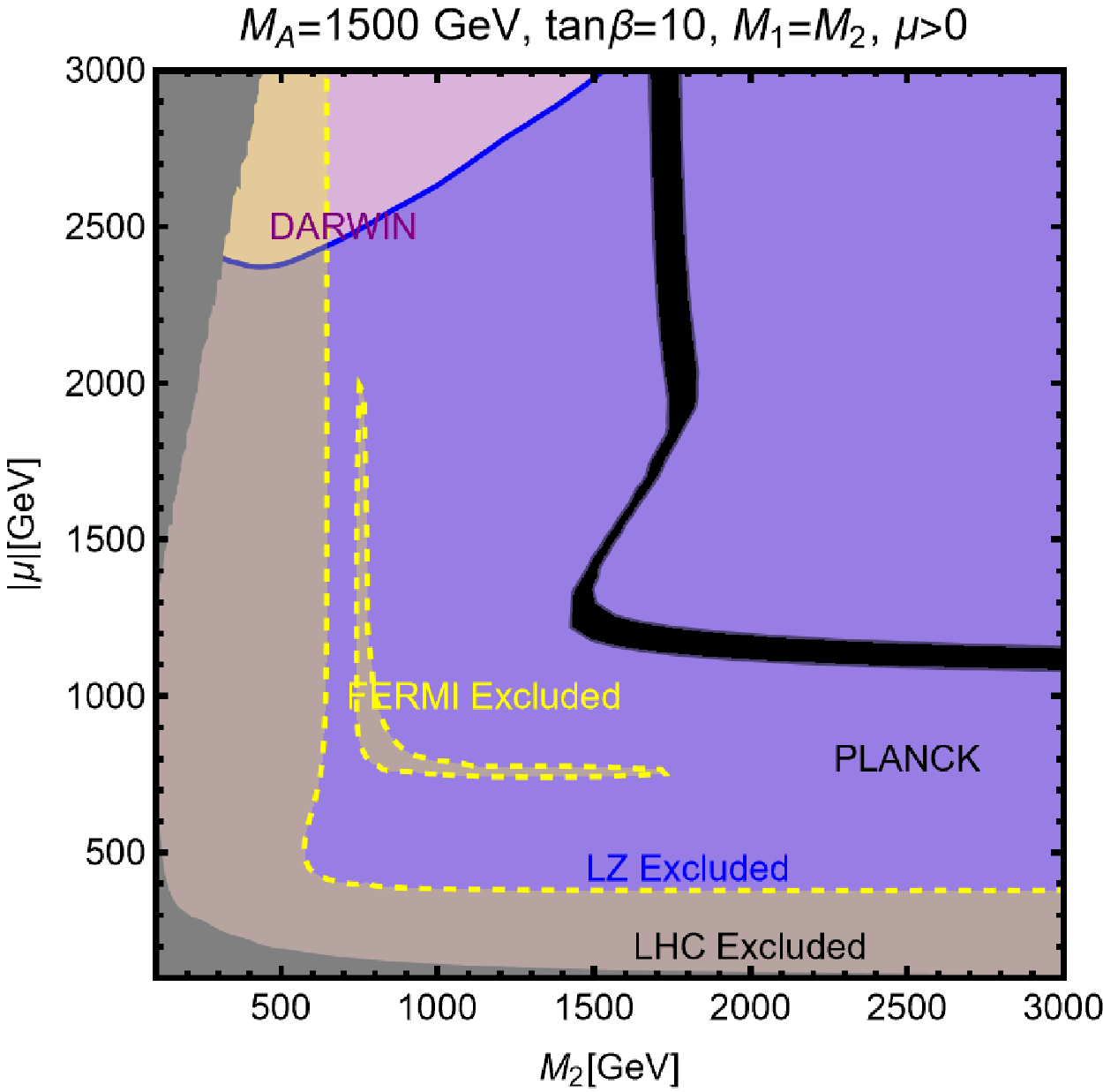}}
    \subfloat{\includegraphics[width=0.5\linewidth]{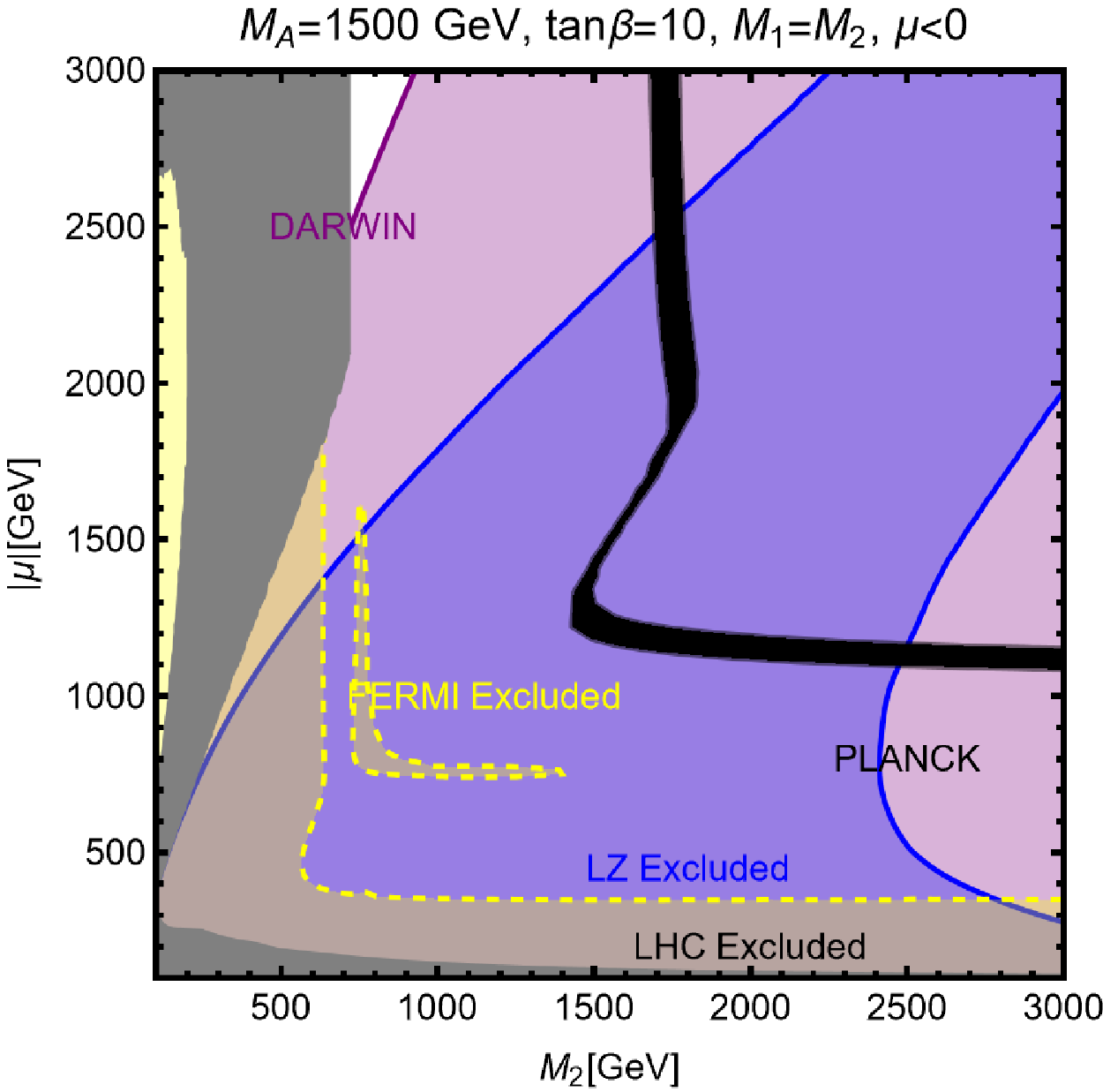}}
    \caption{Summary of DM constraints in the $[M_2,|\mu|]$ plane for the $M_2=M_1$ scenario. The upper panels are for the $(M_A,\tan\beta)$ assignments $(1,\mbox{TeV,5})$,  while the lower panels ones consider $(1.5\,\mbox{TeV},10)$. For each benchmark we have considered both signs for the $\mu$ parameter. In all panels, the black isocontours correspond to regions in which the correct DM relic density is reproduced, the colored contours the present and future sensitivities regions of the LZ (DARWIN) direct detection experiments and the regions marked in gray are those ruled out by LHC searches of charginos/neutralinos.}
    \label{fig:scans_temp}
\vspace*{-2mm}
\end{figure}

The assignment $M_1=M_2$ for the gaugino mass parameters leads to bino and wino states with comparable masses and corresponds to one possible option of the so-called  ``well tempered" LSP neutralino scenario. In this special case,  the resulting combined DM and LHC constraints on the $[M_2,|\mu|]$ parameter space are illustrated by Fig.~\ref{fig:scans_temp}. The benchmark scenarios for the Higgs sector and the sign of the $\mu$ parameter, as well as the color coding for the various constraints, is exactly the same as in Fig.~\ref{fig:scans_gut}. 

Looking first at the relic density contours, one can see that they feature two plateaux: a first one for $|\mu| > M_2 \gtrsim 1\,\mbox{TeV}$ and another one for $M_2 > |\mu| \gtrsim 1\,\mbox{TeV}$. It can be noticed, furthermore, that these relic density curves are not sensitive to variations of the $(M_A,\tan\beta)$ parameter sets. The reason is that in the second case, i.e. for $|\mu|< M_{1,2}$, the limit of a pure higgsino DM, already discussed in the previous subsection, is recovered. In the opposite regime, i.e. for $|\mu| > M_{1,2}$, the $M_1=M_2$ relation imposes a substantial mixing, of the order of the Weinberg angle $\theta_W$,  between the bino and wino components of the mixed bino-wino DM neutralino. Consequently, in both limits, the DM relic density is mostly accounted for by DM annihilation into gauge boson pairs, whose cross sections are not sensitive to the Higgs sector parameters. In addition, as one has $m_{\chi_1^0} \approx m_{\chi_2^0} \approx m_{\chi_1^\pm}$, co-annihilation processes (in particular those involving charginos which have stronger couplings) play an important role. In both cases, a multi-TeV DM neutralino is favored. 

For what concerns the limits from direct detection experiments, they are again very similar to the ones obtained in the previous $M_1 \simeq \frac12 M_2$ case. These very strong constraints can be thus evaded only if blind spots, i.e.  very specific correlations among the model parameters, are realized as also discussed before. Indirect detection constraints (yellow regions) play a complementary role at low DM masses. This is mostly due to the fact that substantial wino and/or higgsino component for the DM induce high annihilation cross-sections into pairs of gauge bosons, in tension with experimental sensitivity.
Notice that, due to Sommerfeld enhancement, also multi-TeV values of the DM mass are potentially subject to constraints from DM indirect detection. As shown e.g. by \cite{benekeID}, the latter sensitively affect the parameter space corresponding to wino-like DM and $\mu >0$. The systematic implementation of Sommerfeld enhancement is beyond the scopes of this work and we hence refer to the dedicated literature. 

\begin{figure}[!ht]
    \centering 
    \subfloat{\includegraphics[width=0.5\linewidth]{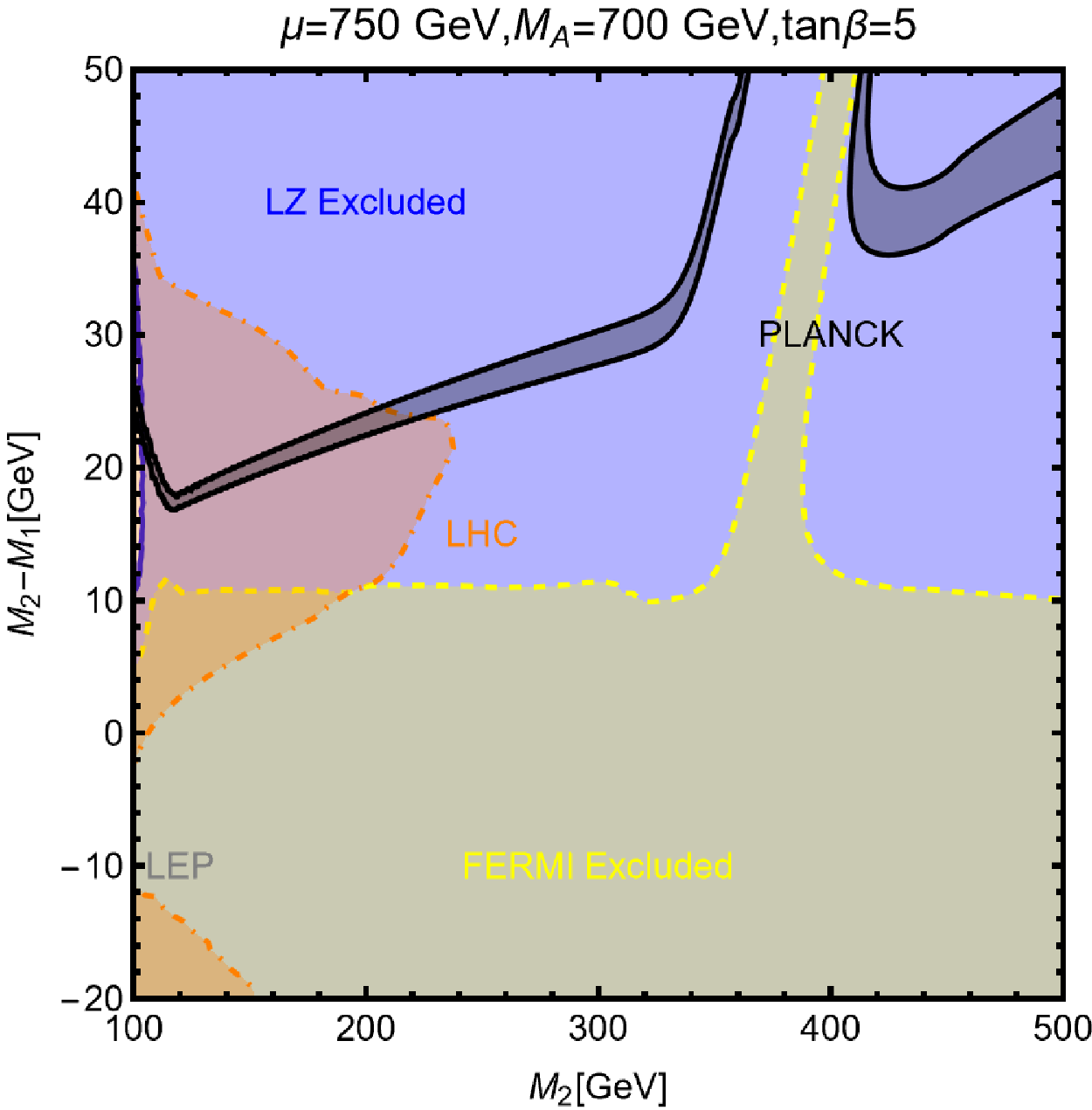}}
    \subfloat{\includegraphics[width=0.5\linewidth]{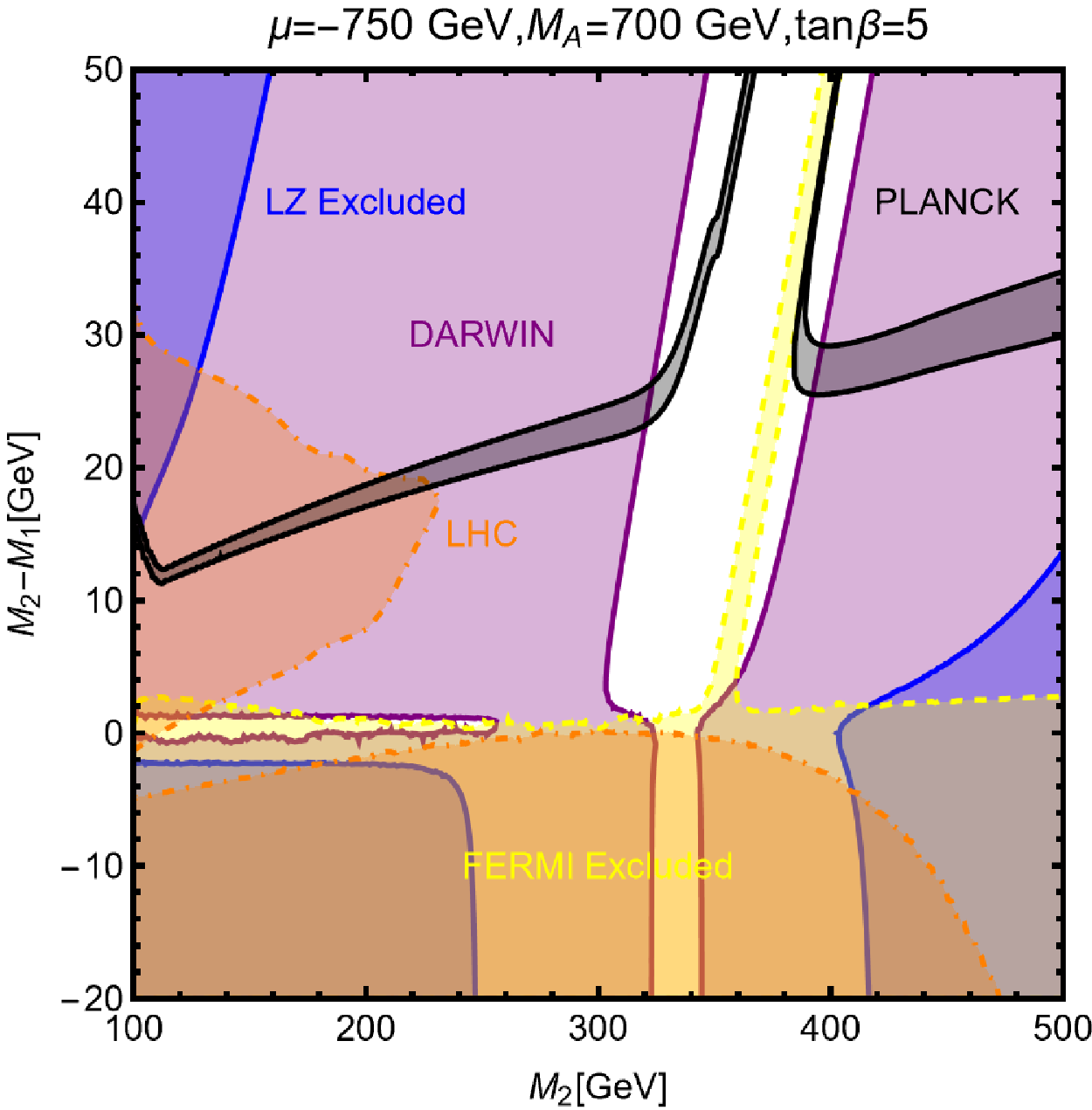}}\\[3mm]
    \subfloat{\includegraphics[width=0.5\linewidth]{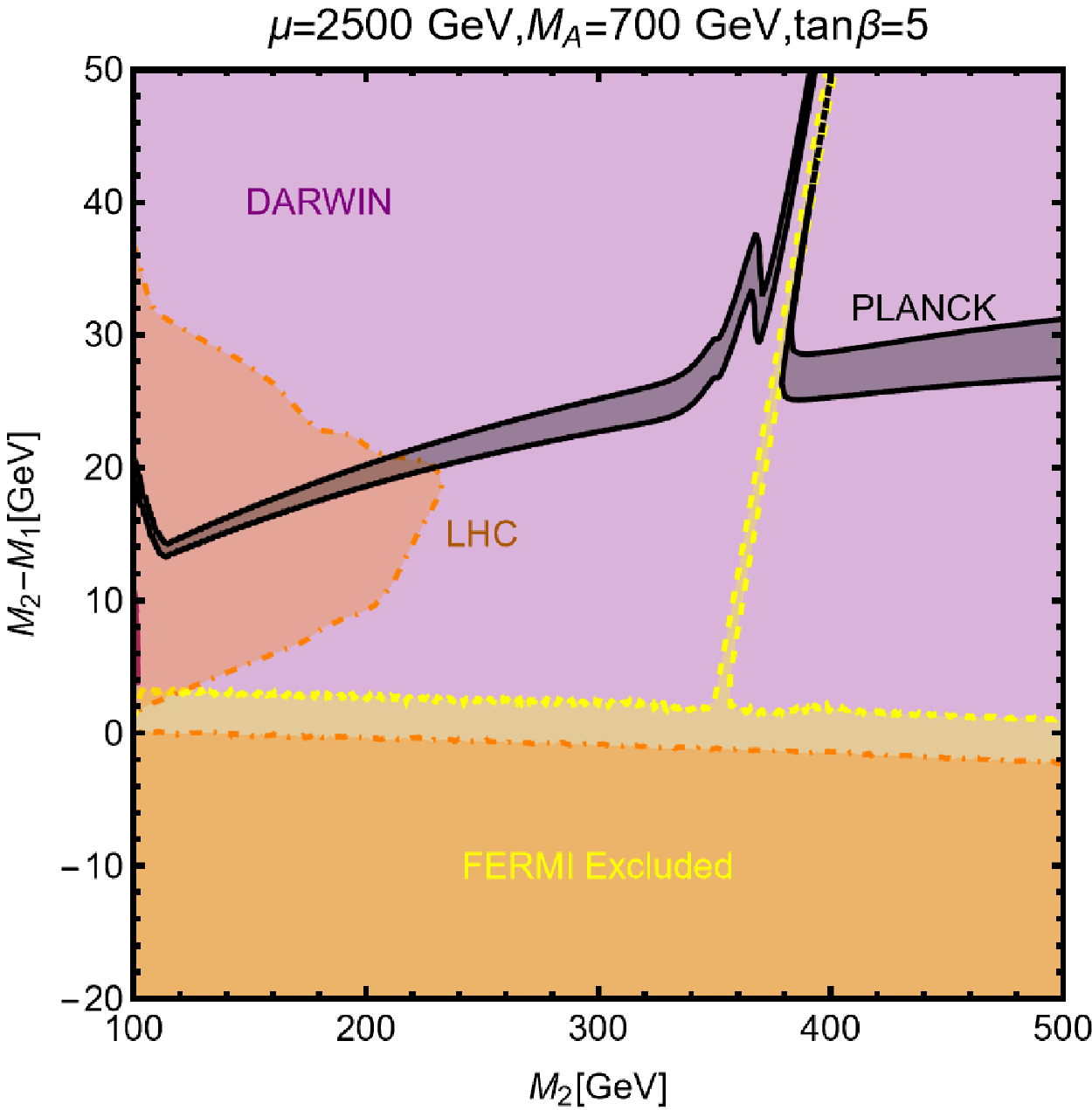}}
    \subfloat{\includegraphics[width=0.5\linewidth]{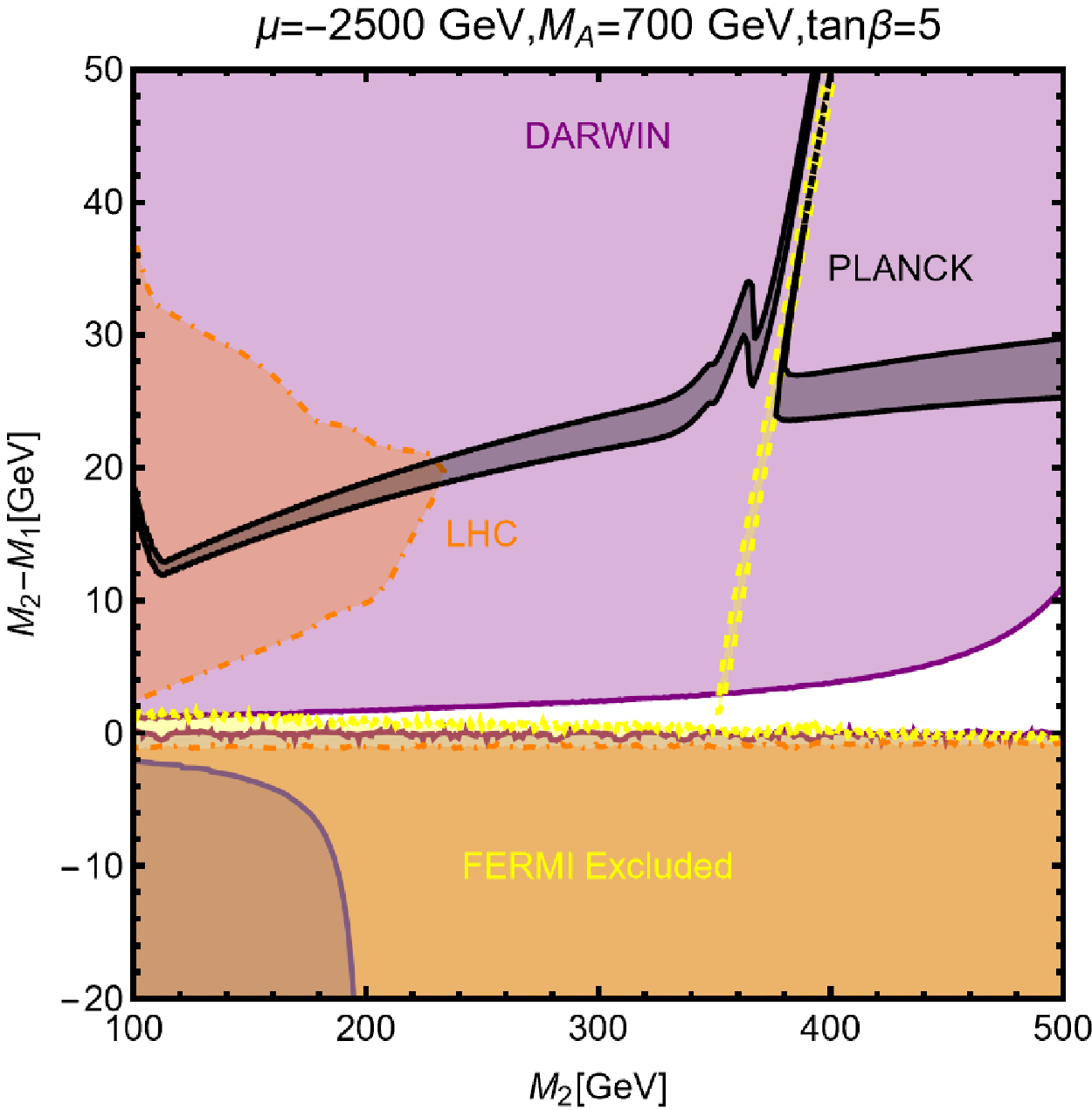}}
    \caption{Summary of constraints, in the $[M_1, (M_2-M_1)]$ parameter space, for some ``well-tempered" wino--like neutralino configurations, identified by the Higgs--higgsino parameter assignments on top of each panel. Following the usual color code, the correct DM thermal relic density is identified by the black isocontours,  the present (future) excluded regions by the LZ (DARWIN) experiments are marked in blue (magenta/purple) and the regions in orange correspond to those excluded from neutralino/chargino searches at the LHC (see main text for details) while, finally, the yellow regions are excluded by Indirect DM searches by the FERMI experiment. }
    \label{fig:scans_temp_bis}
\end{figure}

The picture just depicted can be altered if a slight deviation from the $M_1=M_2$ relation is considered. This is illustrated in Fig.~\ref{fig:scans_temp_bis} in which a difference in the two gaugino mass values   $M_2-M_1 \simeq 5\!-\!10\,\mbox{GeV}$ is assumed. For $|\mu| \gg M_{1,2}$ as chosen in all panels of the figure, this  leads to small splittings between the masses of the gaugino-like next-to-lightest neutralino and lightest chargino states $\chi_2^0$ and $\chi_1^\pm$ and the mass of the LSP neutralino $\chi_1^0$. This would strongly suppress the co-annihilation processes and allows that the correct DM relic density is achieved for lower DM masses. 

As can be seen from the lower  panels of Fig. \ref{fig:scans_temp_bis} with $\mu= \pm 2.5\,\mbox{TeV}$ (and also the upper panel with $\mu=-750$ GeV), the direct detection constraints from the LZ experiment can be evaded for relatively low values of the gaugino mass parameters $M_1$ and $M_2$, and hence LSP masses,  without relying on blind spot configurations. Nevertheless, some of these values are already constrained by LHC searches of charginos and neutralinos given by the orange regions. The portion of parameter space with $M_2 < M_1$, hence wino-like DM, is also ruled out both by DM indirect detection (again because of the very efficient s-wave annihilation cross-section) and both by LHC. For this kind of setup the most effective limits come from searches of disappearing tracks \cite{ATLAS-compress,CMS-compress}. The upgraded sensitivity expected for the future DARWIN experiment would allow to probe a very large part of the remaining viable DM parameter space. 

\subsection{$\mathbf{M_1 \ll M_2}$ and $\mathbf{M_2 \ll M_1}$}

The scenario $M_1 \ll M_2$ is the one in which the bino-like neutralino is lighter than the wino-like neutralino and chargino and, for this reason at least for what concerns DM phenomenology, it has strong similarities with the case $M_1= \frac12 M_2$ discussed before. Hence, most of the results presented in subsection 5.1 apply also in this case. Nevertheless, in the case $M_1 \ll M_2$ and for large values of the higgsino mass parameter,  $|\mu| \ggg M_1$, the LSP can be almost bino--like with no or very small couplings to the $Z$ boson. It is thus no longer constrained by the LEP bounds and, hence, can be very light. We will hence focus the analysis below on the specific $M_1 \lsim \frac12  M_h$ sub-case.

\begin{figure}[!ht]
    \centering
    \subfloat{\includegraphics[width=0.5\linewidth]{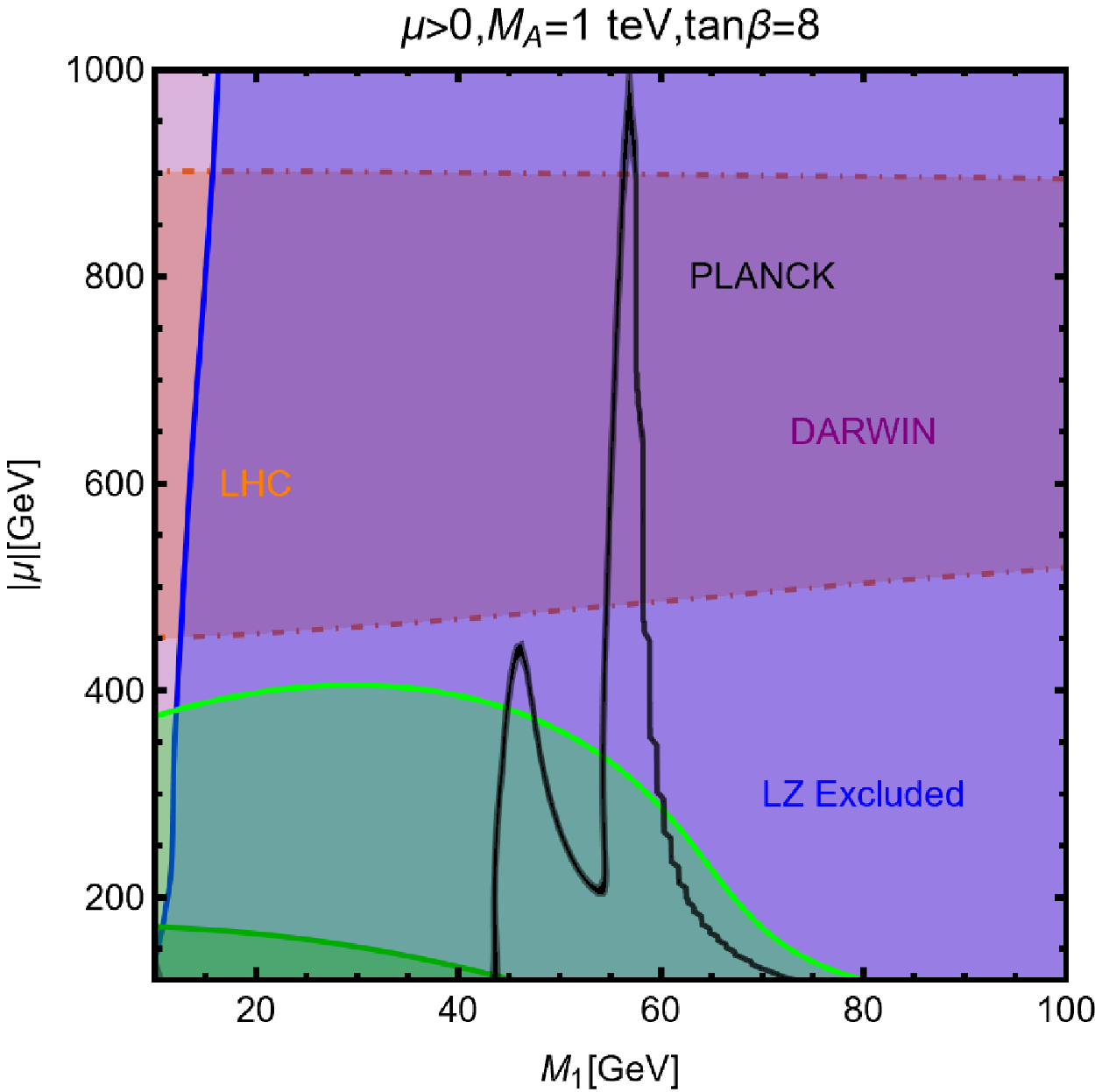}}
    \subfloat{\includegraphics[width=0.5\linewidth]{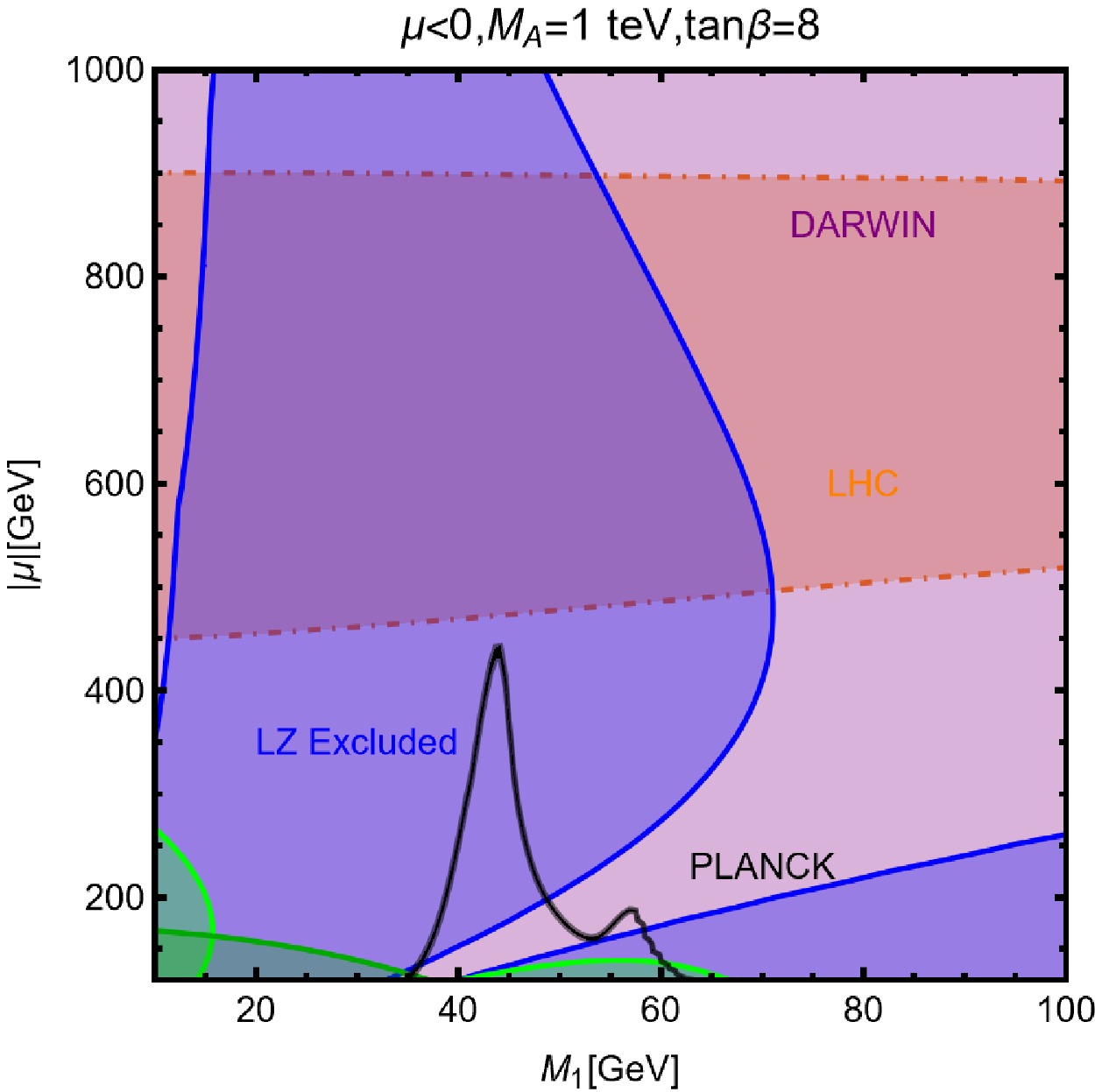}}\\[2mm]
    \subfloat{\includegraphics[width=0.5\linewidth]{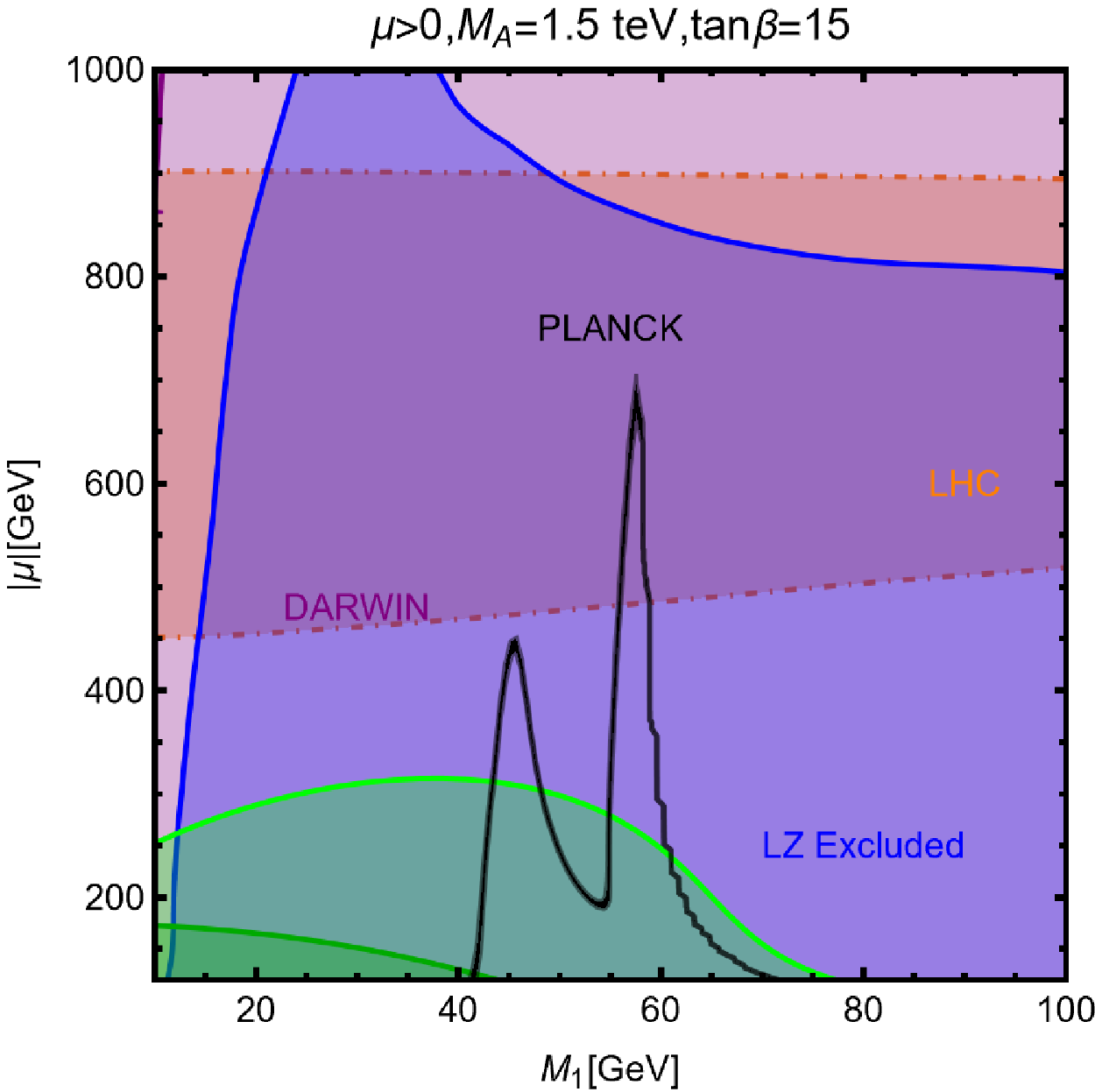}}
    \subfloat{\includegraphics[width=0.5\linewidth]{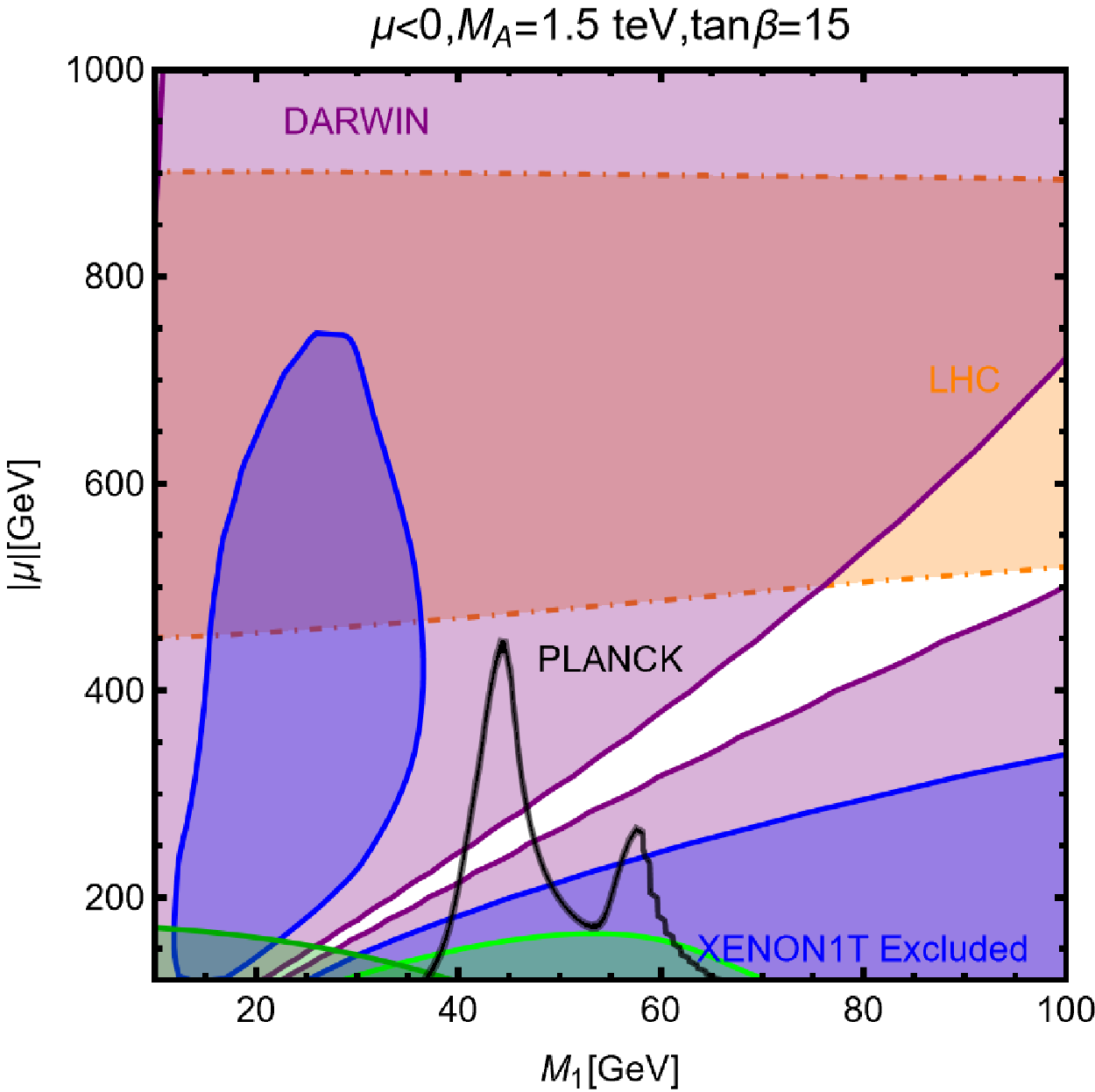}}
    \caption{Summary of constraints for the scenario with a light DM neutralino and the choice $M_1 \ll M_2$, in the $[M_1,|\mu|]$ bidimensional space for the benchmark assignments of the parameters $M_A$, $\tan\beta$ and $\mbox{sign}(\mu)$ reported on top of the four  panels. Following the usual color code, the black isocontours have the correct DM relic density, while the blue (magenta) regions are currently (will be) excluded by the LZ (DARWIN) experiments. In addition, shown in orange are the regions excluded by LHC searches of neutralinos and charginos and,  in green (dark green), the regions excluded by invisible decays of the lightest Higgs ($Z$) boson.}
    \label{fig:scans_light}
\clearpage
\end{figure}

Fig.~\ref{fig:scans_light} shows the combined constraints in the $[M_1,|\mu|]$ plane for the $(M_A,\tan\beta)=(1\,\mbox{TeV},8)$ and  $(M_A,\tan\beta)=(1.5\,\mbox{TeV},15)$ Higgs parameter sets and both signs of $\mu$. 

As already mentioned, the DM is mostly bino  in the $M_1\! \ll \! M_2$ scenario and the DM cosmological relic density is achieved essentially through DM annihilation processes into SM particle final states, via the SM-like Higgs and the $Z$ boson exchanges in the $s$--channel. Bino-like neutralinos have, however, strongly suppressed couplings to these particles so that the correct relic density is achieved only around the $m_{\chi_1^0}\simeq \frac12 M_{Z}$ or $\frac12 M_{h}$ poles.

For $\mu>0$, the two considered benchmark scenarios  are, nevertheless, already ruled out by present constraints from DM direct detection ad exception of very narrow regions around the SM Higgs pole\footnote{Notice that in such regions the conventional numerical computations of the relic density are not completely reliable \cite{relicresonance}.}. For $\mu <0$, direct detection can be evaded in large regions of the parameters space thanks to the occurrence of ``blind spots". However, having ``blind spots" means suppressed DM couplings to the SM--like Higgs boson which will then make it difficult to achieve  the correct relic abundance.  Indeed, as it can be seen from the plots, the PLANCK value of the DM relic density is reproduced only in the $h$ pole region, provided that $|\mu| \lesssim 500\,\mbox{GeV}$. The portion of the $(M_1,\mu)$ bidimensional space highlighted in orange is excluded by LHC searches. For the considered setup, we adopted the limits determined in \cite{atlasHB} for bino DM with higgsino NLSP.

For completeness, we also display in Fig.\ref{fig:scans_light}, the excluded region from the invisible decays of the $h$ boson, measured at the LHC (when combining the ATLAS and CMS values) to be \cite{H-invisible}
\beq
{\rm BR}( h \to {\rm invisible}) \leq 11\%,  
\eeq
and the invisible width of the $Z$ bosons, which is constrained from the very precise LEP measurements to be \cite{PDG} 
\beq
\Gamma( Z \to {\rm invisible}) \leq 2.49~{\rm MeV}.  
\eeq
One notices that the corresponding constraints are not as competitive as the ones from the LZ experiment just discussed before. Contrary to the previous scenario, no limits from DM indirect detection are displayed. This because of the p-wave suppression of the annihilation cross-section of bino-like DM.

Finally, in the $M_2 \ll M_1$ scenario, the DM neutralino is a wino-higgsino admixture for $|\mu|$ values comparable to $M_2$ and a pure wino (higgsino) for large (small) values of $|\mu|$.  The DM relic density is then determined essentially by the gauge interactions. Due to their effectiveness, and further enhanced by Sommerfeld effects \cite{Sommerfeld} and bound state formation \cite{bound-state-formation}, the experimental value of $\Omega h^2$ is matched only for multi-TeV lightest neutralino states. Hence, there is almost no complementarity between the phenomenology of the DM state including the constraints from the relic density and direct detection and the LHC searches for neutralinos and charginos. For this reason, we will not discuss this scenario in detail. 

\subsection{Summary of the DM constraints}

\begin{figure}[!h]
 \vspace*{3mm}
    \centering
    \subfloat{\includegraphics[width=0.5\linewidth]{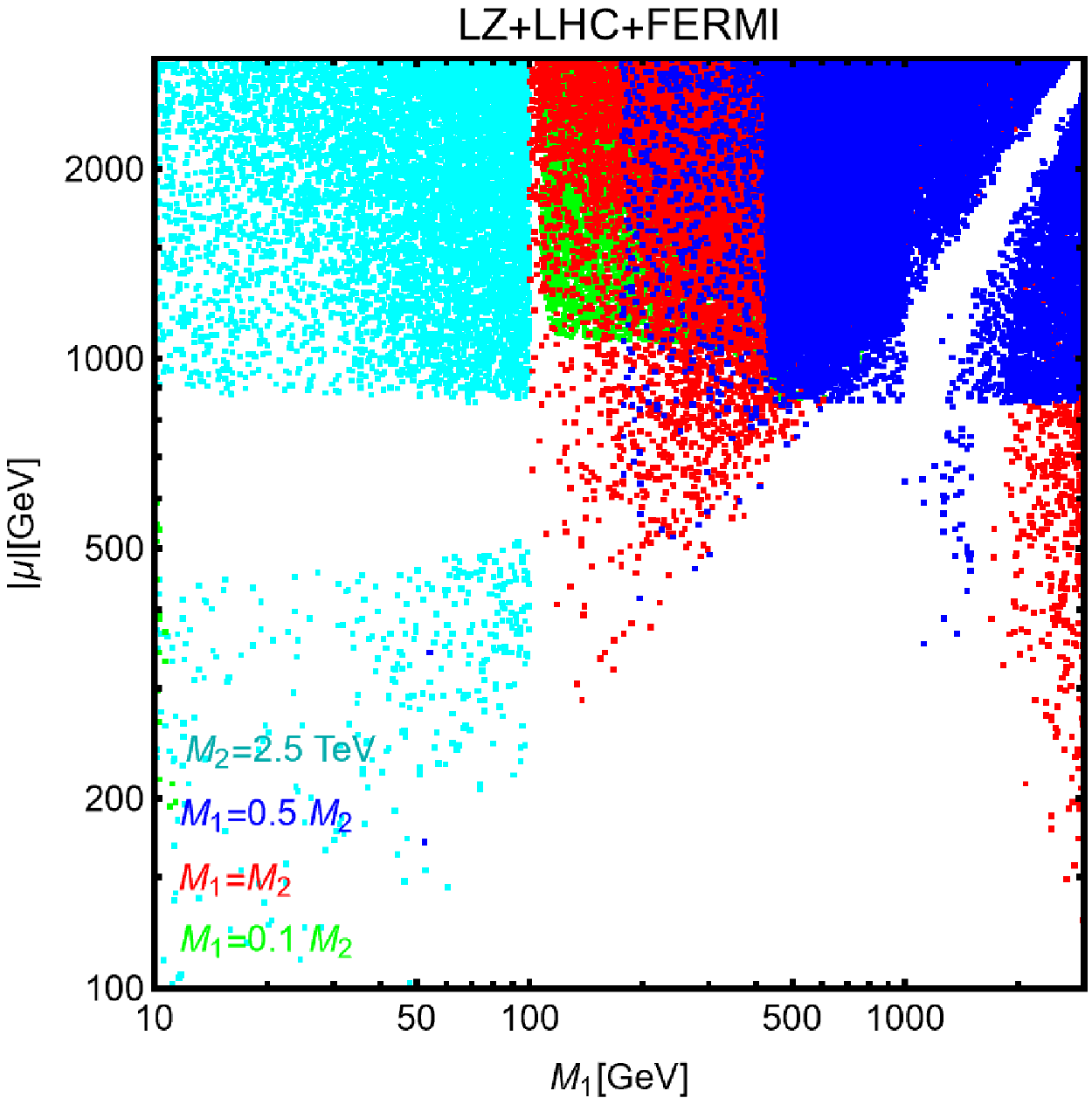}}
    \subfloat{\includegraphics[width=0.5\linewidth]{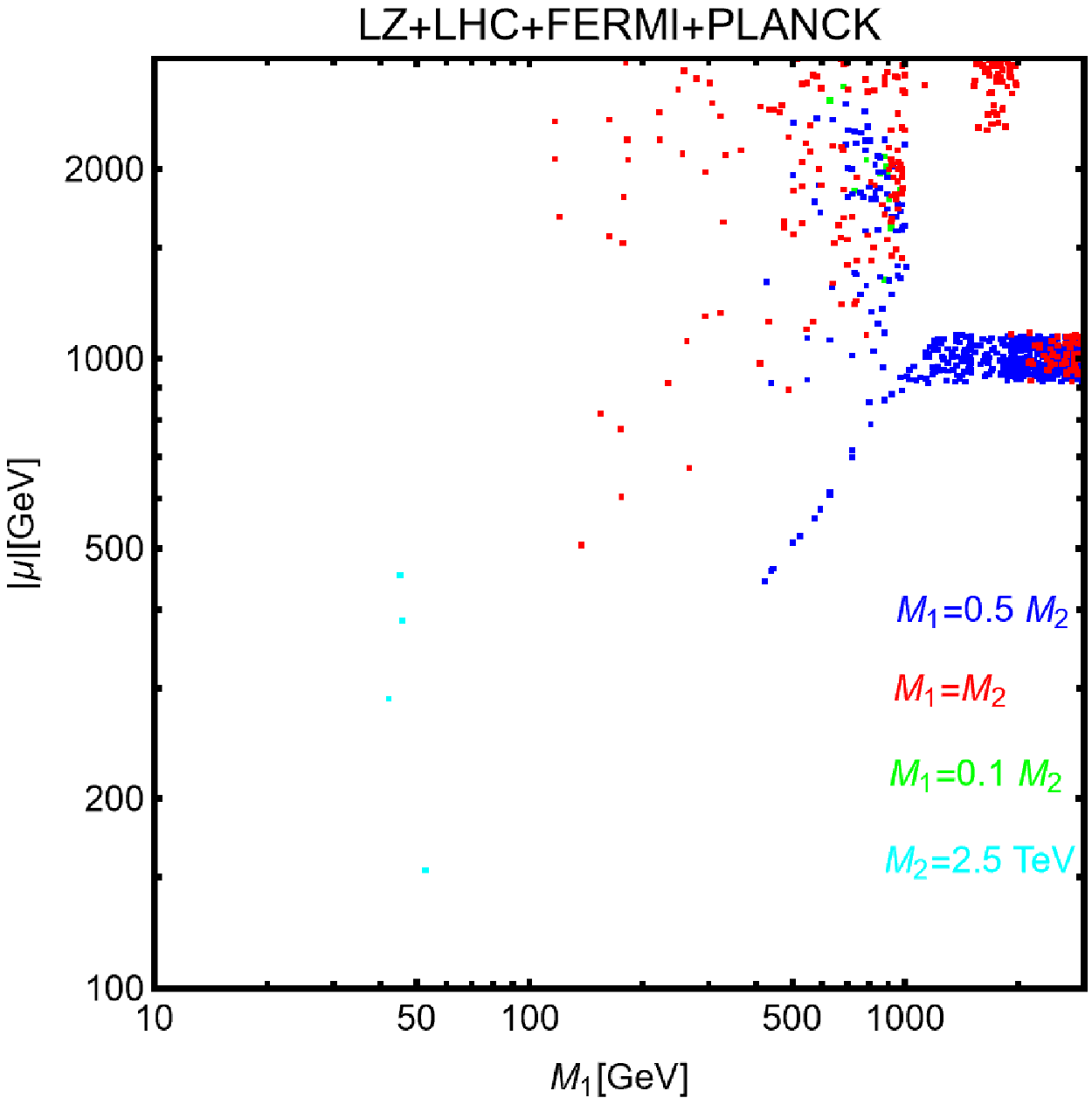}}\\[3mm]
     \subfloat{\includegraphics[width=0.5\linewidth]{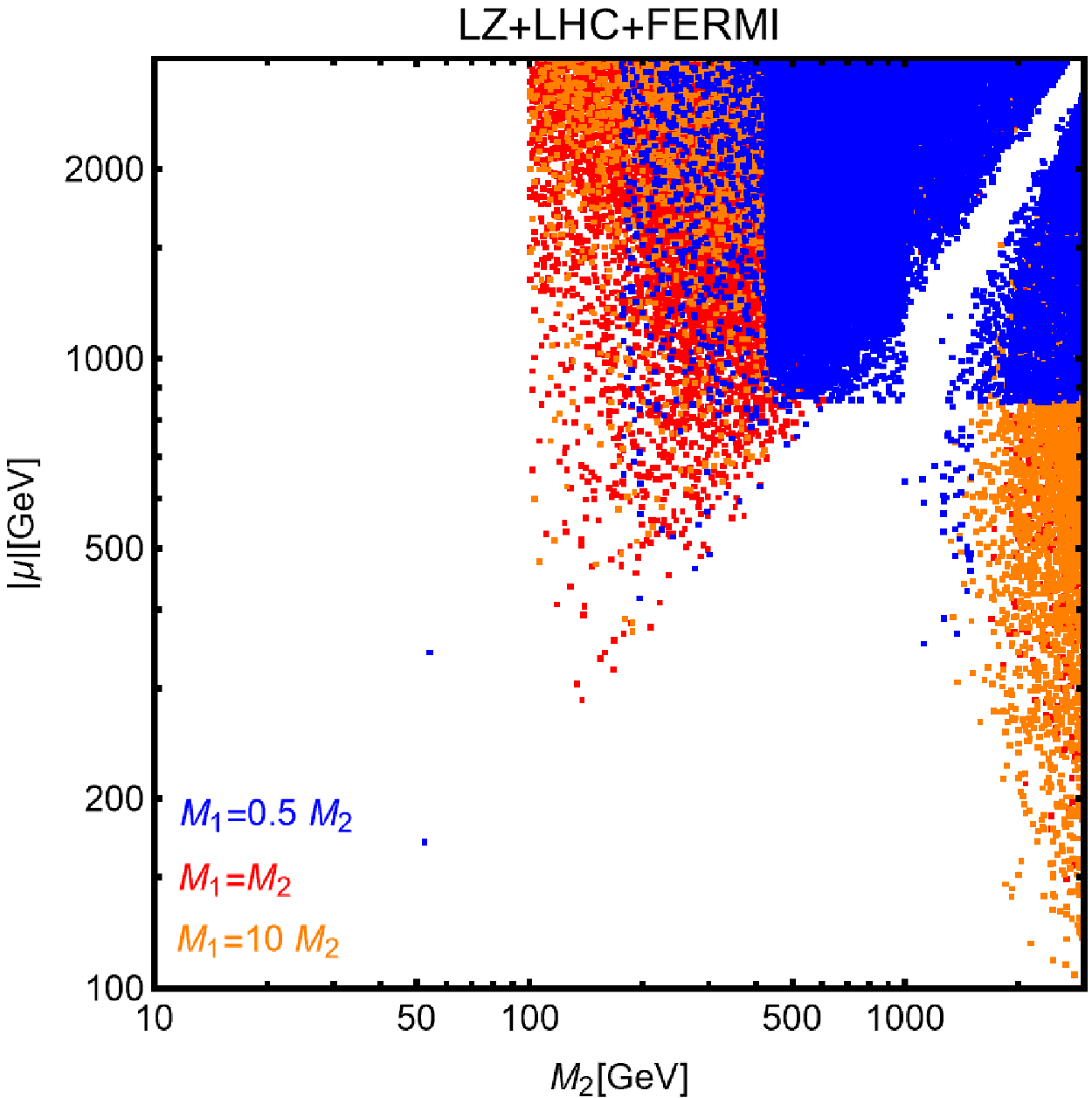}}
    \subfloat{\includegraphics[width=0.5\linewidth]{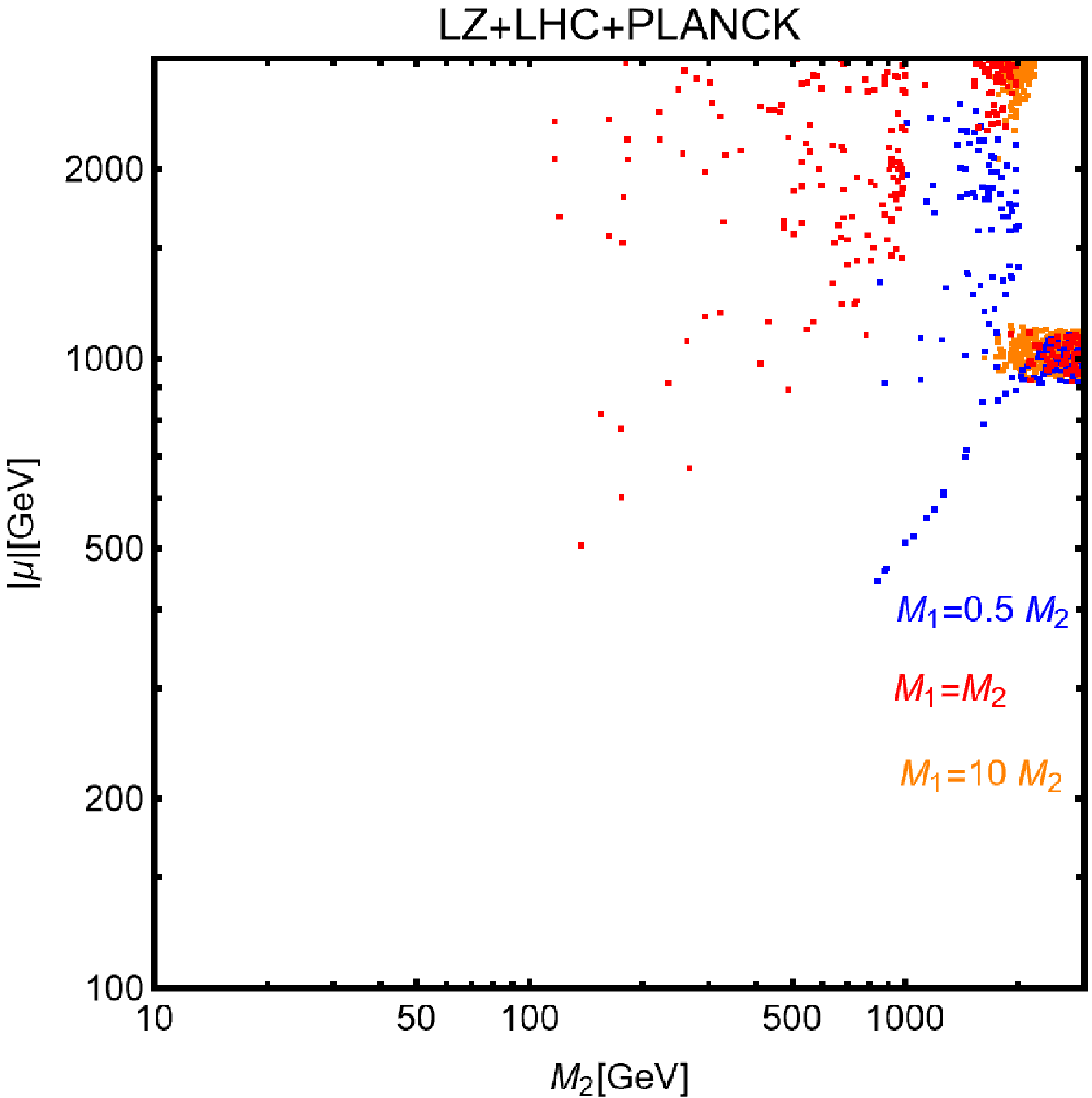}}
    \caption{Summary plots of the combined LHC, direct detection and relic density constraints on the hMSSM. The colored points represent the outcome of different parameter scans (see main text for details) conduced assuming $M_1 =0.1 M_2$ (green points), $M_1 = 0.5 M_2$ (blue points), $M_1 \simeq M_2$ (red points), $M_1 = 10 M_2$ (orange points) and $M_2=3\,\mbox{TeV}$. The left panel shows the points evading constraints from DM direct detection and from LHC while the right panels shows, in addition, the model points complying with the correct relic density.}
    \label{fig:summary}
\clearpage
\end{figure}

We summarize the outcome of the present DM analysis, in the two plots given in Fig.~\ref{fig:summary} which show the outcome of different scans over the hMSSM parameters, namely  $M_1,M_2,\mu$ and $M_A,\tan\beta$. Concerning the  Higgs--higgsino parameters $M_A,\tan\beta$ and $\mu$, we have assumed in all cases  the following ranges of variation:
\begin{equation}
    M_A \in \left[600,2000\right]\,\mbox{GeV}\, , \,\,\,\,\,\mu\in \left[-3,3\right]\,\mbox{TeV},\,\,\,\,\,\,\tan\beta \in \left[1,60\right] \, . 
\end{equation}
For what concerns the gaugino masses $M_{1,2}$ we have considered the following cases:
\begin{align}
    & M_2=10 M_1 \, , \,\,\, M_1 \in \left[10,1000\right]\mbox{GeV}, \nonumber\\
    & M_1 = 0.5 M_2\, , \,\,\, M_2 \in \left[100,3000\right]\,\mbox{GeV} , \nonumber\\
    & M_1=M_2 \pm 50 \mbox{GeV}\, ,  \, \, \, M_2 \in \left[100,3000\right]\,\mbox{GeV} \, , \nonumber\\
    & M_1=10 M_2, \,\,\, M_1 \in \left[100,3000\right]\mbox{GeV},\,\nonumber\\
    & M_1 =\left[10,100\right]\mbox{GeV}\,\,\,M_2=2.5\,\mbox{TeV} . 
\end{align}
The results are presented in the $[M_1,\mu]$ and $[M_2,\mu]$ planes. The lower range for the value $M_2$ and $\mu$ is chosen to automatically account for the LEP limit on the lightest chargino mass, implying $\mbox{min}[M_2,\mu] \gtrsim 100\,\mbox{GeV}$. The panels in the left column of Fig.~\ref{fig:summary} show the model points evading the constraints from DM direct detection, as given by LZ, indirect detection, as given by FERMI, collider constraints from searches of additional Higgs bosons, invisible decays of the $Z/h$ bosons and searches for neutralinos/charginos. The right column shows the model points which, in addition, feature the correct DM relic density according to the WIMP paradigm. The different colors correspond to, respectively, the $M_2=3\,\mbox{TeV}$ (cyan), $M_1 \! = \! 0.1 M_2$ (green), $M_1 \! = \! \frac12 M_2$ (blue),  $M_1  \! \simeq  \! M_2$ (red) and $M_1  \! =  \! 10 M_2$ (orange) configurations.

Focusing on the upper left panel, we see that the $M_1=0.5 M_2$ configuration occupies the smallest region of parameter space. This is due to the limits on LHC searches of chargino/neutralino production which are particularly strong in the case of bino-LSP/wino-NLSP \cite{atlasWZ}. In agreement with the results shown in the previous subsection, Indirect Detection constraints configurations corresponding to light higgsino-like DM while, bino-higgsino mixtures, namely $M_1 \simeq \mu$, are ruled out by LZ\footnote{Notice that, to apply direct and indirect detection limits in the way described above, one has to implicitly assume that each model points complies with the correct DM relic density, possibly via unknown non-thermal mechanism. Alternatively, one could assume that the neutralino DM represents only a fraction of the DM of the universe and rescale accordingly the experimental exclusion bounds.}. In order that light DM evades observational constraints we need to increase the hierarchy between the $M_1$, $M_2$. Assuming the latter parameter to be decoupled at a mass scale of $3\,\mbox{TeV}$, it is possible to access values of the DM mass below 100 GeV. Looking finally at the $M_1 \sim M_2$ case, the model points occupy a relatively large region of parameter space for $M_1>100\,\mbox{GeV}$. While the impact of DM, namely direct and indirect detection, constraints is similar as the previous cases, LHC bounds are weaker due to the compressed neutralino/chargino spectrum. The latter have substantial impact in the case of almost pure wino DM where limits from disappearing tracks apply.


Moving to the bottom left panel, showing the $[M_2,|\mu|]$ bidimensional space, we can assess the viable parameter region for $M_2 =0.1 M_1$. A generic admixture of wino and higgsino components always features s-wave annihilation cross-section which is, hence, constrained by indirect detection. This implies the lower bounds $|\mu| \gtrsim 300\,\mbox{GeV}$ and $M_2\gtrsim 700\,\mbox{GeV}$. Again, the case $M_2 \sim |\mu|$ is ruled by Direct Detection for the whole range of the latter parameters considered in this work. 
As it can be seen from the right-hand side column of Fig.~\ref{fig:summary},  the strict requirement of the conventional thermal paradigm as mechanism to achieve the correct DM relic density has a dramatic impact on the viable parameter space. This is due to the fact that one has to assume very specific and possibly fine-tuned relations between the model parameters. In the case $M_1 = 0.5 M_2$, the correct DM relic density is achieved only for higgsino-like DM with $\mu \simeq 1\,\mbox{TeV} < M_{1,2}$ or for bino-like DM, provided that $m_{\chi_1^0} \simeq \frac12 M_A$. In light of this tight relation, the lower bound on $M_A$ from LHC searches of resonances in the $pp \rightarrow A/H \rightarrow \tau \tau$ channel, serves as a lower bound on the DM mass, translating into the constraint $m_{\chi_1^0} \simeq M_1 \gtrsim 350\,\mbox{GeV}$. In the case $M_1 \simeq M_2$, the occurrence of the well tempered bino-wino regime allows to have $M_1$ as low as 100 GeV, provided that $|\mu|$ is above the TeV range. Furthermore, the correct DM relic density can be achieved in the case of a wino-like DM, with $M_1$ at around 2 TeV. No new viable parameter regions open up for $M_1 \gg M_2$. In this last case, all the experimental constraint are passed for pure higgsino or pure wino DM while the mixed composition for the DM is ruled out by direct detection. Finally, for $M_1 \ll M_2$, an additional small region of viable parameter space is present which  corresponds to the lightest Higgs pole region $m_{\chi_1^0} \simeq M_1 \simeq \frac12  M_h$.  

\section {Conclusions}
\label{sec:6}

The hMSSM, in which the MSSM Higgs sector is parameterized in such a way that the mass of the lightest Higgs boson is automatically set to the value $M_h=125$ GeV measured at the LHC, has become one of the two main  benchmark scenarios used by the ATLAS and CMS collaborations in constraining the parameter space of the model either by interpreting their measurements of the SM-like lightest $h$ couplings to fermions and gauge bosons \cite{H-couplings} or in their searches of the heavier $H,A$ and $H^\pm$ states in various discovery channels. A nice feature of this approach, which tremendously simplifies the experimental and theoretical analyses, is that only two basic input parameters, generally taken to be $\tan\beta$ and $M_A$ exactly like at the tree-level, are then needed in order to describe the Higgs sector. This is particularly true in the case where the masses of scalar quarks or, alternatively, the SUSY-breaking scale $M_S$, are large as suggested by the LHC data.  It is nevertheless useful and in fact mandatory in view of future LHC upgrades, to allow for the possibility of light weakly interacting particles  which are much less constrained by current data and which will be subjects to intensive searches at these upgrades.

This is the main scope of the present paper: to extend the by now traditional hMSSM \cite{Habemus1,Habemus2} in order to incorporate the effects of possibly light charginos $\chi_i^\pm$ and neutralinos $\chi_i^0$ and study the implications at both colliders such as the LHC and in astroparticle physics experiments searching for the dark matter particle which, in our context, is represented by the lightest neutralino $\chi_1^0$. Still assuming large values of the SUSY scale $M_S$ and the gluino mass parameter $M_3$, which implies that squarks and gluinos are too heavy to have any direct impact on Higgs phenomenology, we define an hMSSM with three extra inputs parameters in addition to $\tan\beta$ and $M_A$,  namely the bino, wino and higgsino mass parameters $M_1, M_2$ and $\mu$. In this extended hMSSM benchmark, we then explore in the most comprehensive way all possible effects of charginos and neutralinos and their interplay with the Higgs sector.   

We first start be summarizing the way we implement such an hMSSM scenario in one of the numerical codes that generate the superparticle and Higgs spectra in the MSSM, the program SuSpect \cite{Suspect}, which we then use in the subsequent  numerical analyses (eventually after linking it with codes that calculate production cross sections and decay branching ratios such as HDECAY  \cite{HDECAY} and SDECAY \cite{SDECAY}). 
We then evaluate the one-loop radiative corrections of charginos and neutralinos to the Higgs boson masses and couplings and show that their impact, compared to the hMSSM as traditionally defined, is rather modest for reasonable values of the additional SUSY parameters. In particular, the one-loop corrections to the Higgs boson self-energies  modify the mass of the $h$ boson by less than 1\%, which is well below the admitted theoretical uncertainty in determining this mass in a general MSSM. These corrections have an even smaller impact on the determination of the masses of the heavier CP-even $H$ and charged $H^\pm$ states as well as on the angle $\alpha$. There are also direct corrections that affect the Higgs-$b\bar b$ vertices at large $\tan\beta$  and $\mu$ values, but it turns out that they are also modest if squarks and gluinos are assumed to be sufficiently heavy and, hence, they do not alter the global picture. 

We then study the collider implications of the charginos and neutralinos in this hMSSM. We first summarize their production and detection channels at the LHC and adapt the present experimental constraints on these particles derived by ATLAS and CMS to our hMSSM with 5 parameters, $M_A, \tan\beta, M_1, M_2$ and $\mu$ and delineate the still allowed parameter space. We then discuss all possible sources of impact of these charginos and neutralinos on the MSSM Higgs sector: the $\chi_i^\pm$ contributions to the decays of the neutral $h,H,A$ states into two photons, the direct corrections to heavier $H,A,H^\pm$ production and decays in the main channels, and their impact on the lighter $h$ boson signal strengths as measured at the LHC. We close this collider part by discussing the possible impact of Higgs decays into charginos and neutralinos on present LHC constraints. 

We finally,  perform a complete analysis in the hMSSM of the astroparticle implications of the lightest neutralino, considered as the dark matter particle. In particular, we update the constraints from the relic abundance of this neutralino as well as the constraints and prospects in direct and indirect detection experiments, paying a special attention to the complementarity between these searches and those performed at colliders.

All in all, we show that extending the original hMSSM scenario to incorporate the possibility of light chargino and neutralino states can be done at a minimal cost.  Such an hMSSM can still parameterize in an accurate manner the MSSM Higgs sector and, at the same time, be a very good benchmark in describing the interplay between this sector and the one involving the electroweak gauginos and higgsinos.
This will ease future simultaneous analyses of the two sectors at the next LHC runs, future colliders and astrophysical experiments.


\subsection*{Acknowledgements}

{ We thank Luciano Maiani collaboration at an early stage of this work and Pietro Slavich for discussions on the hMSSM. } 
A.D. is supported by the Estonian Research Council grants MOBTT86 and by the Junta de Andalucia through the Talentia Senior program as well as by the grants A-FQM-211-UGR18, P18-FR-4314 with ERDF and PID2021-128396NB-I00.
HJH and RQX are
supported in part by the National NSF of China (under grants Nos.\,11835005 and 11675086) and by the National Key R\,\&\,D 
Program of China (No.\,2017YFA0402204).\ 



\begin{thebibliography}{99}

\bibitem{H-discovery}
  The ATLAS Collaboration,
  Phys.\ Lett.\ {\bf B716} (2012) 1;
  The CMS Collaboration,
  Phys.\ Lett.\ {\bf B716} (2012) 30.

\bibitem{H-couplings}
  The ATLAS and CMS Collaborations,
  JHEP {\bf 1608} (2016) 045;
  The ATLAS Collaboration,
ATLAS-CONF-2019-005;
  The CMS Collaboration,
JHEP \textbf{01} (2021) 148.


\bibitem{SUSY} 
For reviews on supersymmetric theories, see
M. Drees, R. Godbole and P. Roy,  {\it Theory and phenomenology of sparticles}, World Scientific, 2005; H. Baer and X. Tata, {\it Weak scale Supersymmetry: from superfields to scattering events},  Cambridge U. Press, 2006; S. Martin, hep-ph/9709356; P. Bin\'etruy, {\it Supersymmetry: Theory, Experiment, and Cosmology}, Oxford University Press, 2006.

\bibitem{HaberKane} H.E. Haber and G.L. Kane, Phys. Rept. {\bf 117} (1985)  75.

\bibitem{pMSSM} A.~Djouadi {\it et al.}, The MSSM working Group,  hep-ph/9901246.

\bibitem{HHG}  J. Gunion, H. Haber, G. Kane and S. Dawson, ``The Higgs Hunter's Guide", Reading 1990; A. Djouadi,   Phys.\ Rept.\  {\bf 457} (2008) 1; M. Spira, Prog.\ Part.\ Nucl.\ Phys.\  {\bf 95} (2017) 98. 
  
\bibitem{Anatomy2} A.~Djouadi,
  Phys.\ Rept.\  {\bf 459} (2008) 1.

\bibitem{LHC-searches} Heavy Higgs searches at the LHC have been discussed recently in e.g.: M. Kakizaki et al., 
Int. J. Mod. Phys. {\bf A30} (2015) no.33, 1550192; J. Baglio, A. Djouadi and J. Quevillon, Rept. Prog. Phys. {\bf 79} (2016) no.11, 116201; P. Athron et al. [GAMBIT], Eur. Phys. J. {\bf C77} (2017) no.12, 824; P. Athron et al. [GAMBIT], Eur. Phys. J. {\bf C77} (2017) no.12, 879; 
L.-C.\ L\"{u}, C.\ Du, Y.\ Fang, H.-J.\ He, and H.\ Zhang, 
Phys.\ Lett.\ B {\bf 755} (2016) 509; 
J.\ Ren, R.-Q.\ Xiao, M.\ Zhou, Y.\ Fang, H.-J.\ He, and W.\ Yao,
JHEP {\bf 06} (2018) 090; 
E. Bagnaschi et al., Eur. Phys. J. {\bf C78} (2018) no.3, 256; E. Bagnaschi et al., Eur. Phys. J. {\bf C79} (2019) no.2, 149.

\bibitem{2HDM} G.C.~Branco, P.M.~Ferreira, L.~Lavoura, M.N.~Rebelo, M.~Sher and J.P.~Silva, Phys. Rept. {\bf 516} (2012) 1.

\bibitem{Reviews-cor} For reviews, see: M. Carena and H.  Haber, Prog. Part. Nucl.  Phys. {\bf 50} (2003) 63; S.~Heinemeyer, W.~Hollik and G.~Weiglein, Phys. Rept. {\bf 425} (2006) 265; S. Heinemeyer, Int. Jour. Mod. Phys. {\bf A21} (2006) 2659; B. Allanach et al., JHEP 0409 (2004) 044;
P.~Slavich et al., Eur. Phys. J. C \textbf{81}, no.5, 450 (2021).

\bibitem{Benchmarks} M. Carena et al.,  Eur. Phys. J.C {\bf 26} (2003) 601;
Eur. Phys. J. {\bf C73} (2013) 2552;     E. Bagnaschi et al,  Eur. Phys. J.C {\bf 79} (2019) 7, 617.

\bibitem{Bagnaschi-WG}  E. Bagnaschi et al., Report LHCHXSWG-2015-002.

\bibitem{Habemus1} A. Djouadi, L. Maiani, G. Moreau, A. Polosa, J.
Quevillon and V. Riquer, Eur. Phys. J. {\bf C73} (2013) 2650.

\bibitem{Habemus2}  A. Djouadi, L. Maiani,  A. Polosa, J. Quevillon and V. Riquer, JHEP
{\bf 06} (2015) 168.

\bibitem{Wagner+Lee}  G.~Lee and C.~E.~M.~Wagner,
   Phys.\ Rev.\ D {\bf 92} (2015) no.7,  075032;
   H.~Bahl, S.~Liebler and T.~Stefaniak, Eur. Phys. J. C \textbf{79} (2019) no.3, 279.

\bibitem{Pietro} P. Slavich, talk given at Higgs Days, Santander (Spain), Sept. 2015. 

\bibitem{Leading-cor} J. Ellis, G. Ridolfi and F.~Zwirner, Phys. Lett. {\bf B257} (1991) 83; Y. Okada, M. Yamaguchi and T. Yanagida, Prog. Theor. Phys. {\bf 85} (1991) 1;  H.~Haber and R.~Hempfling, Phys. Rev. Lett. {\bf 66}  (1991) 1815.

\bibitem{Leading-cor-high} M.~Carena, J.R.~Espinosa, M.~Quiros and C.E.~Wagner, Phys.  Lett. {\bf B355} (1995) 209; H.~Haber, R.~Hempfling and A.~Hoang, Z. Phys. {\bf C75} (1997)539; S.~Heinemeyer, W.~Hollik and G.~Weiglein, Phys. Rev. {\bf D58} (1998) 091701; Eur. Phys. J. {\bf C9}  (1999) 343; G. Degrassi, P. Slavich and F. Zwirner, Nucl. Phys. {\bf B611} (2001) 403; A. Brignole, G. Degrassi, P. Slavich and F. Zwirner, Nucl. Phys. {\bf B631} (2002) 195;  Nucl. Phys. {\bf B643} (2002) 79;  S. Martin, Phys. Rev. {\bf D75} (2007) 055005;  P. Kant, R. Harlander, L. Mihaila and M. Steinhauser, JHEP {\bf 1008} (2010) 104;
E.~Bagnaschi, J.~Pardo Vega and P.~Slavich, Eur. Phys. J. C \textbf{77} (2017) no.5, 334. 

\bibitem{h-inverse}
R.~El-Kosseifi, J.~L.~Kneur, G.~Moultaka and D.~Zerwas,
[arXiv:2202.06919 [hep-ph]].

\bibitem{Adam:2021rrw} W.~Adam and I.~Vivarelli,
 Int. J. Mod. Phys. {\bf A37} (2022) 02, 2130022.

\bibitem{HL-LHC} The ATLAS collaboration, ATL-PHYS-PUB-2018-054; the  CMS collaboration, contribution to CSS2013, 1307.7135 [hep-ex]; M. Cepeda, CERN Yellow Report, Monogr. {\bf 7} (2019) 221-584; arXiv:1902.00134; J.~de Blas et al., JHEP {\bf 01} (2020), 139.

\bibitem{R-parity} G. Farrar and P. Fayet, Phys. Lett., {\bf 76B} (1978) 575.

\bibitem{DM-Drees} M. Drees and G. Gerbier, Mini-Review of Dark Matter for the Review of Particle Physics,  arXiv:1204.2373 [hep-ph].

\bibitem{DM-PhysRep}  G. Arcadi, A. Djouadi and M. Raidal,  Phys. Rept. {\bf 842} (2020) 1.

\bibitem{GunionHaber12} J.F. Gunion and H.E. Haber, Nucl. Phys., {\bf B272} (1986) 1;  Nucl. Phys. {\bf B278} (1986) 449.

\bibitem{Suspect} A. Djouadi, J.L. Kneur and G. Moultaka, Comput. Phys. Commun. {\bf 176} (2007)  426.


\bibitem{Arbey} 
A. Arbey et al., Phys. Lett. {\bf  B708} (2012) 162; Phys. Lett. {\bf  B720} (2013) 153;  JHEP {\bf  1209} (2012) 107.

\bibitem{Habemus-early} L. Maiani, A.D. Polosa and V. Riquier,  New J.~Phys. {\bf 14} (2012) 073029; Phys. Lett. {\bf B718} (2012) 465; and Phys. Lett. {\bf B724} (2013) 274; A. Djouadi and J. Quevillon, JHEP {\bf 1310} (2013) 028; G.~Chalons, A.~Djouadi and J.~Quevillon, Phys.\ Lett.\ {\bf B780} (2018) 74;  S.~Liebler, M.~M\"uhlleitner, M.~Spira and M.~Stadelmaier,
  Eur.\ Phys.\ J.\ {\bf C79} (2019) no.1,  65.

\bibitem{H+mass}     A. Brignole, Phys. Lett. {\bf B277} (1992) 313; M. Frank et al., Phys. Rev. {\bf D88} (2013) 055013.

\bibitem{BPMZ} D.M.~Pierce, J.A.~Bagger, K.T.~Matchev and R.-J.~Zhang, Nucl.\ Phys.\ {\bf B491}, 3 (1997).

\bibitem{PV} G.~Passarino and M.~J.~G.~Veltman,
Nucl. Phys. B \textbf{160}, 151-207 (1979).

\bibitem{Warsaw} P.H. Chankowski, S. Pokorski and J. Rosiek,  Nucl. Phys. {\bf B423} (1994) 437.

\bibitem{hffselw} A.~Dabelstein, Nucl.\ Phys.\ {\bf B456} (1995) 25.

\bibitem{hffqcd} See e.g., A.~Djouadi, M.~Spira and P.~M.~Zerwas,  Z.\ Phys.\ {\bf C70}, 427 (1996).

\bibitem{Deltab1+Deltab2}
  M.~Carena, D.~Garcia, U.~Nierste and C.~M.~Wagner,  Nucl.\ Phys.\ {\bf B577}, 88 (2000);  J.~Guasch, P.~H\"afliger and M.~Spira, Phys.\ Rev.\ {\bf D68}, 115001 (2003);
D. Noth and M. Spira, Phys. Rev. Lett. {\bf 101}, 181801
(2008); and JHEP {\bf 1106}, 084 (2011); L. Mihaila and
C. Reisser, JHEP {\bf 1008}, 021 (2010); A. Crivellin
and C. Greub, Phys. Rev. {\bf D87} (2013) 015013 Erra-
tum: [Phys. Rev. {\bf D87} (2013) 079901]; L. Mihaila and
N. Zerf, JHEP {\bf 1705} (2017) 019; M. Ghezzi, S. Glaus,
D. Muuller, T. Schmidt and M. Spira, Eur. Phys. J.
{\bf C81} (2021) no.3, 259.

\bibitem{last-SUSY-paper} A. Arbey et al.,  arXiv:2201.00070 [hep-ph]. 

\bibitem{pp-chi-NLO} W. Beenakker et al., 
Phys. Rev. Lett. {\bf 83} (1999) 3780-3783; J. Fiaschi and M. Klasen, Phys. Rev. {\bf D98} (2018) 5, 055014. 

\bibitem{pp-chi-prospino} W. Beenakker, R. Hopker and M. Spira, e-Print: hep-ph/9611232;  T. Plehn, hep-ph/9809319; M. Spira,  hep-ph/0211145.  

\bibitem{Haber-Gunion3} J.F. Gunion and H.E. Haber, Nucl. Phys. {\bf B307} (1988) 445;
(E) hep-ph/9301205.

\bibitem{chi-decay-old} J.F. Gunion et al., Int. J. Mod. Phys. {\bf A2} (1987) 1145; A.~Bartl, W.~Majerotto and N.~Oshimo,
Phys. Lett. B \textbf{216} (1989) 233; H. Baer, M. Bisset, X. Tata and J. Woodside, Phys. Rev. {\bf D46} (1992) 303; D. Denegri, W. Majerotto and L. Rurua, CMS-NOTE-1997-094, hep-ph/9711357; S. Abdullin et al. (CMS collaboration), J. Phys. {\bf G28} (2002) 469; I. Hinchliffe et al. (ATLAS collaboration), Phys. Rev. {\bf D55} (1997) 5520.

\bibitem{chi-decays} A.K. Datta et al., 
Phys. Rev. {\bf D65} (2002) 015007;  
A.K. Datta et al., Nucl. Phys. {\bf B681} (2004) 31;
A. Djouadi, Y. Mambrini and M. Muhlleitner,  Eur. Phys. J. {\bf C20} (2001) 563;
M.~Bisset, M.~Guchait and S.~Moretti,
Eur. Phys. J. C \textbf{19} (2001), 143-154; 
H.~Baer, A.~Mustafayev, S.~Profumo, A.~Belyaev and X.~Tata,
JHEP \textbf{07} (2005), 065; 
K.~Huitu, R.~Kinnunen, J.~Laamanen, S.~Lehti, S.~Roy and T.~Salminen,
Eur. Phys. J. C \textbf{58} (2008), 591-608;
A. Djouadi et al., JHEP \textbf{07} (2008) 002;
P. Bandyopadhyay et al., JHEP {\bf 03} (2010) 04;
G.~D.~Kribs, A.~Martin, T.~S.~Roy and M.~Spannowsky,
Phys. Rev. D \textbf{82} (2010), 095012;
A.~C.~Fowler and G.~Weiglein,
JHEP \textbf{01} (2010), 108;
S.~Gori, P.~Schwaller and C.~E.~M.~Wagner,
Phys. Rev. D \textbf{83} (2011), 115022; 
F.~Yu, Phys. Rev. D \textbf{90} (2014) no.1, 015009;
T.~Han, S.~Padhi and S.~Su,
Phys. Rev. D \textbf{88} (2013) no.11, 115010;
A.~Arbey, M.~Battaglia and F.~Mahmoudi,
Eur. Phys. J. C \textbf{75} (2015) no.3, 108.


\bibitem{SDECAY} M. Muhlleitner, A. Djouadi  and  Y. Mambrini, Comput. Phys. Commun. {\bf 168} (2005) 46-70.

\bibitem{SUSY-HIT} A. Djouadi, M.M. Muhlleitner and M. Spira,    Acta Phys. Polon. {\bf B38} (2007) 635-644. 


\bibitem{ATLAS:2020qdt}
The ATLAS collaboration, 
ATLAS-CONF-2020-027.

\bibitem{CMS:2020gsy}
The CMS collaboration, 
CMS-PAS-HIG-19-005.


\bibitem{ATLAS:2020zms}
The ATLAS collaboration, 
Phys. Rev. Lett. \textbf{125}, no.5, 051801 (2020).

\bibitem{ATLAS:2020jqj}
The ATLAS collaboration, 
ATLAS-CONF-2020-039.

\bibitem{CMS:2022rbd}
The CMS collaboration,  CMS-PAS-HIG-21-001.

\bibitem{CMS:2019pzc}
The CMS collaboration, 
JHEP \textbf{04}, 171 (2020).

\bibitem{ATLAS:2018rvc}
The ATLAS collaboration,
Eur. Phys. J. C \textbf{78} (2018) no.7, 565. 


\bibitem{ATLAS:2021ayy}
 The ATLAS collaboration, 
ATL-PHYS-PUB-2021-030.



\bibitem{ATLAS:2018eui}
The ATLAS collaboration, 
Phys. Rev. D \textbf{98}, no.9, 092012 (2018);
 Phys. Rev. D \textbf{101}, no.5, 052005 (2020);
 Eur. Phys. J. C \textbf{81}, no.12, 1118 (2021);
 ATLAS-CONF-2021-022;
 Eur. Phys. J. C \textbf{80}, no.8, 691 (2020);
 JHEP \textbf{05}, 071 (2014);
 Eur. Phys. J. C \textbf{80}, no.2, 123 (2020);
 Phys. Rev. D \textbf{93}, no.5, 052002 (2016); 
 Eur. Phys. J. C \textbf{78}, no.12, 995 (2018);
 JHEP \textbf{10}, 096 (2014); 
Eur. Phys. J. C \textbf{78}, no.2, 154 (2018). 

\bibitem{CMS:2014kxj}
The  CMS collaboration, 
Phys. Rev. D \textbf{90}, no.9, 092007 (2014);
 JHEP \textbf{10}, 045 (2021);
arXiv:2106.14246 [hep-ex]; 
 JHEP \textbf{04}, 123 (2021);
 JHEP \textbf{06}, 077 (2011).


\bibitem{CMS-compress} The CMS collaboration, Phys. Lett. \textbf{B806}, 135502 (2020);  arXiv:2111.06296
[hep-ex]. 

\bibitem{ATLAS-compress} ATLAS collaboration, 
Phys. Rev. \textbf{D 101}, 052005 (2020);   ATLAS-CONF-2021-015  (2021); arXiv:2201.02472 [hep-ex]. 


\bibitem{Higgs-prod-decay} LHC Higgs Cross Section Working Group]:  S.~Dittmaier {\it et al.}, arXiv:1101.0593 [hep-ph]; arXiv:1201.3084 [hep-ph]; S. Heinemeyer et al., 
arXiv:1307.1347; J. Baglio and A. Djouadi, JHEP {\bf 03} (2011) 055;  JHEP {\bf 10} (2010) 064.

\bibitem{HDECAY} A. Djouadi, J. Kalinowski and M. Spira. Comput. Phys. Commun. {\bf 108} (1998) 56–74;  A. Djouadi, J. Kalinowski, M. Muehlleitner and M. Spira, Comput. Phys. Commun. {\bf 238} (2019) 214-231. 

\bibitem{Michael} Michael Spira,     Fortsch. Phys. {\bf 46} (1998) 203-284; hep-ph/9510347. See the website: http://tiger.web.psi.ch/proglist.html.

\bibitem{tt-interference} See e.g. 
A.~Djouadi, J.~Ellis, A.~Popov and J.~Quevillon,
JHEP \textbf{03} (2019) 119. 


\bibitem{Hgamma}
A.~Djouadi, V.~Driesen, W.~Hollik and J.~I.~Illana,
Eur. Phys. J. C \textbf{1} (1998), 149-162;
A. Djouadi, Eur. Phys. J. {\bf C73} (2013) 2498;
A. Djouadi, J. Quevillon and R. Vega-Morales, Phys. Lett. {\bf B757} (2016) 412.  

\bibitem{HZp} 
A.~Djouadi, V.~Driesen, W.~Hollik and A.~Kraft,
Eur. Phys. J. C \textbf{1} (1998), 163.


\bibitem{Griest:1987qv}
K.~Griest and H.~E.~Haber,
Phys.\ Rev.\ D \textbf{37}, 719 (1988).

\bibitem{Gunion:1988yc}
J.~F.~Gunion and H.~E.~Haber,
Nucl.\ Phys.\ B \textbf{307}, 445 (1988)
[erratum: Nucl. Phys. B \textbf{402}, 569 (1993)].

\bibitem{Djouadi:1992pu}
A.~Djouadi, J.~Kalinowski, and P.~M.~Zerwas,
Z.\ Phys.\ C \textbf{57}, 569-584 (1993).

\bibitem{Djouadi:1996pj}
A.~Djouadi, J.~Kalinowski, P.~Ohmann, and P.~Zerwas,
Z.\ Phys.\ C \textbf{74}, 93-111 (1997).




\bibitem{LSP-DM}
J.~R.~Ellis, J.~S.~Hagelin, D.~V.~Nanopoulos, K.~A.~Olive and M.~Srednicki,
Nucl. Phys. B \textbf{238} (1984), 453-476;
H.~Goldberg,
Phys. Rev. Lett. \textbf{50} (1983), 1419;
K.~Griest,
Phys. Rev. D \textbf{38} (1988), 2357.


\bibitem{bulk}
M.~Drees and M.~M.~Nojiri,
Phys. Rev. D \textbf{47} (1993), 376-408.

\bibitem{Planck:2018vyg}
N.~Aghanim \textit{et al.} [Planck],
Astron. Astrophys. \textbf{641} (2020), A6
[erratum: Astron. Astrophys. \textbf{652} (2021), C4].


\bibitem{coanih-stau}
J.~R.~Ellis, T.~Falk and K.~A.~Olive,
Phys. Lett. B \textbf{444} (1998), 367-372;
J.~R.~Ellis, T.~Falk, K.~A.~Olive and M.~Srednicki,
Astropart. Phys. \textbf{13} (2000), 181-213;
M.~E.~Gomez, G.~Lazarides and C.~Pallis;
Phys. Lett. B \textbf{487} (2000), 313-320;
H.~Baer, C.~Balazs and A.~Belyaev,
JHEP \textbf{03} (2002), 042.

\bibitem{coanih-stop}
C.~Boehm, A.~Djouadi and M.~Drees,
Phys. Rev. D \textbf{62} (2000), 035012;
A. Djouadi, M.  Drees and J.L. Kneur,  JHEP \textbf{08} (2001) 055;
J.~R.~Ellis, K.~A.~Olive and Y.~Santoso,
Astropart. Phys. \textbf{18} (2003), 395-432;
R.~L.~Arnowitt, B.~Dutta and Y.~Santoso,
Nucl. Phys. B \textbf{606} (2001), 59-83;
J.~Ellis, K.~A.~Olive and J.~Zheng,
Eur. Phys. J. C \textbf{74} (2014), 2947.


\bibitem{Hportal}
G.~Arcadi,
Eur. Phys. J. C \textbf{78} (2018) no.10, 864;
G.~Arcadi, M.~Dutra, P.~Ghosh, M.~Lindner, Y.~Mambrini, M.~Pierre, S.~Profumo and F.~S.~Queiroz,
Eur. Phys. J. C \textbf{78} (2018) no.3, 203;
M.~Escudero, A.~Berlin, D.~Hooper and M.~X.~Lin,
JCAP \textbf{12} (2016), 029,
L.~Calibbi, A.~Mariotti and P.~Tziveloglou,
JHEP \textbf{10} (2015), 116;
G.~Arcadi, Y.~Mambrini and F.~Richard,
JCAP \textbf{03} (2015), 018.

\bibitem{Hpole}
K.~Griest and D.~Seckel,
Phys. Rev. D \textbf{43} (1991), 3191-3203,
H. Baer, A. Belyaev, T. Krupovnickas and A.
Mustafayev, JHEP \textbf{ 0406} (2004) 044; A. Djouadi, M.
Drees and J.L. Kneur, Phys. Lett. \textbf{B624} (2005) 60.

\bibitem{limit-LSP}
A.~Hryczuk, R.~Iengo and P.~Ullio,
JHEP \textbf{03} (2011), 069;
L.~Roszkowski, E.~M.~Sessolo and A.~J.~Williams,
JHEP \textbf{02} (2015), 014;
M.~Beneke et al., 
JHEP \textbf{03} (2016), 119;
M.~Beneke et al., 
JHEP \textbf{01} (2017), 002;
L.~Roszkowski, E.~M.~Sessolo and S.~Trojanowski,
Rept. Prog. Phys. \textbf{81} (2018) no.6, 066201.



\bibitem{tempered}
N.~Arkani-Hamed, A.~Delgado and G.~F.~Giudice,
Nucl. Phys. B \textbf{741} (2006), 108-130;
H.~Baer, A.~Mustafayev, E.~K.~Park and X.~Tata,
JCAP \textbf{01} (2007), 017;
N.~Bernal, A.~Djouadi and P.~Slavich,
JHEP \textbf{07} (2007), 016;
M.~Badziak, M.~Olechowski and P.~Szczerbiak,
Phys. Lett. B \textbf{770} (2017), 226-235;
S.~Profumo, T.~Stefaniak and L.~Stephenson Haskins,
Phys. Rev. D \textbf{96} (2017) no.5, 055018.


\bibitem{f-values}
J.~M.~Alarcon, J.~Martin Camalich and J.~A.~Oller,
Annals Phys. \textbf{336} (2013), 413-461;
A.~Crivellin, M.~Hoferichter and M.~Procura,
Phys. Rev. D \textbf{89} (2014), 054021;
B.~Kubis, J.~Ruiz de Elvira, M.~Hoferichter and U.~G.~Mei\ss{}ner,
PoS \textbf{CD15} (2015), 021;
M.~Hoferichter, P.~Klos, J.~Men\'endez and A.~Schwenk,
Phys. Rev. Lett. \textbf{119} (2017) no.18, 181803.

\bibitem{LZ:2022ufs}
J.~Aalbers \textit{et al.} [LZ],
[arXiv:2207.03764 [hep-ex]].


\bibitem{XENON:2018voc}
E.~Aprile \textit{et al.} [XENON],
Phys. Rev. Lett. \textbf{121} (2018) no.11, 111302.

\bibitem{XENON:2019rxp}
E.~Aprile \textit{et al.} [XENON],
Phys. Rev. Lett. \textbf{122} (2019) no.14, 141301.


\bibitem{MicroMegas}
G.~Belanger, F.~Boudjema, A.~Pukhov and A.~Semenov,
Comput. Phys. Commun. \textbf{185} (2014), 960-985;
Comput. Phys. Commun. \textbf{149} (2002) 103-120;
G.~Belanger, F.~Boudjema, P.~Brun, A.~Pukhov, S.~Rosier-Lees, P.~Salati and A.~Semenov,
Comput. Phys. Commun. \textbf{182} (2011), 842-856.

\bibitem{XENONnT}
E.~Aprile \textit{et al.} [XENON],
JCAP \textbf{11} (2020), 031.

\bibitem{LZ}
K.~Pushkin [LZ],
Nucl. Instrum. Meth. A \textbf{936} (2019), 162-165.


\bibitem{DARWIN}
J.~Aalbers \textit{et al.} [DARWIN],
JCAP \textbf{11} (2016), 017.

\bibitem{blind-spots}
P.~Huang and C.~E.~M.~Wagner,
Phys. Rev. D \textbf{90} (2014) no.1, 015018;
J.~R.~Ellis, A.~Ferstl and K.~A.~Olive,
Phys. Rev. D \textbf{63} (2001), 065016;
H.~Baer, A.~Mustafayev, E.~K.~Park and X.~Tata,
JCAP \textbf{01} (2007), 017.

\bibitem{relicresonance}
M.~Duch and B.~Grzadkowski,
JHEP \textbf{09} (2017), 159;
T.~Binder, T.~Bringmann, M.~Gustafsson and A.~Hryczuk,
Phys. Rev. D \textbf{96} (2017) no.11, 115010.

\bibitem{atlasHB}
The ATLAS collaboration,
Phys. Rev. D \textbf{104} (2021) no.11, 112010.

\bibitem{H-invisible}
The ATLAS collaboration,
Phys. Rev. Lett. \textbf{122} (2019) no.23, 231801;
Phys. Lett. B \textbf{829} (2022), 137066.

\bibitem{PDG}
P.A.~Zyla \textit{et al.} [Particle Data Group],
PTEP \textbf{2020} (2020) no.8, 083C01.


\bibitem{Sommerfeld}
M.~Beneke, R.~Szafron and K.~Urban,
JHEP \textbf{02} (2021), 020,
A.~Hryczuk and R.~Iengo,
JHEP \textbf{01} (2012), 163. 


\bibitem{bound-state-formation}
B.~von Harling and K.~Petraki,
JCAP \textbf{12} (2014), 033,
A.~Mitridate, M.~Redi, J.~Smirnov and A.~Strumia,
JCAP \textbf{05} (2017), 006,
T.~Binder, L.~Covi and K.~Mukaida,
Phys. Rev. D \textbf{98} (2018) no.11, 115023;
J.~Harz and K.~Petraki,
JHEP \textbf{07} (2018), 096,
T.~Binder, B.~Blobel, J.~Harz and K.~Mukaida,
JHEP \textbf{09} (2020), 086.

\bibitem{fermi}
M.~L.~Ahnen \textit{et al.} [MAGIC and Fermi-LAT],
JCAP \textbf{02} (2016), 039;
S.~Hoof, A.~Geringer-Sameth and R.~Trotta,
JCAP \textbf{02} (2020), 012.


\bibitem{benekeID}
M.~Beneke et al., 
JHEP \textbf{01} (2017), 002.


\bibitem{atlasWZ}
The ATLAS collaboration,
Phys. Rev. D \textbf{104} (2021) no.11, 112010; 
arXiv:2204.13072.


\end{thebibliography}
\end{document}